\documentclass[12pt]{article}%
\usepackage{amsmath}
\usepackage{amsfonts}
\usepackage{amssymb}
\usepackage{graphicx}%
\providecommand{\U}[1]{\protect\rule{.1in}{.1in}}

\begin{document}

\title{Gravitation and General Relativity at King's College London}
\author{D C Robinson\\Mathematics Department\\King's College London\\Strand, London WC2R 2LS\\United Kingdom\\david.c.robinson@kcl.ac.uk}
\maketitle

ABSTRACT: \ This essay concerns the study of gravitation and general
relativity at King's College London (KCL). \ It covers developments since the
nineteenth century but its main focus is on the quarter of a century beginning
in 1955. \ At King's research in the twenty-five years from 1955 was dominated
initially by the study of gravitational waves and then by the investigation of
the classical and quantum aspects of black holes. \ While general relativity
has been studied extensively by both physicists and mathematicians, most of
the work at King's described here was undertaken in the mathematics
department.\newpage

\section{Introduction}

This essay is an account of the study of gravitation and general relativity at
King's College London (KCL) and the contributions by mathematicians and
physicists who were, at one time or another, associated with the college. \ It
covers a period of about 150 years from the nineteenth century until the last
quarter of the twentieth century. \ Beginning with a brief account of the
foundation of the College in the 19th century and the establishment of the
professorships of mathematics and natural science it concludes in the early
1980's when both the College and the theoretical physics research undertaken
there were changing.

During the nineteenth century the College was small and often struggled.
\ Nevertheless there were a number of people associated with it who, in
different ways, made memorable contributions to the development of our
understanding of gravity. \ These include, as a professor, one of the greatest
theoretical physicists, James Clerk Maxwell and, as a schoolboy, the
outstanding mathematician William Kingdon Clifford. \ Others who at some time
were members of the College also contributed to nineteenth century
astronomical and gravitational physics. \ Aspects of their work are briefly discussed.

The early twentieth century saw the College more firmly established in terms
of both student numbers and its financial position. \ In addition to the two
founding colleges, University College and King's, new colleges in the
University of London were created and intercollegiate activity became
increasingly important. \ Departments at King's were re-organized; mathematics
and physics departments were introduced and the physics department in
particular attracted some outstanding people. \ By the time Einstein visited
the College in 1921 the popular press had published articles about his new
theory of gravity and his visit was a notable event. \ Although the general
theory of relativity was still little understood by most physicists it was
already being studied at King's; William Wilson in the physics department had
published work on the theory in 1918.

Subsequently, and for most of the twentieth century, research at King's on
general relativity was carried in the mathematics department beginning with
the work of George Jeffery in the 1920s. \ \ People working on the subject
would have been appointed as applied mathematicians but that term was broadly
interpreted and included mathematical or theoretical physics and cosmology.
\ Most of those working on general relativity at King's in the inter-war years
also carried out research on other topics. \ Indeed they often did their most
important work in other areas. \ Their gravitational research, even if it was
not always of enduring significance, ensured a continuity of interest in
general relativity in the College. \ This was not without importance,
particularly in the period, from about 1930 to 1950, when general relativity
was little studied and exciting developments in quantum mechanics were of
central interest to theoretical physicists. \ Jeffery became noted for his
early translations of important relativity papers and his research on plane
gravitational waves. \ George Temple and George McVittie, whose work included
research on cosmology and general relativity, came to the mathematics
department in the 1930s but spent years away during the second world war.
\ For them, as for others, the war disrupted everything. \ Many people were
absent from King's for long periods and for a time the College had to be
evacuated from London.

In the decade after the end of the war there was a slow but steady renewal of
interest in general relativity. \ New centres of research into the subject
were established and one of these was at King's. \ In 1954 Hermann Bondi
arrived to replace Temple as the professor of applied mathematics and he
quickly formed a research group consisting initially of himself, Clive
Kilmister who was already at King's and Felix Pirani who arrived the following
year. \ The work of the relativity group at King's, from Bondi's arrival until
the beginning of the 1970s, forms the central part of this essay.

Formation of a group devoted to the study of gravitation was a new and major
development for the still small mathematics department. \ It coincided with
the end of a long quiet period in the study of general relativity. The Berne
conference in 1955 marked the end of that period while the renewal of the
subject was celebrated by the Chapel Hill conference in 1957. \ Both meetings
were important for Bondi and Pirani and they had a big influence on the way
the King's group developed.

Bondi was a man of many parts and his influence was widely felt, not only in
the expanding world of general relativity, but also in the wider scientific
and political community. \ During his time at King's the general relativity
group became an important international research centre attracting many short
and long term visitors. \ In the Bondi years old uncertainties about
gravitational radiation were resolved and new and fruitful insights were
developed. \ Work on gravitational radiation goes back to Einstein and his
foundational work on the subject in 1916 and 1918. \ However in subsequent
years confusion about the existence of gravitational waves arose. The major
advances in the study of gravitational waves at King's (and elsewhere of
course) were made by Bondi, Pirani, their students and a number of visitors
and research associates. \ The latter included Andrzej Trautman, Ivor
Robinson, Ray Sachs, Josh Goldberg, Roger Penrose and Ted Newman.

The 1960s saw the discovery of quasars, the cosmic black body radiation,
pulsars and the naissance of experimental work on gravity. \ These brought
general relativity back in from the cold, slowly but steadily back into the
mainstream of astrophysics and physics. \ The discovery of quasars and the
black body radiation sounded, for most, the death knell for the steady state
theory of the universe. Bondi, Pirani and others at King's who had been
supporters of the theory had regarded the fact that the theory made
predictions which could be disproved as one of its strengths so they quietly
accepted this. \ The work on gravitational radiation and these new
developments led into the radically new work of Roger Penrose on the global
structure of space-time, gravitational collapse and space-time singularities.
\ Together with Stephen Hawking, Robert Geroch and others Penrose introduced
the innovative use of topology and modern differential geometry into
relativity. \ A distinctly post-Einstein era had arrived.

The new astrophysical observations encouraged the study of black holes.
\ Vigorously promoted by John Wheeler at Princeton, they became a topic of
central interest in general relativity. \ The group at King's responded to the
new theoretical developments in different ways, as did relativists in general.
Some were enthusiastic about them, others had reservations. \ As the 1960s
progressed while Bondi become increasingly involved in external activities the
relativity group expanded. \ Three younger people, Peter Szekeres, Michael
Crampin and Ray d'Inverno, took up mathematics lectureships in the second half
of the 1960s and pursued research on gravitational waves, applications of
modern differential geometry and algebraic computing applied to general relativity.

There were further changes to the King's group in the early 1970s. \ Between
1970 and 1973 Szekeres, Crampin and d'Inverno left to be replaced in the
relativity group by myself and Paul Davies. \ Bondi left in 1971, although he
retained formal links with the College. \ He was replaced by John G.Taylor
whose primary research area then was quantum field theory. \ He also worked on
neural networks and subsequently on supersymmetry and string theory. \ Chris
Isham, a quantum field theorist with active interests in general relativity,
joined the department in 1973 and contributed significantly to the relativity group.

A second period of vigorous activity at King's took place in the 1970s. \ Many
in the King's group became centrally concerned with the theory of black holes
and particularly with the new insights being produced by Stephen Hawking and
others. \ Research was undertaken into both the classical regime, with
astrophysical bodies in mind, and the quantum mechanical regime. \ I was
involved in research on uniqueness theorems for classical black holes.
\ Davies' and Isham's work mainly concerned studies of quantum field theory in
curved space-times and other aspects of quantum field theory and gravity. As
in the 1960s there were post-docs and visitors who made major contributions to
these lines of research. \ They included Henning M\"{u}ller zum Hagen, Steve
Fulling, Steve Christensen, Mike Duff, Stanley Deser and Larry Ford.
\ Research students also did important work, particularly in the newly
developing area of quantum field theory in curved space-times.

By the early 1980s Isham and Davies had left King's. \ While work on general
relativity and topics such as Penrose's twistor theory, pursued at King's by
Stephen Huggett and Andrew Hodges, continued the dominant theoretical physics
research interest at King's began to move towards supersymmetry, supergravity
and then string theory.. \ Activity in supersymmetry and string theory at
King's had begun in the 1970s and in support of this Peter West was appointed
to a mathematics lectureship in 1978. \ He and Kellog Stelle, at King's for a
year in 1977-78, undertook pioneering work on these topics. \ Research at
King's in these areas is a different story and it will be only touched upon in
this essay. \ In the mid 1980s a reorganization of the University of London
led to major changes in the mathematics and other departments and Pirani and
Kilmister retired.

This essay about work on gravity and general relativity is generally, although
not always strictly, chronologically ordered. \ Attention centres on the work
done at King's but other related research is often discussed. \ More complete
accounts of the latter can be found in the cited reviews and books. \ The
period 1955 to 1965 was such an interesting and innovative time, at King's and
elsewhere, that the research done then is discussed in more detail, both
technical and non-technical, than the work done in other periods. \ Since I
was a member of the King's relativity group from 1970 onwards my account of
the period should be regarded as being, in part at least, my personal reminiscences.

I have retained the conventions and notations of the original works as far as
possible. \ They differ from paper to paper but little beyond metrics is
displayed here so that should not be a problem. \ Lifetime dates are given
only for people who worked or studied at King's.

\section{The nineteenth century}

\subsection{The College}

King's College London was inaugurated in 1828 with the support of King George
\textrm{IV}, various Church of England bishops, clergy and laity, and
Hanovarian politicians such as the Duke of Wellington and Sir Robert Peel. \ A
Royal Charter of Incorporation was acquired in 1829.

At the beginning of the nineteenth century there were only eight universities
in the British Isles and it was increasingly felt that a large metropolis such
as London should have its own university. \ As a consequence University
College London (as it is now known) came into being. In a reaction to both the
creation of that avowedly secular institution and the turbulent times
conservative religious and political leaders supported the establishment of a
second university in London to be named in honour of the King. \ This new
institution was to be divided into a senior department and a junior
department. \ The latter, also known as King's College School, moved to
Wimbledon in 1897 and eventually ceased to be associated with the College.

The College opened on the Strand in October 8, 1831. \ Apart from a period
during the Second World War when most of the staff and students were evacuated
to Bristol, Birmingham or, in the case of the medical school, to Glasgow
(Huelin 1978), the Strand has always been the main College site and
mathematics and physics have always been located there. \ The governors and
most of the professors had to be members of the Church of England but in
practice King's functioned, as far as its teaching and students were
concerned, in a manner not unlike that of University College. \ Regarded by
some nineteenth century observers as an Anglican backwater the College was not
as socially reactionary as that might suggest. \ It pioneered evening classes
to artisans and others, Thomas Hardy being one of the beneficiaries (Hearnshaw
1929). \ In addition members of its staff were instrumental in the founding of
the first British College, Queen's College, expressly for the education of
women. \ However, religious tests on staff remained for a long time and were
not finally abolished, except within the theology department, until the King's
College London Act of 1903.

The first professors were appointed in 1830, one in each of classics,
mathematics, English and history. \ A further eight, in financially less
advantageous subjects, were appointed subsequently. \ The Rev. Thomas G. Hall
(1803-1881), who had been fifth wrangler\footnote{The results of the Cambridge
Mathematical Tripos were published in order of merit from 1753 until 1909.
\ They were divided by degree class, Wranglers were in the top class.} at
Cambridge in 1824 and was a fellow and tutor of Magdalene College Cambridge,
was appointed to the chair of mathematics and held this position until
1869\footnote{The Professor of Medieval History, Fossey J.C. Hearnshaw, in his
centenary history of King's, wrote that Hall "continued modestly, faithfully
and inconspicuously to occupy (rather than fill) for the next thirty-nine
years" his position. (Hearnshaw 1929). \ However it should be noted that when
Queen's College opened in 1848 Hall was one of the early lecturers, his course
in mathematics being a novelty for women at that time (Rice 1996).}. \ For a
few years Hall also had to deliver some history lectures. \ Approximately
seven months later, early in 1831, the first professor of natural and
experimental philosophy, including astronomy, was appointed. The professor was
the Rev. Henry Moseley (1801-1872), a thirty year-old Cambridge mathematician
who had been seventh wrangler in 1826. \ Moseley, much more involved in
original work than Hall, was particularly interested in mechanics and such
problems as the stability of ships \ While at King's his teaching included
lectures on astronomy which were eventually published in a number of editions
(Moseley 1839). \ He also wrote on astro-theology. \ Moseley held the Chair
until 1844 and also acted as College chaplain from 1831 to 1833.

Each year the courses and costs, in guineas and shillings, were listed in the
King's College calendar\footnote{\ The King's College Calendars and the Annual
Reports of the Delegacy presented to the Court and the Senate of the
University of London are the sources of College and departmental data..
\par
There are two extended accounts of the history of the College (Hearnshaw 1929,
Huelin 1978). \ \ A survey can be found online in the Wikipedia History of
King's College London.}. \ For instance the 1834-35 calendar records that

\begin{quotation}
The Rev.T.G. Hall M.A. of Magdalene College Cambridge will commence a morning
course of lectures in the first week of October and will continue them for
five days in the week during each term. \ They will embrace all the branches
of mathematics usually taught in the Universities, and in the order which may
be found most expedient. \ The fee to Occasional Students is 4\textit{l}.4s
for each of the three terms or 10\textit{l}.10s if the student enters for the
whole year.
\end{quotation}

and

\begin{quotation}
the Rev.H. Moseley, M.A. of St. John's College, Cambridge, will, on Tuesday,
the 7th of October, commence an afternoon course of lectures. \ Fee for the
course 2\textit{l}.2s.
\end{quotation}

At Moseley's request experimental philosophy was given a separate chair in
1834, the first occupant being Charles Wheatstone. \ His first course,
advertised for 1835, was to be on the philosophy of sound and the fee was only
1$l$.1s.

Hall's three year mathematics syllabus included the study of Newton's
Principia, sections 1, 2, 3, 9 and 11, hydrostatics, optics and astronomy.
\ Mathematics was one of the subjects in the general course of systematic
study but natural philosophy was not and depended on occasional students for
its continuance. \ As holders of professorships had no guaranteed salaries but
were remunerated by a portion (usually three quarters) of the fees received
from their pupils\textit{ }(Hearnshaw 1929) this caused some difficulties
between Hall and Moseley (Rice 1996). \ However the establishment of civil
engineering in 1838, with engineering students required to take courses in
natural philosophy, eased Moseley's situation. \ In 1854 the then Principal,
the Rev. R. W. Jelf, a specialist on the approaching end of the world and the
eternity of future punishment, suggested to the College Council that the
causes of unity and economics could be advanced by amalgamating the
professorship of natural philosophy with that of manufacturing, art and
machinery. \ Amalgamation took place to be followed by separation again in
July 1860.

Initially neither King's nor University College could award degrees but in
1836 the University of London was created as an examining body with the right
to confer degrees on students of both Colleges and other approved
institutions. \ By the late nineteenth century its responsibilities were not
merely metropolitan but indeed imperial (Thompson 1990). \ With the University
of London Act of 1898 this degree-granting body was formally acknowledged as a
teaching institution with both internal and external examiners. \ Within a few
years it was the largest university, in student numbers, in the British Isles.

The relationship between its constituent colleges and the University of London
has changed a number of times over the years as indeed has the constitution
and governance of King's. \ Today the Colleges of the University effectively
function as large autonomous universities but until comparatively recently
most were fairly small. \ Intercollegiate teaching and examining declined in
stages from the 1960's onward but research interactions remain important.

Over the years King's has had its ups and downs, as have many academic
institutions. \ During the nineteenth century it often struggled to attract
students to its senior department and on occasion it had serious difficulty
paying its bills. \ Indeed towards the end of the nineteenth century it
imposed a tax on its staff because of financial difficulties; appeals were
allowed and were made. \ Nevertheless it has had sufficient academic strength
to survive such challenges and it has been the home of much excellent original
research, scholarship, investigative studies and pedagogy. \ It has been
fortunate to attract some figures of outstanding stature, including a number
who have had an important influence on our understanding of gravitation.

\subsection{Gravitation and Astronomy at King's}

Throughout the nineteenth century science in the British Isles was dominated
by the University of Cambridge and its graduates. \ Passing the Mathematical
Tripos examination as a highly placed Wrangler ensured a leading place in
academia or professions such as law. \ However the examination did not
encourage original thinking or the development of a research ethos in British
Universities.\ \ Despite its dead hand, applied mathematics and physics in the
British Isles flourished but research in pure mathematics\ was comparatively
weak (Gray 2006; Rice 2006). \ The majority of students in the senior
department at King's were either preparing themselves for professional
examinations or for entry to Oxford or Cambridge. With some notable exceptions
original contributions in the nineteenth century to mathematics and physics by
those who were, at some time, members of the College cannot be said to have
been overly distinguished. \ The exceptions include contributions related to
gravitation and the development of relativity, in particular those made by
James Clerk Maxwell (1831-1879).

Aged 29 Maxwell was appointed as the fourth Professor of Natural philosophy
and Astronomy in 1860\textit{. \ }Maxwell's time at King's has been recorded
in numerous biographies from the first, coauthored by his friend Lewis
Campbell (Campbell and Garnett 1882) to more recent ones e.g. (Tolstoy 1981;
Mahon 2003). \ Reports by physicists detailing Maxwell's teaching and research
activities at King's are contained in an insightful article by Cyril Domb
(Domb 1980) and a lecture by John Randall (Randall 1963), the latter being one
of a series given at King's in commemoration of Maxwell's tenure of his chair
there (Domb 1963). \ Maxwell's chair was in the department of applied sciences
described as providing a general education, of a practical nature, suitable
for young men intending to be engaged in commercial and agricultural pursuits,
professional employments such as civil and military engineering, surveying,
architecture and the higher branches of manufacturing arts. \ Maxwell was a
conscientious, well organized and thoughtful teacher although, according to
some accounts, not a great one. \ His teaching and scientific work tended to
occupy his mornings. \ One evening a week, as part of his duties, he gave
lectures to working men. \ Maxwell's courses were up-to-date, covering
mechanics, optics, electricity and magnetism and gravitation. \ \ He also
conducted experiments, assisted by his wife, at home. \ His time at King's is
particularly noteworthy because it was then that he produced his most
important scientific papers. \ As William Niven, editor of Maxwell's collected
scientific papers (Niven 1890), writes in the preface,

\begin{quotation}
Maxwell was a professor at King's College from 1860 to 1865 and this period of
his life is distinguished by the production of his most important papers.
\end{quotation}

Maxwell had begun to study the work of Michael Faraday while he was at
Cambridge. \ He aimed to obtain a mathematical formulation of electricity and
magnetism that incorporated Faraday's ideas and results. \ In the resulting
paper "On Faraday's Lines of Force" (Maxwell 1856) he moved away from the
approach, established since the time of Newton, of forces acting at a
distance. \ \ After his move to London in 1860 Maxwell followed up this work
with two papers (Maxwell 1861; 1865). \ In the first, published in four parts
in the Philosophical Magazine, he introduced a mechanical model with electric
and magnetic energy residing in an elastic vortex medium ether. \ It suggested
to him the concept of a "displacement current" and the modification of
Amp\`{e}re's law. \ He deduced that, in his own italized words \textit{Light
consists in the transverse undulations of the same medium which is the cause
of electric and magnetic phenomena }(Tolstoy 1981).

In his next paper on the subject "A Dynamical Theory of the Electromagnetic
Field", read to the Royal Society in 1864 and published in 1865, Maxwell left
behind his complicated model of 1861. \ He produced a mathematical theory that
demonstrated how the electromagnetic theory of light followed simply from his
field equations. \ He wrote that this was "a theory of the
\textit{Electromagnetic Field} because it has to do with the space in the
neighbourhood of the electric or magnetic bodies, and it may be called a
\textit{Dynamical Theory, }because it assumes that in the space there is
matter in motion, by which the observed electromagnetic phenomena are
produced". \ For Maxwell the energy of electromagnetic phenomena was now
mechanical energy residing in the electromagnetic field both in electrified
and magnetic bodies and in the space surrounding them (Tolstoy 1981).

Maxwell's two great papers on electricity and magnetism not only laid the
foundations of electromagnetic theory as we know it today but also changed the
point of view which had dominated since the times of Isaac Newton. They
replaced the concept of forces acting at a distance with the notion of forces
transmitted via fields. \ They demonstrated that light was an electromagnetic
phenomenon. \ They demonstrated that electromagnetic waves existed, a
conclusion verified experimentally by Heinrich Hertz in 1888, were transverse
and moved with the speed of light. \ The speed of light now became of
universal significance. \ Maxwell's work initiated the desire for the
unification of physical forces which persist to this day.

Some years earlier Maxwell had discussed with Faraday the latter's suggestion
that gravity could be mediated by lines of force and Maxwell had found these
ideas exciting. \ Before concluding "A Dynamical Theory of the Electromagnetic
Field" he turned his attention to gravitation. \ In a brief section, entitled
"Note on the Attraction of Gravitation" Maxwell commented on the similarities
between the inverse square laws of Newtonian gravity and those of
electromagnetism, and the differences - gravity is attractive whereas unlike
charges attract but like charges repel. \ In his own words

\begin{quotation}
After tracing to the action of the surrounding medium both the magnetic and
the electric attractions and repulsions, and finding them to depend on the
inverse square of the distance, we are naturally led to inquire whether the
attraction of gravitation, which follows the same law of the distance, is not
also traceable to the action of a surrounding medium.

Gravitation differs from magnetism and electricity in this; that the bodies
concerned are all of the same kind, instead of being of opposite signs, like
magnetic poles and electrified bodies, and that the force between the bodies
is an attraction and not a repulsion, as is the case between like electric and
magnetic bodies.
\end{quotation}

Applying his field theory formulation of electromagnetism to gravity, making
the appropriate sign changes, he saw that this led to a negative energy
density for the gravitational field. \ After a brief computation, using the
analogy with two magnetic poles, Maxwell arrived at an expression for the
intrinsic energy $E$ of the field surrounding two gravitating bodies:
$E=C-\sum\frac{1}{8\pi}R^{2}dV$, where $C$ is a constant and $R$ is, as he put
it, the resultant gravitating force. \ He observed that the intrinsic energy
of gravitation must therefore be less when ever there is a resultant
gravitating force. \ With his model of energy being carried by stresses and
strains in the luminiferous ether he could not see how such a medium could
have negative energy. \ The minus sign convinced Maxwell that he could not
construct an acceptable field formulation for gravity analogous to that for
electromagnetism\footnote{Subsequently Oliver Heaviside, to whom the modern
formulation of Maxwell's equations are due, investigated a Maxwellian vector
formulation of gravity and it is still occasionally discussed (Havas 1979).}.
\ He ended this section by writing,

\begin{quotation}
As energy is essentially positive, it is impossible for any part of space to
have negative intrinsic energy. \ Hence those parts of space in which there is
no resultant force, such as the points of equilibrium in the space between the
different bodies of a system, and within the substance of each body, must have
an intrinsic energy per unit of volume greater than $\frac{1}{8\pi}R^{2}$,
where $R$ is the greatest possible value of the intensity of gravitating force
in any part of the universe.

The assumption, therefore, that gravitation arises from the action of the
surrounding medium in the way pointed out, leads to the conclusion that every
part of this medium possesses, when undisturbed, an enormous intrinsic energy,
and that the presence of dense bodies influences the medium so as to diminish
this energy wherever there is a resultant attraction.

As I am unable to understand in what way a medium can possess such properties,
I cannot go any further in this direction in searching for the cause of gravitation.
\end{quotation}

As is well known, Maxwell's work exerted a major influence on Albert Einstein
in his formulation of the special theory of relativity and consequently
general relativity. \ Einstein acknowledged his debt to Maxwell on a number of
occasions. \ He wrote in his autobiographical notes (Einstein 1949)

\begin{quotation}
The most fascinating subject at the time that I was a student was Maxwell's
theory. \ What made this theory appear revolutionary was the transition from
forces at a distance to fields as fundamental variables. The incorporation of
optics into the theory of electromagnetism, with its relation of the speed of
light to the electric and magnetic absolute system of units as well as the
relation of the refraction coefficient to the dielectric constant, the
qualitative relation between the reflection coefficient and the metallic
conductivity of the body - it was like a revelation. \ Aside from the
transition to field theory, i.e. the expression of the elementary \ laws
through differential equations, Maxwell needed only one single hypothetical
step - the introduction of the electric displacement current in the vacuum and
in the dielectrica and its magnetic effect, an innovation which was almost
prescribed by the formal properties of the differential equations....
\end{quotation}

Einstein also remarked on the similarity of the pair Faraday-Maxwell to the
pair Galileo-Newton - Faraday and Galileo grasping relations intuitively,
Maxwell and Newton exactly formulating and quantitatively applying them
(Schilpp 1970).

Maxwell resigned his professorship in order to return to Glenlair, his estate
in Scotland, early in 1865, although he agreed to return to London to teach
his evening classes the following winter (Mahon 2003; Domb 1963). \ \ By that
time the total King's student body, including occasional students and the
school, numbered 1490. \ William Grylls Adams (1836-1915), younger brother of
the astronomer John Couch Adams whose papers he coedited, had been appointed
to assist with the teaching of natural philosophy in 1863 and he was appointed
as Maxwell's successor. In the annual report presented to the Court of
Governors and Proprietors in April 1865 it is recorded that

\begin{quotation}
The Department of Applied Sciences is undoubtedly prospering.... The Council
regrets to state that J.C.Maxwell Esq. the distinguished Professor of Natural
Philosophy has resigned his office. \ They have elected Mr William Grylls
Adams, Fellow of St John's College, Cambridge, late lecturer on the same
subject, to the office of Professor. \ The lectureship thus vacated will not
be filled up; Professor Adams undertaking to attend 5 days in the week instead
of three, and then to do the whole work unaided\textit{.}
\end{quotation}

Towards the end of the nineteenth century faculties began to be modernized and
departments were re-organized. \ When Adams retired in 1905 natural philosophy
was split into physics, with its own professorship, and applied mathematics
which became the responsibility of the professor of mathematics (Rice 1996).

Neither Grylls Adams, who had a significant career\footnote{In 1876, Adams and
his student Richard Evans Day showed that illuminating a junction between
selenium and platinum had a photovoltaic effect. Such production of
electricity from light led to the modern solar cell.}, nor any other
nineteenth century King's (or, arguably, British) mathematician or physicist
matched James Clerk Maxwell's achievements. \ However, as far as gravitation
and astronomy are concerned, three people who were associated with King's
should be briefly mentioned although their contributions were made elsewhere.

The first is William Kingdon Clifford (1845-1879) who at 15 won a mathematical
and classical scholarship to the department of general literature and science
at King's. \ While at King's, from 1860 until 1863, he excelled in
mathematics, classics, English literature and gymnastics and published his
first mathematical paper (Clifford 1863). \ He went to Cambridge from King's,
a not unusual progression in those times. \ While there he participated in the
December 1870 observations of the solar eclipse in Sicily. \ A notable feature
of this expedition was that the ship on which the expedition was travelling
from Naples, the \textit{Psyche}, was wrecked before it reached landfall on
Sicily. \ Everyone reached land safely. \ Clifford was a member of the group
led by Grylls Adams and was one of those responsible for the weather affected
polarization observations (Adams 1871). \ Possibly his thoughts about the
geometry of space encouraged him to participate in this expedition (Galindo
and Cervantes-Cota 2018). \ Earlier, in March of that year, Clifford had read
a brief paper to the Cambridge Philosophical Society. \ Influenced by work on
non-Euclidean geometry, particularly that of Bernhard Riemann some of whose
work he translated into English (Clifford 1873), Clifford had speculated about
the nature of space (Clifford 1876). \ In his paper he wrote:

\begin{quotation}
I hold in fact

(1) That small portions of space \textit{are} in fact of a nature analogous to
little hills on a surface which is on the average flat; namely, that the
ordinary laws of geometry are not valid in them.

(2) That this property of being curved or distorted is continually being
passed on from one portion of space to another after the manner of a wave.

(3) That this variation of the curvature of space is what really happens in
that phenomenon which we call the\textit{ motion of matter}, whether
ponderable or ethereal.

(4) That in the physical world nothing else takes place but this variation,
subject (possibly) to the laws of continuity.
\end{quotation}

These observations now seem remarkably prescient. However to go beyond the
late nineteenth and early twentieth century speculations about the physical
role of non-Euclidean geometries (Kragh 2012) it would require Albert
Einstein, and his incorporation of both space and time into a single
geometrical and dynamical entity, before these qualitative remarks were given
substantive quantitative form.

Karl Pearson (1857-1936), better known for his career at University College
London and his role in establishing mathematical statistics, deputized for the
professor of mathematics at King's in 1881. \ At that time his interests were
broad and included philosophy, physics and applied mathematics. \ Possibly as
a consequence of completing, after Clifford's early death, the unfinished
volume of the latter's "The Common Sense of the Exact Sciences"\textit{
}(Clifford 1885)\textit{ }Pearson wrote "The Grammar of Science" (Pearson
1892). \ This was one of the first books Albert Einstein recommended in 1902
to his friends in their little discussion group in Berne, the so-called
Akademie Olympia. \ The book included discussions of a number of topics which
were important to Einstein in his development of the theories of relativity.

The third man was Ralph Allen Sampson (1866-1939) who was a lecturer in
mathematics from 1889 until 1891. \ At King's he worked mainly on
hydrodynamics but after he returned to Cambridge in 1891 he was occupied with
\ astronomical spectroscopy. In his early days he advanced the hypothesis \ of
radiative equilibrium in a star's interior. \ Subsequently he developed a
dynamical theory of the four largest satellites of Jupiter in work which was
significant in its time. \ He became Astronomer Royal of Scotland in 1910.
\ Like Grylls Adams he was one of the editors of the papers of his Cambridge
tutor John Couch Adams (Greaves 1940). \ The crater Sampson on the moon is
named after him.

\section{The first half of the twentieth century}

In 1900 the University of London became a federal institution and its two
founding Colleges, University and King's Colleges, became Schools of the
University with King's becoming completely incorporated in 1908. \ Upon Grylls
Adams' retirement in 1905 his chair was offered to Ernest Rutherford, then at
McGill University in Canada. \ However he was told by J\ J.\ Thompson that
funds and facilities at King's were inadequate so he declined and eventually
went to Manchester (Huelin 1978). \ Over time the financial position of the
College improved and in the next quarter of a century the physics department
attracted figures like the Nobel prize winners George Barkla, Owen Richardson
and Edward Appleton.

Student numbers had remained small; in 1917-18 there were just 1,775 King's
students. \ However the end of the first World War saw the student body
boosted to 3,879 by the 1919-1920 session. \ By that time the nineteenth
century world view of space and time had been radically altered in the
space-time theory of special relativity. \ Newtonian dynamics had been
modified and unified with Maxwellian electrodynamics. \ Furthermore, on
November 25, 1915 Einstein had been able to bring to a successful conclusion,
in a final whirlwind of activity, his attempts to formulate a new relativistic
theory of gravitation and its field equations (Einstein 1915). \ The action at
a distance Newtonian gravity was now the non-relativistic limit of Einstein's
field theory of general relativity.

In 1921, when only a few British physicists or mathematicians had mastered his
new ideas on gravitation, Einstein outlined the theory's development for
Nature (Einstein 1921a)\textit{. \ }Acknowledging the role of Faraday and
Maxwell in his thinking he wrote:

\begin{quotation}
The entire development starts off from, and is dominated by, the idea of
Faraday and Maxwell, according to which all physical processes involve a
continuity of action (as opposed to action at a distance), or, in the language
of mathematics, they are expressed by partial differential equations.
\end{quotation}

Einstein visited England for the first time that year. \ This was more than
just a scientific event. \ The febrile post World War One atmosphere of the
time meant it had considerable political significance. \ After the famous 1919
Royal Society and Royal Astronomical Society expeditions to Sobral and
Principe (Dyson 1920), resulting in solar eclipse measurements supporting the
predictions of Einstein's theory of gravity about the deflection of light,
general relativity had become a topic of widespread interest, even for the
popular press. \ Einstein's visit included two lectures, one at Manchester
University and then one at King's (Clark 1973). \ The lectures were reported
in an unsigned article in Nature (Anon. 1921). \ Einstein's London host was
the Liberal politician and educationalist Viscount Richard Haldane; his talk
at King's was arranged by the College Principal Ernest Barker and the
professor of German Henry Atkins. \ On 13 June Einstein delivered his lecture
at King's, in German and without notes, to a capacity audience in the Great
Hall. \ It was entitled "The Development and Present Position of the Theory of
Relativity" (Einstein 1921b). \ With memories of the First World War still raw
Einstein began tactfully

\begin{quotation}
It is a special joy for me to be able to speak in the capital of the country
from where the most important basic ideas of theoretical physics were brought
into the world. \ I think of the theories of the motion of masses and of
gravitation, which Newton gave us, and of the concept of the electromagnetic
field by Faraday and Maxwell, which provided physics with a new foundation.
One may well say that the theory of relativity brought a kind of conclusion to
Maxwell's and Lorentz's grand framework of ideas by trying to extend the
physics of fields to all of its phenomena, gravitation included.
\end{quotation}

\ Einstein went on to emphasize that the theory had no speculative origin but
aimed to adapt theory to physical facts. \ He then outlined the basic
principles underlying both the special and general theory of relativity and
pointed out the consequent changes to the notions of space, time, gravitation
and geometry. \ Einstein ended his talk by briefly outlining ideas of Ernst
Mach about the origin of the inertia of bodies and by asserting that their
incorporation into general relativity could be achieved if the universe was
spatially closed. \ The talk was warmly received and enthusiastically reported
by the national press. \ However some British scientists, like Ralph Sampson
who was present at the lecture, although impressed were apparently still
worried about what they thought were extraneous metaphysical overtones (Clark
1973). \ To commemorate his visit Einstein donated copies of about fifty of
his papers, written over the previous twenty years, to the College.

Albert Einstein's ground breaking papers in the second decade of the twentieth
century on general relativity and its consequences were forcefully promoted in
Britain, during the years immediately after the first World War, by Arthur
Eddington. \ Eddington was by far the most influential British worker in this
area between the two world wars, particularly in the 1920's (S\'{a}nchez-Ron
1992). \ Most of the people at King's who wrote papers on general relativity
before the middle of the century had been taught, or influenced by him, at
Cambridge. \ His book "The Mathematical Theory of Relativity" (Eddington 1923)
was the standard reference work, in Britain at least. \ This period was
dominated by the new quantum physics and the contribution of other British
physicists and mathematicians to the development of general relativity was not
large. \ In addition, in comparison with continental Europe and the United
States, British research on differential geometry was not strong and made
little contribution to the mathematics relevant to the theory. \ On the whole,
cosmology and astrophysics apart, the study of relativity was just one aspect
of most interested people's research. \ This was certainly the situation at
King's where most of the small number of people interested in general
relativity did their important work in other areas. \ While it cannot be
claimed that this was a time when major advances occurred at King's some
people did make contributions which are still significant today.

Despite the recent war, and the newness of the theory, research into general
relativity had already commenced at King's by the time of Einstein's 1921
visit. \ William Wilson (1875-1965) had been appointed to an assistant
lectureship in the King's physics department in 1906 and remained in the
department until 1921 (Temple and Flint 1967). \ Wilson came to King's after
spending time studying mathematics and physics at the University of Leipzig.
\ There he obtained a thorough grounding in Hamiltonian mechanics from Carl
Neumann. \ Possibly most widely known for the Wilson-Sommerfeld quantum
conditions of the old quantum theory, independently discovered by Wilson in
1915-16 and soon after by Arnold Sommerfeld, and for extending them to include
the effect of an electromagnetic potential, Wilson had broad interests and
these included general relativity. \ In 1918 he delivered a paper to the
Physical Society of London\footnote{A forerunner of the Institute of Physics.}
(Wilson 1918) in which he showed how the equations of motion of a particle in
special relativity, or as he termed it, Minkowskian or the old relativity,
could be put in Hamiltonian form. \ He then demonstrated that it was a simple
matter to generalize this result to the new or general relativity. \ This was
one of the earliest papers fleshing out the structure of the new general
theory of relativity. After leaving King's to take up a professorship in the
University of London at Bedford College for Women (later Bedford College),
Wilson continued to have an interest in relativity. \ His paper "Relativity
and wave mechanics" (Wilson 1928) included an interpretation of the equations
of motion of charged particles in four dimensions as geodesics in five
dimensions. \ He noted that after he had developed this point of view his
attention had been drawn to the (now well-known) five dimensional formulation
of Oskar Klein (Klein 1928) of which he had been unaware.

After Wilson, and for most of the rest of the twentieth century, research into
general relativity at King's was carried out in\ the mathematics department.
\ By 1922, when George Barker Jeffery (1891-1957) was appointed as the sixth
professor of mathematics, the other academic members of staff numbered six and
included another professor, of pure mathematics, and the College secretary
Sydney Shovelton. \ Jeffery had been a student at the King's College school -
the Strand School as it was known by then. \ He was an applied mathematician
with a predilection for exact solutions and he worked on a range of subjects
including elasticity, hydrodynamics and general relativity. \ He was also
considered to be an inspiring teacher. \ \ Like Eddington, Jeffery was a
Quaker. \ He had been briefly imprisoned as a conscientious objector during
the Great War. \ By the time he came to King's from University College he had
already published a few papers on relativity and been in correspondence about
relativity with Einstein. \ He had obtained a reference from Einstein when he
applied for the professorship at King's.

At King's Jeffery wrote his only book, a textbook on relativity "Relativity
for physics students" (Jeffery 1924). \ This was a collection of his lectures,
his inaugural lecture in the chair of mathematics and lectures to King's
physics students. In this period, with W. Perrett, he published two
translations into English from the German. \ The first was of lectures by
Einstein (Einstein 1922). \ The second became, for many years, the definitive
English translation of papers by Albert Einstein, Hendrik Lorentz, Hermann
Weyl and Hermann Minkowski. \ It also included notes by Arnold Sommerfeld.
\ These were mostly from the Des Relativitatsprinzip published by Teubner.
\ This collection "The Principle of Relativity" (Perrett and Jeffery 1923)
made some of the most important papers on both special and general relativity
widely available in English for the first time. \ It is still in print and
includes Einstein's 1916 review of general relativity (Einstein 1916a). \ This
was Einstein's first review of his new theory and it aimed to make his results
more accessible to mathematicians and physicists (Sauer 2004). \ The review
contains the basic framework of general relativity on which much future
research would be based. \ It also contains certain aspects which would be
sources of future uncertainties or problems not least in the study of
gravitational radiation.

In this review Einstein explained how his 1905 theory of special theory of
relativity was generalized to include gravity. \ He noted the fundamental role
of the long known empirical fact, recently experimentally verified to high
accuracy by Lor\'{a}nd E\^{o}tv\"{o}s, that the gravitational field imparts
the same acceleration to all bodies. \ Various formulations of this statement
are now called the principle of equivalence\footnote{Broadly, the weak
principle of equivalence for test bodies, the strong principle of equivalence
for massive bodies. \ Modern investigations and experimental work have led to
more precise formulations (Will 1993).}. \ He pointed out the equivalence of
the local effects of gravitation and uniform acceleration and argued that

\begin{quotation}
the laws of physics must be of such a nature that they apply to systems of
reference in any kind of motion... the general laws of nature are to be
expressed by equations which hold good for all systems of coordinates, that
is, are covariant with respect to any substitution whatever (generally covariant).
\end{quotation}

He then explained that the flat metric, or line element, of special relativity
$ds^{2}=-dX_{1}^{2}-dX_{2}^{2}-dX_{3}^{2}+dX_{4}^{2}$ will be replaced by the
curved metric of general relativity $ds^{2}=\sum_{\tau\sigma}g_{\sigma\tau
}dx_{\sigma}dx_{\tau}$, where the metric components are functions determined
by and determining the gravitational field. \ Einstein then discussed, at some
length, the tensor algebra and calculus of Gregorio Ricci-Curbastro and his
student Tullio Levi-Civita. \ This was the mathematics, unfamiliar to most
physicists, used in the new theory of gravitation. \ He pointed out the
significance of the Riemann-Christoffel tensor which vanishes if and only if
the metric is flat. \ Finally he outlined the theory and some of its
consequences. \ Einstein did not do this covariantly but imposed the
coordinate condition, $\sqrt{-g}=1$, where $g$ is the determinant of the
space-time metric. \ Presumably he did this in order to make unfamiliar and
complicated expressions more readily understandable. \ However, in the light
of his discussion of general covariance, it was not the most fortunate of
steps. \ Coordinate conditions were the source of many future confusions, not
least for Einstein himself. \ He discussed the gravitational field equations
for the vacuum (empty space or source-free) case as well for systems with
matter (non-gravitational) sources. \ He derived Newtonian gravity as "a first
approximation" of his theory and concluded by discussing the gravitational red
shift, the bending of light, and the precession of the perihelion of mercury.

Einstein dealt with conservation of energy and momentum for the gravitational
field by deriving his pseudo-tensor $t_{\sigma}^{\alpha}.$ \ This was the
first of a number of non-tensorial stress-energy-momentum pseudotensors that
would be introduced because a local energy-momentum tensor for the
gravitational field does not exist. \ By virtue of his field equations it
satisfied the local "conservation law", ($t_{\sigma}^{\alpha}+T_{\sigma
}^{\alpha}$)$,_{\alpha}=0$, where the comma denotes partial differentiation
and $T_{\sigma}^{\alpha}$ denotes the energy-momentum tensor of
non-gravitational fields. \ For an asymptotically flat system, where
asymptotically Minkowskian coordinates are available, the total
energy-momentum of the system can be obtained by using it. \ In more than a
century now a large amount of time has been spent by relativists studying the
construction and use of various energy-momentum pseudotensors and
investigating possible tensorial alternatives. \ Past confusions about their
use have now been dealt with. \ The useful quantities all give the same
results for the total energy-momentum of asymptotically flat systems and can
also be used to construct quasi-local energy-momentum expressions (Chen 2018).

Interestingly nowhere in his review does Einstein write down the covariant
form of his field equations\footnote{Einstein introduced the cosmological
constant $\Lambda$ only subsequently and the field equations then became
$G_{\mu\nu}+\Lambda g_{\mu\nu}=kT_{\mu\upsilon}$.}%
\[
G_{\mu\nu}=kT_{\mu\upsilon}%
\]
where $G_{\mu\nu},T_{\mu\upsilon},k,$denote, respectively, the components of
the Einstein tensor, the components of the non-gravitational energy-momentum
tensor and the coupling constant which ensures the Newtonian limit.

Jeffery was only at King's for a short time. \ He returned to University
College in 1924 and subsequently published little mathematics. \ Education
became his primary interest and eventually he became the Director of the
Institute of Education in the University. (Titchmarsh 1958). \ However fairly
soon after returning to University College he did write, with O.R. Baldwin, a
significant paper on plane gravitational waves (Baldwin and Jeffery 1926).

In 1916 and 1918 Einstein had studied gravitational waves in the context of
his new theory of general relativity (Einstein 1916b; Einstein 1918). \ These
papers were landmarks but they also initiated a long saga of confusion,
controversy, clarification and, most recently, measurement. \ This story has
become widely known, for instance through the publications of Daniel
Kennefick, (Kennefick 1999; 2005; 2007) and more recent accounts
(Cervantes-Cota 2016; Chen 2017; Blum 2018). \ Some of it, particularly those
aspects related to research at King's, bear repetition here.

The existence and importance of electromagnetic wave solutions of Maxwell's
equations meant that Einstein naturally and quickly investigated the existence
of gravitational wave solutions of his new field equations. \ Various people,
including Maxwell and Clifford, had mooted the idea of gravitational waves
analogous to electromagnetic waves. \ However it was not until Einstein had
finally formulated the equations of his theory of general relativity that the
possibility of gravitational waves could be satisfactorily
addressed\footnote{\ At first Einstein had thought that gravitational waves
might not exist, possibly because there was no gravitational analogue of the
electromagnetic radiation emitted by dipoles.}. \ Because the field equations
of general relativity are non-linear, and considerably more complicated than
Maxwell's equations, investigation of the topic seemed difficult without the
use of approximations. \ In that case it was important to choose the right
approximation scheme for the study of waves. \ Following a suggestion of the
Dutch astronomer Willem de Sitter Einstein employed one which is still in use
today (Einstein 1916b). He considered systems where the gravitational field
would be weak so that the space-time metric could be represented by the
Minkowski metric of special relativity plus a small deviation term. \ Assuming
that the perturbing term was sufficiently small that only terms of the first
order in it were significant his field equations became linear field equations
for the perturbing term. \ Moreover the deviation from the flat Minkowski
metric was determined by solutions of the Minkowski space-time wave equation
just as electromagnetic waves were. \ Einstein quickly and correctly concluded
that gravitational waves existed and propagated with the speed of light. \ He
then used his linear \ approximation to the full theory to investigate plane
gravitational waves, that is waves with flat wave fronts like the wave fronts
of radiation far from an isolated source. \ He classified them into three
types, subsequently termed transverse-transverse, longitudinal-longitudinal
and longitudinal-transverse by Hermann Weyl, a description Weyl included in
his influential book (Weyl 1922). \ The first two types were familiar from
electromagnetic and sound waves respectively, the third type was new.
\ Einstein then computed the energy transported by the three types using his
pseudotensor. \ He concluded, incorrectly, that only waves of the third type
transported energy; he thought the first two were artifacts of his choice of
coordinates. \ Einstein had made a calculational error. \ Moreover he had run
into trouble with his use of coordinates and preferred coordinate conditions,
the pseudo-tensor and the computation of energy transport using it. \ Problems
like these and those arising from the choice and use of inappropriate
approximation schemes, were to bedevil the study of gravitational waves for
decades. \ The Finnish physicist Gunnar Nordstr\^{o}m found that his own
calculations did not lead to the conclusions that Einstein had reached. \ His
intervention helped Einstein understand where he had gone wrong and in 1918 he
published a follow up paper, repeating the main points of the 1916 paper, but
correcting his mistakes (Einstein 1918). \ He now concluded that no energy was
propagated by the longitudinal-longitudinal and longitudinal-transverse waves
because they were merely artifacts of the choice of coordinates. \ The
corresponding metrics were just the flat Minkowski metric written in terms of
"oscillating" coordinates. \ The energy propagating waves belonged to the
transverse-transverse class. \ As in electromagnetism gravitational waves were
transverse waves, that is the oscillations were transverse to the direction of
wave propagation. \ Another key calculation in this paper was Einstein's
derivation of the famous (mass) quadrupole formula for gravitational
radiation. \ In general relativity monopole radiation is forbidden as a result
of mass conservation, dipole radiation is absent as a result of momentum
conservation but quadrupole radiation is permitted. \ The quadrupole formula
describes how the energy loss per unit time, when gravitational waves are
emitted from a system of masses, is related to the third time derivative of
the source quadrupole moment.

Eddington had not been convinced by Einstein's proof that gravitational waves,
in the linear approximation, propagated with the speed of light so he
undertook to re-investigate this question (Eddington 1922). \ He was\textit{
}able to confirm that Einstein was right about this and that the
longitudinal-longitudinal and longitudinal-transverse waves were spurious.
\ Importantly he eliminated the spurious wave solutions by computing the
curvature of the metric, that is the Riemann tensor, which is zero if and only
if the metric is flat. \ This is the coordinate independent method of arriving
at the conclusion that, as he put it, "they are merely sinuosities in the
coordinate system". \ He also considered the Einstein-Maxwell equations, that
is gravity coupled to the electromagnetic field outside any source. \ He
confirmed, in this context, that the electromagnetic waves were of also of the
transverse-transverse type. \ As a bonus in the paper he also checked and
corrected, by a numerical factor of two, Einstein's quadrupole formula.

At that time exact solutions of Einstein's equations were few in number so the
use of approximation schemes was natural and remains so for physically
complicated systems. \ However, understanding of the equations was in its
infancy and there often was, as there remained for a long time, uncertainty
about the relationship between approximate and exact solutions. \ In
particular questions remained about the existence of gravitational wave
solutions for Einstein's equations when approximations were not made. \ This
led Baldwin and Jeffery to consider exact plane wave solutions of the full
Einstein and Einstein-Maxwell equations. \ While such solutions do not play
the same role in a non-linear theory as they do in a linear theory where they
can be added, and while they are in themselves not physically realistic, any
wave front, far from a physically realistic isolated source, tends,
increasingly with increasing distance, to become planar. \ Such solutions
should exist in a sensible field theory and if they did not there would be a
real question mark over the theory. \ Baldwin and Jeffery took Eddington's
paper as their starting point but now they imposed his symmetry assumptions on
the full space-time metric and not just the perturbing term of the linearized
theory. \ In their coordinate system the metric components were a function of
a single variable and the Einstein field equations reduced to ordinary
differential equations. \ They showed that Eddington's small amplitude results
also held in the full theory and only the transverse-transverse waves were non-flat.

In the transverse-transverse case the space-time metric in their coordinates,
still considered useful today, was taken to be%
\[
ds^{2}=-(dx_{1})^{2}+g_{22}(dx_{2})^{2}+2g_{23}dx_{2}dx_{3}+g_{33}(dx_{3}%
)^{2}+(dx_{4})^{2}.
\]
They considered waves propagating with velocity $V$ in the negative direction
of $x_{1}$ and so assumed that the non-constant metric components were all
functions only of $x_{1}+Vx_{4}$. \ They found that for non-flat solutions
$V=1$, that is in their units $V$ had to be the speed of light. \ By
considering the solutions of the field equations they demonstrated that the
metric became singular after a finite time. \ They suggested this meant that
infinite plane waves could not be propagated without change of the wave-form.
\ A proper understanding of the global structure of exact plane wave solutions
would have to wait until the 1950's; nevertheless Baldwin and Jeffery's work
was a noteworthy pioneering contribution to their study.

At about the same time the Harvard mathematician Hans Brinkmann was
investigating a problem in local differential geometry, the conformal mappings
of Einstein spaces (Brinkmann 1923; 1925). \ In the course of his analysis he
considered four dimensional spaces satisfying Einstein's empty space-field
equations. \ Brinkmann's papers were rediscovered in the 1950's. \ It was not
until then that it was realized that solutions found by Brinkmann (now termed
plane fronted with parallel rays following Wolfgang Kundt and Jurgen Ehlers)
included ones which could be identified as plane gravitational wave metrics.
\ His work, and the work of Baldwin and Jeffery, had been forgotten by almost everyone.

Jeffery's influence at King's remained even after he had left. \ One of his
post-graduate students John Combridge (1897-1986), whom Jeffery had supervised
at King's for a MSc degree, was appointed as an assistant lecturer in 1926.
Before the mid thirties only the senior members of the mathematics department
had much time for research and junior members like Combridge carried a heavy
teaching load. \ \ After completing his MSc, and while he was at The Royal
College of Science\footnote{The Royal College of Science merged with the Royal
School of Mines and \ the City \& Guilds College to form Imperial College
London in 1907.}, he published three papers on relativity but after coming to
King's little more. However Combridge remained interested in the subject and
corresponded extensively with\ Eddington, an association begun when Combridge
was an undergraduate at Cambridge. \ In 1937 he became the assistant secretary
of the College and eventually registrar (Kilmister 1988). \ After Combridge
retired in 1962 King's published an edited version of a bibliography of over
1700 relativity papers Combridge had accumulated from 1921-1937 (Combridge
1965). \ Although this reflects his personal interests it is still of use to historians.

The appointment of George Temple (1901-1992) as the 8th professor in 1932
marked a turning point for the mathematics department. \ It had reached a low
ebb and needed to be revitalized in both its teaching and research. \ As head
of department he and the algebraic geometer John Semple, appointed four years
later, were able to modernize the department's teaching and turn it into one
with a strong emphasis on research (Kilmister 1994; 1995).

Temple's research included work on gravitation. \ Two of his early papers were
an investigation of Alfred North Whitehead's theory of gravity (Whitehead
1922) and a generalization where the role of Minkowski space-time in the
theory was played instead by a maximally symmetric space-time of constant
curvature, like de Sitter space-time (Temple 1923; 1924). \ Although, for
observational reasons, now very dead (Gibbons and Will 2008) in its time, and
for decades afterwards, Whitehead's theory remained of interest to some as a
possible alternative to general relativity. \ In the 1920's, in particular,
there were those who found it attractive because it retained the role of the
flat background space-time metric of special relativity in determining causal
relationships. \ Matter fields coupled only to a second "physical metric".
\ Whitehead's theory yielded the same predictions as general relativity for
light bending, gravitational redshift and the precession of the perihelion of
mercury without requiring the complete world view and machinery of Einsteinian
gravity. \ Eddington quickly realized the reason the predictions of the two
theories coincided was that, when the cosmological constant is zero, the
Schwarzschild solution is an exact solution in both cases (Eddington 1924).
\ Subsequently Temple published a number of papers related to gravity, now
within the context of general relativity. \ These led to his PhD, awarded by
the City and Guilds College.

After periods at Cambridge with Eddington and then back at Imperial Temple
moved to King's. \ By that time his main interests had moved, first to
analysis and then to the new quantum theory. \ However he continued to have
some interest in relativity and subsequently published several papers in that
area. \ These included the introduction of suitable coordinates to discuss
astronomical optics in general relativity (Temple 1938) and an influential
discussion of the contemporary state of relativistic cosmology (Temple 1939).

Temple spent the second world war at the Royal Aircraft Establishment at
Farnborough and did not return to King's until 1945. \ From 1948 until 1950 he
was also Principal Scientific Advisor to the Ministry of Civil Aviation. \ At
King's he had an active research group studying supersonic flow and
hydrodynamics. His attention then turned to generalized functions and he
continued research on that topic after moving to Oxford in 1953. \ Following
his retirement in 1968 he wrote a book "100 years of mathematics" which,
although chronologically ordered, essentially ranged over the wide number of
areas he himself\ had actively pursued. \ These included aspects of tensor
calculus, differential geometry and relativity (Temple 1981). \ After the
death of his wife in 1979 he became a Benedictine monk.

In the years between the first and second world wars George McVittie
(1904-1988), more than anyone else, devoted his research at King's to general
relativity and cosmology\footnote{McVittie's life and work have been explored
a number of times (McVittie 1978; MacCallum 1989; Sanchez-Ron 2005) and
further details about his career and research, beyond the outline presented
here, can be found there.}. \ McVittie began his university studies in
Edinburgh where he started his post doctoral research with Edmund Whittaker.
\ Following a practice not uncommon at that time, Whittaker sent him to
further his doctoral studies under the supervision of Eddington at Cambridge.
\ For his PhD, awarded in 1930, McVittie worked on unified field theories of
gravitation and electromagnetism and published a couple of papers on the topic
(McVittie 1929a; 1929b). \ In later years he commented in an interview

\begin{quotation}
Then there was more unified theory and more and more theoretical
solutions...and I began to say to myself `There is no way out of this
multitude. \ There is no reason for preferring one rather than another'...And
it then occurred to me, slowly, that there is surely a way of getting some
order into this confusion, and that is to look at the observational data, and
pick out things by that criterion, and not by what seems reasonable or
mathematically elegant... (McVittie 1978).
\end{quotation}

This was an attitude which McVittie would hold to throughout his career
leading to him being termed "empiriciste irr\'{e}ducible" (Mavrid\`{e}s 1973).
The tag "uncompromising empiricist" appears to have been one which McVittie
was happy to embrace.

After completing his PhD McVittie held appointments\ for brief periods at
Liverpool, Edinburgh and Leeds. \ As he said in the interview above

\begin{quotation}
One was always supposed to get rid of one's good men to another university.
This is what it amounted to, in many cases at least....Promotions had to be
obtained normally by moving to another university, rather than being promoted
in your own.
\end{quotation}

The idea of building up a strong, sustainable group did not seem to exist at
that time, or for some time after, in many British mathematics departments.

In the nineteen thirties McVittie worked on various topics in cosmology and
astrophysics. \ \ One paper in particular from this time, "The mass-particle
in an expanding universe" (McVittie 1933), is of some current interest.
\ Combined in it are McVittie's interest in exact spherically symmetric and
cosmological solutions of Einstein's equations and his life long view that, in
the appropriate circumstances, the cosmological constant should be retained.
\ In this paper McVittie aimed to reconcile, as he put it, the two types of
metrics used for the universe, the Schwarzschild metric for discussing the
motion of the planets around the sun and cosmological metrics of the
Lema\^{\i}tre and de Sitter classes. \ Techniques for analyzing the global
properties of exact solutions, even spherically symmetric ones like those
constructed by McVittie, were not well developed until the second half of the
twentieth century. \ McVittie's solutions were not well understood for a long
time but in recent years McVittie's metrics have been shown to include regular
black holes in expanding universes (Kaloper 2010; Lake and Abdelqader 2011;
Nolan 2017).

McVittie took up a readership in the King's mathematics department in 1936.
\ There were now nearly 2,500 students attending the College with just over
400 students being taught by the mathematics department. \ The department
\ still had only one professor of applied mathematics and one professor of
pure mathematics (Temple and Semple) but there were now eleven academic
members of staff. McVittie's duties consisted mainly of undergraduate teaching
and research; there was only a handful of post-graduate (MSc.) students.

In 1937 he published a little book, "Cosmological Theory" (McVittie 1937).
\ The key to McVittie's thinking lies in his continuing emphasis on relating
theory to observation. \ This was particularly notable at a time when, in
Britain at least, some discussions of cosmology were drifting into the realms
of philosophy. \ This point is illustrated by the fact that the first chapter
of this book dealt with extra-galactic nebulae, stellar magnitudes and their
distances. \ Only after those topics had been dealt with did it move on to the
tools needed to deal with the analysis of models of the expanding universe.
\ The book concluded with a chapter on his own version of Edward Milne's
kinematical relativity where he tried to relate his approach to observations.
\ In this book McVittie gave the definition of a "radius of the visible
universe" which was later called by Wolfgang Rindler "the particle horizon".

In 1939 McVittie and Temple participated in a joint meeting of the Royal
Astronomical Society and the Physical Society of London. \ The aim of the
meeting was to review the current state of cosmology. \ McVittie was chosen to
review the observational situation (McVittie 1939) and Temple the theoretical
(Temple 1939). \ Interestingly both McVittie and Temple presented Milne's
kinematical relativity and relativistic cosmology as equal competitors, as far
as observations were concerned, with McVittie noting that, at that time,
observations could not discriminate between them (Gale 2015).

The second world war caused great destruction and disruption. \ The
mathematics department moved for a time to Bristol and during the London Blitz
the College, which had been taken over by the Auxiliary Fire Service, was
damaged (Huelin 1978). \ People like Combridge continued to run the College,
outside of and then back in London, as best they could. \ Others, like Temple
and McVittie, were absent on leave engaged in war-related work. \ McVittie
became the founder and head of the meteorology group at Bletchley Park. \ This
group was actually involved in important and useful cryptographic work
deciphering enemy weather messages. \ That work stimulated his interest in
hydrodynamics and gas dynamics and he subsequently wrote a number of papers in
these areas and made novel use of his knowledge of tensor calculus and
relativity (MacCallum 1989).

Although he was heavily involved in war work McVittie managed to write a few
papers on cosmology. \ A couple of his war time relativity and cosmology
papers were, as he put it, just quarrels with Milne and his student Arthur
Walker about kinematical relativity. \ However they also included a paper,
arising from these altercations with Milne, concerning descriptions of events,
coordinates and the regraduation of observers' clocks (McVittie 1946). \ One
of his conclusions was that arbitrary regraduation of an observer's clock
merely implied a coordinate transformation within space-time. \ While such a
conclusion would now be regarded as unsurprising, aspects of the paper proved
to be of some subsequent interest. \ McVittie later regretted that he had not
pursued the topic. \ Working on these papers gave him some relief from his
cryptographic work. \ He was later to say

\begin{quotation}
There were periods when I thought I would go crazy if I went on dealing with
these ciphers (McVittie 1978).
\end{quotation}

At the end of the war Temple and McVittie returned to King's. \ There McVittie
heard about the steady state model of the universe directly from Hermann
Bondi. \ Bondi was still based in Cambridge but in 1948 he stopped in London
on his way to a conference in Edinburgh where he and Thomas Gold planned to
announce their theory\footnote{The steady state theory was introduced by Bondi
and Gold, and separately, in a field theory version, by Hoyle (Bondi and Gold
1948; Hoyle 1948).}. \ According to this theory the universe does not change
in appearance over time (the perfect cosmological principle) and the universe
has neither beginning nor end. \ However it accepted that the universe was
expanding \ and hence required the continuous creation of matter in order that
the density of the universe did not decrease. \ It gave rise, in Britain at
least, to heated controversy during the 1950's and early 1960's, and it was
vigorously defended by its three authors. \ Many must have felt that at that
time cosmology merited the observation of Lev Landau "Cosmologists are often
in error, but never in doubt" (Kragh 1996). \ McVittie's immediate response to
the theory was not enthusiastic. \ He thought it was more restrictive than
general relativity and he found the creation process mysterious. \ For a while
he could not avoid being attracted by some aspects of the model, as he had
been attracted earlier by the deductive approach of Milne. \ However in
subsequent years his enthusiasm dwindled as he increasingly felt that the
model did not agree with observations.

In 1948 McVittie left King's to move up the ladder once again. \ \ He became
the first professorial head of the mathematics department at Queen Mary
College, another college of the University of London. \ There he was now able
to engage in more postgraduate teaching and took on his first research
student, Clive Kilmister. \ \ Unsympathetic to the steady state model, then
all the rage in some quarters in Britain (McVittie's forthrightness did not
always win him friends) and uncomfortable at Queen Mary, he moved to the
University of Illinois in 1952. \ His career flourished there and a few years
later he wrote "General Relativity and Cosmology" (McVittie 1956). This
influential book featured that aspect of his research which had become so
prominent, the derivation of predictions from cosmological models and their
comparison with observations. \ On his retirement in 1972 he returned to
England and continued to be active at the University of Kent. \ In 1984 a
minor planet, formerly 2417, was renamed "McVittie" by the International
Astronomical Union.

By the end of the first half of the twentieth century the status of general
relativity, amongst physicists in Europe and the United States at least, was
not high. \ Little progress appeared to have been made, and few had worked in
the field in the previous twenty five years (Eisenstaedt 1986; 1989a; 2006).
\ The small number of papers that were being published were too often detached
from experiment, observation and the rest of physics\footnote{According to
Eisenstaedt (1989a) at the time of Einstein's death in 1955 there were 10,000
annual references in Physics Abstracts but the relativity output was only 30
papers per year.}. \ The focus of theoretical physics was very much still on
quantum mechanics and the developing quantum field theory. \ The general view
was that, cosmology apart, general relativity predicted only small corrections
to Newtonian gravity. \ Furthermore cosmology was still widely viewed as not
being a proper part of physics. \ The mathematical techniques used in general
relativity were not particularly difficult but were different from those most
physicists needed and used. \ They saw no reason to spend time and energy
equipping themselves with them. \ That general relativity was in the doldrums
in the early 1950's, particularly in the United States, has been forcefully
attested to by workers in the field such as Ted Newman (Newman 2005), and
Bryce DeWitt (DeWitt B. 2009). \ However perhaps the benign neglect noted by
Subrahmanyan Chandrasekhar (Chandrasekhar 1979) was the experience of most,
particularly outside of the United States where the subject also had a home in
a number of mathematics departments.

While all might have been rather quiet on the physics front for many of the
inter-war years important developments in mathematics, which would later
impact on general relativity, were taking place. \ Differential geometers like
Tullio Levi-Civita, Jan Schouten\ and \'{E}lie Cartan had taken an interest in
Einstein's work from the beginning and this interest influenced and stimulated
the development of differential geometry. \ In the 1930's the mathematicians
Oswald Veblen, Henry Whitehead and Hassler Whitney laid the foundations of
modern global differential geometry. \ Their work became important later in
the study of the global structure of space-time manifolds. \ By the late
1930's the modern approach to differential geometry was starting to be
introduced into relativity by people like Cartan's student Andr\'{e}
Lichnerowicz (Lichnerowicz 1992).

At mid-century, despite the apparent low point it had reached, general
relativity was about to blossom again (Blum 2015; 2017). \ In the ten years
after 1945 recovery from the massive dislocations and damage of the war was
accompanied by the formation of new relativity research groups in places like
Syracuse, Princeton, Hamburg, Warsaw and London. \ New people entered the
field and in the 1950's they and their groups began to make important
contributions to the teaching and development of the subject. \ Experimental
tests began to be devised and carried out (Will 1993) and the naissance of
experimental gravity physics took place (Pebbles 2017). \ Slowly but steadily
the study of relativistic gravity and general relativity began to return to
the mainstream of physics (Schutz 2012).

\section{The Bondi group and waves}

\subsection{The early members}

Hermann Bondi (1919-2005) was appointed as the tenth professor of mathematics
in 1954. \ He replaced George Temple who had left to take up the Sedleian
Professorship of Natural Philosophy at Oxford. \ Semple, the ninth and only
other professor, replaced Temple as head of department. \ The size of the
College and the department remained about the same as in McVittie's time but
now there were 18 post-graduate students. \ In 1955 Bondi brought his former
Cambridge student,\ Felix Pirani (1928-2015), to King's as a lecturer in
mathematics. \ Bondi, Pirani and Clive Kilmister (1924-2010), who was already
lecturing there, formed the original core of what became a highly productive
and influential relativity group. \ Its major contributions in the 1950s and
60s were to the study of gravitational radiation. \ These three, in various
ways, played major roles in the King's relativity group during much of its
most creative period. \ Their careers before they came to King's provide
interesting insights into the times and their formative backgrounds.

Bondi was born and raised in Vienna and his life and career are fully
discussed in his autobiography (Bondi 1990a), American Institute of Physics
oral history (Bondi 1978) and Royal Society obituary (Roxburgh 2007).
\ Concerns about the turbulent situation in Austria and the increasing
anti-Semitism, particularly in university circles in Vienna, together with a
meeting with Eddington, led to him becoming an undergraduate at Trinity
College Cambridge in 1937. \ In 1940, being an Austrian
national\footnote{Bondi did not become a British citizen until 1946.}, he was
interned by the British and eventually transported to a camp in Quebec. \ On
the first night of his internment he met Tommy Gold and later Alfred Schild.
\ Both men would play significant roles in Bondi's future career. \ During his
internment he did his first teaching, which he enjoyed, and also a little
research. \ In 1941 he was released and was able to return to England. \ While
he had been interned he had been awarded his B.A. by Cambridge so he returned
as a research student with Harold Jeffreys as his supervisor. \ Jeffreys gave
him a problem connected with waves on the surface of water and within a few
months this led to his first paper (Bondi 1942). \ After seven months as
Jeffrey's student he was enlisted to undertake radar related research for the
Admiralty. \ As Bondi later put it

\begin{quotation}
there was a very short time from my being behind barbed wire \ because I was
so "dangerous" to my being behind barbed wire because the work I did was so
secret (Bondi 1978).
\end{quotation}

He was eventually reunited with Gold and for a period they shared a house in
Surrey where they were periodically joined by the leader of their small
Admiralty group Fred Hoyle. \ They worked long hours but were able to discuss
physics and astrophysics in their free time. \ Bondi felt that he received a
second education from the somewhat older Hoyle. \ As a result of this
interaction Bondi became interested in a problem that Hoyle and Raymond
Lyttleton had worked on (Hoyle and Lyttleton 1939). \ He constructed a
detailed mathematical description of the accretion of matter onto stars as
they passed through gas and dust clouds in the interstellar medium. \ This
work, which showed that the accretion rate was much greater than had been
previously estimated and that the interstellar medium played an important role
in stellar dynamics, led to a paper with Hoyle (Bondi and Hoyle 1944) and
election to a Trinity College Research Fellowship in 1943. \ In those days
this was considered so grand that he did not continue working towards a PhD.
\ Bondi later continued this line of research after he had returned to
Cambridge. He studied spherically symmetric accretion onto a point mass and
computed the accretion rate (Bondi 1952a), work often now used in the context
of neutron stars and black holes. \ Although it contains many simplifying
assumptions, subsequent research has shown that Bondi-Hoyle-Lyttleton
accretion is broadly correct (Edgar 2004). \ Their work is still widely used
and is regarded as laying the foundation of accretion theory. \ All this
research was carried out within the context of Newtonian gravity and
throughout his career Bondi's thinking about gravity was strongly influenced
by Newtonian ideas.

While at Cambridge he published on a wide range of topics including fluid
motion, waves on the surface of water, geophysics, the solar corona, rigid
body mechanics and electromagnetism. \ He also did further work, some with his
wife Christine, on stellar structure. \ Bondi's mathematical forte was
differential equations although he later wrote

\begin{quotation}
I have always liked the idea of experiments and explanation that could be
qualitative rather than dependent on the exact figures, that could be sketched
out rather than drawn with perfection, and yet convey the information (Bondi 1990a).
\end{quotation}

He felt he had powerful tools to hand and if he was interested in a problem he
was willing to apply these to it (Bondi 1978). \ Becoming more interested in
general relativity he published a paper on relativistic gravity,
"Spherically\ symmetric models in general relativity" (Bondi 1947). \ In this
paper Bondi applied Einstein's field equations to pressure-free spherically
symmetric systems of particles, derived the equations of motion and described
the various properties of the systems. \ As in his later work on gravitational
radiation he paid careful attention to the physical interpretation of the
coordinates he used. \ The paper is now regarded as a classic. It clarified
and extended previous work of Georges LeMa\^{\i}tre and Richard Tolman in ways
that initiated a number of lines of work which are still pursued today,
including inhomogeneous cosmological models, shell crossing, and aspects of
gravitational collapse.

About that time Bondi was asked by the Royal Astronomical Society to write a
report on the state of cosmology. \ He was surprised because he felt he had
only a superficial knowledge of the field. \ The report (Bondi 1948) was
regarded as masterly. \ It led him to write his book "Cosmology" (Bondi
1952b). \ This brief book was extremely influential and gives an excellent
idea of the state of cosmology at the time. \ It aimed to treat cosmology as a
branch of physics in its own right, not a universally held point of view at
the time. \ The book covers basic principles, observational evidence and the
then current cosmological theories including Newtonian and general
relativistic cosmologies, various theories of Eddington, Dirac and Jordan,
kinematic relativity and the steady-state theory. \ One consequence of the
steady state theory, and his work on cosmology, was that he became much more
widely known. \ A second edition of Cosmology was published in 1960, with an
additional chapter entitled "The Present Position in Cosmology". In this he
summarized the changes that had taken place since the first edition. \ Later
in life Bondi would be slightly irked at being identified as a cosmologist. He
considered that he did not produce any new research of substance in cosmology
after the mid fifties when he began to concentrate on general relativity. He
considered his work at King's on gravitational radiation much more important.

Felix Pirani was born in London and attended schools in England, Australia,
New Zealand and Canada where his family settled in 1941 (Robinson 2016). \ His
first scientific paper, "Use of the Hartmann formula", was written in
collaboration with his optics teacher, A Willena Foster, while he was still an
undergraduate at the University of Western Ontario (Foster and Pirani 1948).
\ This was a few paragraphs exhibiting a simplified method of making
calculations of spectral wavelengths.\ \ \ After graduating he moved to the
University of Toronto where he was introduced to general relativity and tensor
calculus by Leopold Infeld and Infeld's former student Alfred Schild. \ He was
just starting his doctoral work, after completing his master's degree in 1949,
when he and Schild attended the second Canadian Mathematical Congress in
Vancouver and heard Paul Dirac's lectures on the quantization of Lagrangian
field theories (Dirac 1950). \ Schild immediately realized that the techniques
described by Dirac could be applied to general relativity. \ Later that summer
Schild moved to the Carnegie Institute of Technology in Pittsburgh. \ Infeld,
the head of their department to whom Pirani had been assigned as a PhD
student, agreed that Pirani could go to Carnegie with Schild as his first
doctoral student. \ Research leading to two of the early papers on the
constrained Hamiltonian formalism of general relativity formed the basis of
Pirani's doctorate (Pirani 1951). \ In the first of these papers a Hamiltonian
formulation of general relativity is constructed with quantization of the
theory in mind (Pirani and Schild 1950). \ The second, co-authored with Schild
and Skinner, contains a discussion of the constraints arising in the
formulation and explicit expressions are obtained for them (Pirani 1952).
\ While this work was being carried out similar research was being undertaken
by Peter Bergmann and his group at Syracuse University. Bergmann's group
developed this line of work for many years, the Carnegie people did not.

Schild had met Bondi when they were both interned in Canada. \ With Schild's
support Pirani obtained a National Research Council of Canada post-doctoral
fellowship to work with Bondi at Cambridge University. \ At Bondi's suggestion
he enrolled for a second doctorate with the result that he ended up with two
doctorates (while Bondi had none).\ \ At Cambridge Pirani's research changed
direction. \ He did discuss his Carnegie work with Paul Dirac who was quite
interested in it. \ However Pirani could not see how to handle the complicated
constraints of the Hamiltonian formalism found at Carnegie and Bondi was not
interested in that line of research. On the other hand Pirani found Bondi's
cosmology book and the steady state theory very exciting. \ For a long time he
was an enthusiastic supporter of the steady state model of the Universe,
possibly more Hoyle's version than that of Bondi and Gold. \ His work on the
steady state theory and other topics in relativistic gravity led to a number
of minor papers and to his Cambridge PhD. \ It was no secret that there was
often an unspoken religious or anti-religious impulse amongst many
cosmologists. \ Bryce DeWitt later wrote that

\begin{quotation}
In the early days of the so-called steady-state theory of the universe,
everyone knew (though no one ever said so in print) that the model was
motivated by antireligious sentiment. \ When evidence for the Big Bang began
to accumulate, the steady-state theory nearly collapsed (a mutilated version
of it has been kept alive) and the Vatican became ecstatic (DeWitt B.
2005)\footnote{Reproduced from Physics Today 2005, \textbf{58}:32-34, with the
permission of the American Institute of Physics.}.
\end{quotation}

That might be something of an overstatement but Pirani later said that,
"people didn't want there to be a beginning" and that certainly was one of his motivations.

In one of the papers reporting some his Cambridge work he coined the word
"gravitino" later re-invented, with quite a different meaning, in the context
of supergravity theory. \ The Pirani gravitinos did not interact with normal
matter, had zero rest-mass and negative energy, and were created at the same
time as "normal matter". \ Their introduction was an attempt to develop a more
specific description of continual creation of material, postulated in the
steady state model, which was consistent with conservation of 4-momentum at
each creation event. \ The possibility that the gravitino could be identified
with a negative energy neutrino was also raised. (Pirani 1955a; Kragh 1996).
\ It was not really the proposal of a realistic physical mechanism and went no
further. \ Pirani later recalled that at the time he was somewhat surprised
that it was accepted for publication.

After Cambridge Pirani spent a year, 1954-55, as a research associate at the
Dublin Institute for Advanced Studies whose permanent members then included
Erwin Schr\H{o}dinger, Cornelius Lanczos and John Synge. \ This was not a
surprising move. \ Synge was a distinguished geometer with significant
interests in physical applications including relativity. \ He had written a
book on tensor calculus and its applications with Schild (Synge and Schild
1949) and had been responsible for Infeld's move to Toronto. \ The world of
relativity was then very small. \ \ Pirani was especially influenced by Synge
and by proof-reading Synge's book on special relativity (Synge 1955). \ Later
Pirani again assisted Synge by proof-reading the latter's book on general
relativity (Synge 1960). \ By that time the influence was two way. \ Indeed in
his preface to the general relativity book Synge writes

\begin{quotation}
Dr. Pirani introduced me to the transport law of Fermi which plays an
important part in the book, and my attempt to turn Riemannian geometry into
observational physics (measure the Riemann tensor!) originated largely in
discussions with him...
\end{quotation}

It was in Dublin that two key events occurred which led to Pirani beginning
his most important body of research. He had been encouraged by Schild to
become a reviewer for Mathematical Reviews and in Dublin he reviewed a paper
by McVittie (McVittie 1955). \ In his review (Pirani 1955b) Pirani noted that
according to the paper gravitational waves were said to exist when the
solutions of the empty space-time Einstein field equations were both time
dependent and solutions of the wave equation. \ Pirani pointed out that this
definition was not invariant . \ He also commented that it did not seem to
have any physical significance since McVittie had to repeatedly rejected
metrics which satisfied these conditions but then turned out to be flat, or
transformable into time independent ones, or both, and hence were metrics
which could not be identified as genuine gravitational wave solutions.
\ Pirani concluded that McVittie had not been able to identify metrics
acceptable as gravitational wave solutions.

Reviewing this paper made him aware of some of the tortuous history of
attempts to investigate gravitational waves. \ He realized that the key to
satisfactory investigations was to focus on invariant or covariant quantities
and thus to avoid misleading coordinate dependent conclusions. \ At about the
same time he came across a review of a paper in which the Soviet physicist
Aleksei Petrov classified the Weyl tensor using classical methods of linear
and multilinear algebra (Petrov 1954). \ The paper itself was in Russian,
which was not one of Pirani's languages. \ However there was sufficient in the
review for Pirani to be able to think that Petrov's work on the algebraic
structure of the Weyl tensor might be used to invariantly define gravitational
radiation. \ He had recently been reading about the special algebraic
structure of plane electromagnetic waves in the proofs of Synge's special
relativity book and he thought that by using Petrov's results he might be able
to deal with gravity in an invariant way.

In 1955 Pirani moved to King's where he completed and published a paper whose
genesis lay in his analysis of the problems he had found in McVittie's work.
\ Belatedly and rather unnecessarily he took and passed his Cambridge PhD
examination in 1956, George Temple being one of the examiners (Pirani 1957a).
\ After he managed to obtain a copy of Petrov's paper a colleague at King's
made a rough translation of it for him and this assisted his production of his
second important paper of the period (Pirani 1957b).

The third founding member of Bondi's group, Clive Kilmister, was born in
Epping, an outer suburb of London, and raised and educated in London
(Silvester 2010). \ He obtained his first degree from Queen Mary College in
1942. \ The College was evacuated to Cambridge for a few years and some of his
time was spent there. \ He then saw military service in the Royal Artillery.
\ Subsequently, on the recommendation of C.P. (Charles) Snow, he worked on the
development of radar for three years. \ Returning to Queen Mary College in
1947 Kilmister completed a Master's degree and then, under the supervision of
George McVittie, a PhD in 1950. \ His research topic entitled "The Use of
Quaternions in Wave-Tensor Calculus" dealt with some of Eddington's ideas and
included the writing of Eddington's E-numbers in terms of quaternions
(Kilmister 1949; 1951). \ This was a rather surprising area of research.
McVittie had long thought that Eddington had lost touch with real physics.
\ According to Kilmister, McVittie considered Eddington's last works
"scandalous", and he hoped Kilmister would "get rid of this scandal" (private
communication). \ \ Kilmister retained a life-long, heterodox fascination with
Eddington's work and became known for his elaboration and elucidation of such
works as Eddington's "Fundamental theory".

McVittie's relationship with Temple and King's facilitated Kilmister's
appointment as an assistant lecturer in the King's mathematics department in
1950. \ He remained at King's until his retirement in 1984. \ When he arrived
no relativity was being done in the department. \ Most of the research by the
applied mathematicians was devoted to fluid mechanics, aerodynamics and so on.
\ Kilmister himself did no relativity before Bondi arrived, pursuing instead
his primary and life-long interest in foundational questions. \ However in
1954, together with Geoffrey Stephenson, he started the King's relativity
seminars and these were to continue at King's for more than two decades.
\ Over the years people from the other colleges of the University, like
William Bonner from Queen Elizabeth College and Gerald Whitrow and Tom Kibble
from Imperial College, also attended the seminar.

Kilmister was delighted when Bondi and Pirani were appointed, "everything
changed" he later said although his personal work on gravity was never more
than a small part of his research. \ The latter did include explorations of
alternatives to general relativity. \ For example he undertook one of the
early investigations of what later tended to be called Yang's theory of
gravity (Kilmister and Newman 1961). \ In 1974 Chen-Ning Yang presented this
theory as an affine gauge theory, with a Yang-Mills type action (Yang 1974).
\ Solutions of Einstein's vacuum field equations are also solutions of Yang's
theory but the converse is not necessarily the case. \ Together with his
student Alan Thompson Kilmister studied the classical field equations, as
weakened field equations for general relativity (Thompson 1962; Kilmister
1966). \ Interest in this theory from the point of view of quantum gravity
diminished when Kellogg Stelle showed that while curvature squared gravity was
perturbatively renormalizable it had a problem with ghosts, the norms of some
states became negative (Stelle 1977; 1978).

Kilmister was elected to the committee of the International Society on General
Relativity from 1971 to 1974 but by then his many other activities were
crowding out his interest in general relativity \footnote{\ Kilmister held the
additional position of Gresham Professor of Geometry from 1972 until 1988, one
of his predecessors being Karl Pearson and one of his successors being Roger
Penrose.
\par
Gresham Professorships are separate from the University of London and the
professors have been delivering free public lectures within the City of London
since 1597 (Wilson 2017).}. \ Kilmister wrote, coauthored or edited about
twenty books on a wide a variety of topics. \ Amongst these was an eclectic
collection of papers on general relativity (Kilmister 1973). \ It included
papers by Bondi and Pirani as well as ones by Riemann, Clifford, Penrose,
Fock, Oppenheimer and Snyder, Pound and Rebka, Infeld and, of course,
Einstein. \ Kilmister's research was usually at an angle to the work of other
members of the group but he and Bondi had an interest in philosophy and
together they wrote a glowing review of Karl Popper's "The logic of scientific
discovery" (Bondi and Kilmister 1959). \ Much later Kilmister recalled that

\begin{quotation}
Popper put forward the view that scientific statements are hypotheses which
the scientist puts forward with the intention of refuting. \ In order to be
valid they must be refutable. \ Both of us were impressed with this point of
view, Bondi particularly because it fitted in so well with the steady-state
model of the universe. \ The latter put forward falsifiable propositions, it
made predictions which could be refuted, as in due course they were (private communication).
\end{quotation}

One of the group's earliest research associates, from 1958 until 1960, was
Dennis Sciama (1926-1999) who came from a fellowship at Cambridge (Ellis and
Penrose 2010). \ Sciama was originally Dirac's research student, completing
his PhD in 1953, but had attended Bondi's cosmology lectures at Cambridge.
\ He knew Bondi, Gold, Hoyle and Pirani well and had interests in common with
them. \ His research encompassed both steady-state cosmology and Mach's
principle. \ In fact his thesis was entitled "Mach's principle and the origin
of inertia". While at King's he published a book which explained these ideas
in terms accessible to the general public (Sciama 1959)\footnote{The tradition
of writing books, aimed at the general public, about relativity and cosmology
became a common one at King's with Hermann Bondi, Felix Pirani, Paul Davies
and John G. Taylor all writing such works.}. \ He also pursued his interest in
generalizations of general relativity. \ Probably the most well-known of that
line of work is the one in which materials with intrinsic spin are described
by the use of connections with torsion (Sciama 1962). It turned out that this
had also been investigated by \'{E}lie Cartan and Tom Kibble and is now
sometimes known as the ECSK (for Einstein, Cartan, Sciama and Kibble) theory.
\ In 1961 Sciama returned to Cambridge and there built up a major group and
supervised some remarkable students. \ Interaction with King's during the
1960's continued with some of his first cohort of Cambridge students attending
lectures at King's and members of both groups travelling backward and forwards
between London and Cambridge to attend seminars and study groups (Ellis 2014).
\ Sciama, like Pirani and Bondi, continued to be interested in Mach's
principle. \ They all gave up on the steady state cosmological model in the
mid 1960s, in the face of the observational evidence, but until then Sciama
was an active and vigorous supporter of the model.

\subsection{The early years}

When Hermann Bondi took up his appointment at King's on his mind was the
selection of a research area where a small group could make an impact.
\ Fortuitously he attended the 1955 conference in Berne, organized by
Andr\'{e} Mercier and Wolfgang Pauli and presided over by the latter, marking
the fiftieth anniversary of special relativity (Mercier and Kervaire 1956).
\ William McCrea who reported on the conference in Nature commented that this
was probably the first international conference ever to be devoted to
relativity (McCrea 1955). \ The conference, held three months after Einstein's
death, was attended by about ninety people, mostly Europeans, many of whom had
worked on general relativity or been colleagues of Einstein in the past. \ The
talks were given in German, French and English. \ \ Bondi considered himself a
comparative novice in the field and thought himself fortunate to have been
invited. \ He gave two talks, the primary one on the steady state theory
(Bondi 1956a) and a secondary one (Bondi 1956b), for which he did not submit a
manuscript, on a paper he'd recently written with Gold (Bondi and Gold 1955).
\ The primary talk consisted of a qualitative discussion of observational
tests of the theory and what he termed evolutionary theories. \ The paper with
Gold concluded that a uniformly accelerated charge radiated. \ It also
included a discussion of a charged particle statically supported in a
gravitational field which they explained was consistent with the principle of equivalence.

Also present at the conference were a few younger people who subsequently made
significant contributions to the subject. \ They included Pirani who spoke
about work he had done in Dublin on the definition of inertial systems in
general relativity (Pirani 1956a). \ Pirani briefly outlined ideas some of
which he would soon develop in important ways. \ He was motivated by ideas
incorporated in Mach's principle and listed a number of formulations of it.
\ The only one he considered relevant to his paper was

\begin{quotation}
(4) The local reference frames in which NEWTON's laws are approximately valid
(without the introduction of Coriolis or centrifugal forces) are those frames
which are approximately non-rotating relative to the distant stars.
\end{quotation}

Pirani's conclusion was that

\begin{quotation}
as far as general relativity is concerned, (4) is an accident, not a
fundamental law - an empirical result which is only approximately confirmed by
theory, and this only when the gravitational field is slowly varying in space
and time.
\end{quotation}

The main interest of this talk now lies in Pirani's focus on the description
of observations relative to a single observer and the use he made of the
equation of geodesic deviation, orthonormal tetrads and
Fermi-Walker\ propagation to define local reference (inertial) frames.

Detailed discussions of topics mentioned in the talk, together with further
results, were subsequently published in a longer paper (Pirani 1956b). These
included an analysis of Fermi-Walker transport in Schwarzschild space-time and
the motion of a spinning particle. \ Pirani showed that a vector transported
around a circular path underwent a secular rotation which was a combination of
the special relativistic Thomas precession and an inertial drag due to the
Schwarzschild mass. \ This led him to conclude that the local inertial frame
determined by local experiments was, in general, not exactly fixed relative to
the distant stars. \ In a section entitled "A Simple Model Gyroscope" he
demonstrated that the Pauli-Luba\'{n}ski vector\ of a spinning particle was
Fermi propagated and observed that a spinning test particle would have a fixed
angular momentum relative to Fermi-propagated axes. \ This confirmed for him
that Fermi-Walker transport was the relativistic analogue of the
transportation of space-axes so that they had fixed direction in the absolute
space of Newtonian theory.

Pirani's interest in Mach's principle was not at all uncommon at that time.
\ The influence of philosopher and physicist Ernst Mach's thought on
Einstein's early work is well known; Einstein coined the term Mach's
principle. \ It has become a catch all name for attempts to relate quantities
such as inertia, local inertial frames and local physical laws to the large
scale structure of the universe (Barbour and Pfister 1995). \ Pirani wrote to
Einstein about it in 1954 but much to his dismay Einstein's reply (Einstein
1954) was lost while it was being copied. \ By then Einstein had become
disenchanted with the principle and these days only a comparatively small
number of people actively engage with some formulation of it. \ By the time
Bondi and Samuel discussed ten versions of the principle (Bondi and Samuel
1997) Pirani had little belief in its usefulness.

The Berne conference cannot be described as having been particularly forward
looking but it gave people like Bondi and Pirani a chance to meet workers in
the field and to get an overview of what was going on. \ Years later Bondi
wrote in his autobiography that there was still confusion about whether
general relativity predicted the existence of gravitational waves or not. \ A
variety of opinions were expressed, one by Nathan Rosen in his talk on
gravitational waves. \ Rosen had been one of Einstein's assistants in the
1930s at the Institute for Advanced Study in Princeton and together they had
written a paper on gravitational waves (Einstein and Rosen 1937). \ The
interesting and amusing history of this paper is now well known (Kennefick
2005; 2007). \ Einstein submitted the paper to \textit{The Physical Review}.
\ Its title was "Do Gravitational Waves Exist?". \ The paper's answer was no.
\ Einstein and Rosen had found an exact solution of Einstein's empty space
field equations which they identified as a plane wave solution. \ However
their metric was not regular everywhere and this led them to conclude that
plane wave solutions were necessarily somewhere singular. \ The mathematics of
global differential geometry was not well formulated until the 1930s and the
application of its techniques to general relativity did not reach maturity for
decades after that. \ Before then there was often confusion about coordinate
and genuine singularities with many assuming, as did Einstein and Rosen, that
one coordinate system should cover all of a space-time manifold. \ Upon
receiving an anonymous referee's report, now known to have come from the
cosmologist Howard Percy Robertson, disagreeing with the conclusion Einstein
took umbrage, possibly because of his unfamiliarity with the journal's
refereeing procedure. \ He sent a new version of the paper to the
\textit{Journal of the Franklin Institute} where it was published along with
papers on topics like economic trends in manufacturing and sales and the
visibility of various type fields. \ The new version did in fact take on board
comments from Robertson. \ The first part of the paper dealt with the
linearized theory and the second part reinterpreted results from the original
paper. \ In the second part metrics of the form%
\[
ds^{2}=-A(dx^{1})^{2}-B(dx^{2})^{2}-C(dx^{3})^{2}+A(dx^{4})^{2},
\]
where $A,B$ and $C$ were taken to be positive functions of $x^{1}$ and
$x^{4\text{ }}$only, were considered. Their solution was now identified as a
cylindrical, not plane, wave solution, with the singularities on the axis of
symmetry where, as was well understood, they could represent idealizations of
sources. \ In fact unknown to the authors the metric had already been
published by the Austrian Guido Beck (Beck 1925). \ Einstein concluded the
section on cylindrical waves by writing

\begin{quotation}
Progressive waves therefore produce a secular change in the metric.

This is related to the fact that the waves transport energy, which is bound up
with a systematic change in time of a gravitating mass localized in the axis...
\end{quotation}

Unfortunately this paper did not end Einstein's wobbling about waves and energy.

The published version of their paper, in particular the second part, had been
rewritten by Einstein after Rosen had left Princeton for the Soviet Union.
\ In fact Einstein added a note at the end of the paper saying

\begin{quotation}
The second part of this paper was considerably altered by me after the
departure of Mr. Rosen for Russia since we had originally interpreted our
formula results erroneously...
\end{quotation}

When he saw the published version Rosen was dissatisfied with it. \ He
published his own - plane wave - paper in a Soviet journal (Rosen 1937).

In the written account of his Berne talk Rosen\ referred back to his work with
Einstein and to his own 1937 paper and reiterated his view that there were no
solutions of the exact equations corresponding to the monochromatic plane wave
solutions of the linear equations. \ Rosen also wrote that calculations of
pseudo-tensors for exact cylindrical wave solutions led him to conclude that
cylindrical waves carried no energy or momentum and that these results fitted
in with the conjecture that a physical system could not radiate gravitational
energy (Rosen 1956). \ The conjecture referred to by Rosen was made by Adrian
Scheidegger who had worked with Leopold Infeld in Toronto, as his research
student, on equations of motion (Scheidegger 1953; Infeld 1951). \ Rosen would
adhere to his view for decades and a similar attitude to gravitational
radiation would also be retained for a long time by his successor as
Einstein's assistant, Infeld.

In his autobiography Bondi noted that, after one of the confused discussions
in Berne about gravitational waves, Marcus Fierz, Professor at the
Eidgen\"{o}ssische Technische Hochschule Z\"{u}rich, (ETH Zurich), took him
aside and told him that the problem of gravitational waves was ready for
solution, and Bondi was the person to solve it (Bondi 1990a). \ Bondi took
this to heart.

\subsection{Gravitational radiation}

The years following the Berne conference saw a major attack on problems
related to gravitational radiation by members of the King's group. \ Between
1956 and 1965 a series of important papers and reviews were published by
members of, and visitors to, Bondi's group. \ \ An influential series of
lectures by Andrzej Trautman were given at King's in 1958\ and then reproduced
and circulated (Trautman 1958a). \ Reports of research from this period were
made at some of the early General Relativity and Gravitation conferences such
as those at Chapel Hill in 1957 (DeWitt C. 1957), Royaumont in 1959 (Bondi
1962; Penrose 1962; Pirani 1962a) and Warsaw in 1962 (Bondi 1964a; Penrose
1964a; Sachs 1964a). \ Ray Sachs and Roger Penrose lectured at the 1963 Les
Houches summer school on "Relativity, groups and topology" (Penrose 1964b;
Sachs 1964b) and two very influential reviews about gravitational radiation
were written by Pirani (Pirani 1962b; Pirani 1962c). \ Much of this work was
discussed when Trautman, Pirani and Bondi gave their 1964 Brandeis summer
school lectures (Deser \ and Ford 1965a).

\subsection{Bondi, Pirani and the Chapel Hill conference}

Both Bondi and Pirani attended the conference, at the University of North
Carolina, Chapel Hill, in January 1957. \ Organized by Bryce and C\'{e}cile
DeWitt\footnote{C\'{e}cile DeWitt-Morette}this was only the second
international conference devoted to general relativity. \ It was smaller than
the Berne conference, with only 45 listed participants and only ten of the
listed participants not based in the United States. \ It included most of the
leading workers in the field from the United States as well as more junior
people, including research students, and it was much more forward looking than
Berne had been. \ The proceedings were subsequently and speedily circulated to
the participants and included a record of\ discussions following the talks
(DeWitt C. 1957)\footnote{These are interesting and illuminating but the
editor, C\'{e}cile DeWitt, cautioned in her foreword that it should not be
believed that the report gave a perfectly true picture of the conference.}.
\ A subset of the talks was published in Reviews of Modern Physics (DeWitt B.
1957). \ The conference is now seen as a watershed in the history of general
relativity. \ The proceedings, with some additions, are now available on the
internet\footnote{The additions \ include a chapter "The Chapel Hill
Conference in Context" by one of the editors Dean Rickles, brief biographies
of the participants and an expanded version of remarks by Richard Feynman on
the reality of gravitational waves (DeWitt, Rickles 2011).}. \ A regular
series of triennial GR (General Relativity and Gravitation) international
conferences was later established. \ Berne is now labelled GR0, Chapel Hill
GR1 and the most recent, held in Valencia in 2019, GR22. \ The first few
conferences had the most distinctive impact because they were held when the
field was emerging from a period of slumber, the number of relativists was
still quite small and international meetings were much less frequent than they
are today.

The Chapel Hill programme covered a broad range of topics, including cosmology
where the steady state theory was vigorously discussed - Gold and Sciama were
also there. \ It was divided into two broad headings: unquantized (where Bondi
and Pirani made most of their contributions) and quantized general relativity.
\ Bondi chaired the session on gravitational radiation. \ In his introductory
remarks he noted the analogy between electromagnetic and gravitational waves.
\ \ However in Bondi's view the analogy

\begin{quotation}
doesn't go very far, holding only to the very questionable extent to which the
equations are similar. \ The cardinal feature of electromagnetic radiation is
that when radiation is produced the radiator loses an amount of energy which
is independent of the location of the absorbers. \ With gravitational
radiation, on the other hand, we still do not know whether a gravitational
radiator transmits energy whether there is a near receiver or not.

Gravitational radiation, by definition, must transmit information; and this
information must be something new.
\end{quotation}

Over the next few years Bondi would repeatedly return to and develop these
comments. \ He was willing to take less for granted than many others who
perhaps relied more heavily on that highly useful, but malleable, concept -
physical intuition. \ Many of the participants at the conference had enduring
memories of Bondi mimicking someone, suddenly and unpredictably, waving two
dumbbells about while asking if he was transmitting, gravitationally, energy
and information about what he was doing. \ Bondi then reported on research,
primarily by his first research student at King's, Leslie Marder, on
cylindrically symmetric waves and the transmission of energy. \ Bondi's talk
and comments reflected his uncertainty about gravitational waves at that time.
\ Marder's work became the first in a series of 16 papers by various authors,
"Gravitational waves in General Relativity \textrm{I-XVI}", published by the
Royal Society of London, between 1958 and 2004\footnote{Listed in the
appendix.}. \ \ All, bar one, were papers by authors who'd been in Bondi's
group at some time; the final paper in the series was by Bondi himself.

Bondi also gave a talk about negative mass and his paper on this topic was one
of those ready in time to be published in the Reviews of Modern Physics
collection (Bondi 1957a). \ Giving a talk on this topic might seem slightly
strange but Bondi was still working his way into the subject and produced an
interesting paper. \ He discussed the various types of mass defined in
Newtonian gravity and general relativity. \ He considered the axially
symmetric, static two body problem and showed that equilibrium situations
could not occur for bodies with positive mass. \ He then considered some
implications of allowing negative mass by investigating solutions of
Einstein's equations with uniformly accelerating pairs of bodies, one of
positive mass and one of negative mass. \ Aspects of this paper were used in
studies of black holes over a decade later.

Pirani gave two talks at the conference. In his first talk he outlined aspects
of his soon to be published paper on an invariant definition of gravitational
radiation (Pirani 1957b). \ This was his first paper specifically on
gravitational waves. \ Pirani, unlike Bondi, was not agnostic about
gravitational waves and he never had any reservations about gravitational
radiation. \ At that time, in common with many others, he thought that the
achievement of a proper understanding of them was a natural part of the
programme to quantize general relativity. \ In his paper Pirani considered the
non-linear empty space Einstein equations. \ He aimed to answer the question,
what is the covariant definition of gravitational radiation in general
relativity? \ He looked to the theory of electromagnetism for guidance in the
non-linear and more complicated gravitational case. Many, but not all, aspects
of his answer were eventually accepted and his paper significantly influenced
much subsequent work on gravitational waves and exact solutions of Einstein's equations.

Pirani wanted a definition that did not depend on any specific motion of an
observer, particular coordinate systems, coordinate conditions or the weakness
of fields. \ Two basic assumptions underlay his definition. \ The first, very
natural in the light of his previous work on measurement and the equation of
geodesic deviation (Pirani 1956b), was that gravitational radiation is
characterized by the Riemann tensor. \ By the principle of equivalence only
variations in the gravitational field, not the field itself, produce real
physical effects and it is the Riemann tensor which describes such variations.
\ The second, motivated by electromagnetic theory and general considerations,
was that gravitational radiation is propagated with the fundamental velocity
(the velocity of light) in empty space-time. \ He argued that these two
assumptions characterized gravitational radiation completely. \ Using
conditions taken from Lichnerowicz's recently published book (Lichnerowicz
1955), ensuring that the space-time was physically and mathematically
satisfactory, Pirani was led to conclude that a gravitational wave-front, such
as would result whenever a gravitational radiation source was switched on or
off, manifested itself as a discontinuity in the empty space Riemann tensor
across a null 3-surface. \ After investigating the physical effects of the
discontinuities, in an invariant way using the equation of geodesic deviation,
he was able to demonstrate the transverse nature of gravitational radiation.

While reading the proofs of the chapter entitled\textrm{ "}The electromagnetic
field in vacuo" in Synge's book on special relativity (Synge 1955) Pirani had
been struck by the fact that pure electromagnetic radiation (for example a
plane wave field) is represented by a null field and is algebraically
distinguished from more general fields such as those due to a system of
stationary or moving charges. \ \ Pirani formed the view that the
gravitational radiation field might be distinguished in an analogous way.
\ However the algebraic structure of the gravitational field is more
complicated than the electromagnetic one. \ Fortunately, as has already been
mentioned, he had read a review of Petrov's paper on the classification of the
Weyl tensor. \ In his paper Pirani made a ground breaking application of it to
gravitational radiation and, assisted by his reading of earlier work by Ruse
and others (Ruse 1946), made Petrov's scheme widely known, for the first time,
to the English language scientific community.

This classification was based on the fact that\ in four space-time dimensions
the algebraic symmetries of the empty space Riemann tensor (Weyl tensor)
enable it to be identified, pointwise, with a real, traceless $6\times6$
matrix. \ If this matrix has respectively 6, 4 or 2 distinct eigenbivectors,
then the Petrov type of the Weyl tensor is respectively \textrm{I},\textrm{ II
}or\textrm{ III}. \ The following refinement of this classification was
subsequently made by others. \ Let the real $6\times6$ matrix be replaced by a
complex $3\times3$ matrix. Then if, in type \textrm{I}, two of the three
possible eigenvalues are equal it is called type \textrm{D}, (Petrov type
\textrm{I} degenerate). \ If, in Petrov type \textrm{II}, the eigenvalue is
zero it is called type \textrm{N} (Petrov type \textrm{II}
null)\textrm{\footnote{Alternative formulations of the classification scheme
were developed later (Pirani 1962b; 1962c). \ These are usually more useful
but Petrov's is the one Pirani knew about in 1957.}}. \ Weyl tensors of type
\textrm{I} are termed algebraically general and those of the other types
algebraically special. \ The similar, but simpler algebraic classification of
the electromagnetic field tensor is made into two types, general and null,
analogous to the gravitational \textrm{I} and \textrm{N.}

Using Petrov's algebraic classification of the Weyl tensor into three types,
and working by analogy with the electromagnetic field, Pirani constructed a
detailed argument leading to the conclusion that only two of the three Petrov
types should be counted as radiation. \ He concluded that gravitational
radiation was present, at any empty space-time event, if the Riemann tensor
was of Petrov Type \textrm{II }or \textrm{III} but not if it was Petrov Type
\textrm{I}. \ He also observed that the difference between the no radiation
type and one of the radiation types could be made to correspond to the
discontinuity across a null 3-surface. \ In other words Pirani identified
(pure) radiation with algebraically special fields. \ This observation was
quickly seen to be an oversimplification (Kerr 2009), and misleading to a
degree, as Pirani would acknowledge (Pirani 1962b; 1962c).\ \ Nevertheless
this landmark paper quickly led to many new developments in both the theory of
gravitational radiation and studies of the solution space of Einstein's equations.

Pirani's second Chapel Hill talk "Measurement of classical gravitational
fields" was based on aspects of his recently published paper (Pirani 1956b).
\ As has already been mentioned this paper included results he'd outlined in
Berne. \ However reviewing McVittie's paper had sharpened his focus, as the
introduction makes clear. \ In this paper Pirani wrote

\begin{quotation}
A difficulty in general relativity theory is the lack of what might be called
a theory of measurement. \ One learns that all coordinate systems are
equivalent to one another, but one does not learn\ systematically how to
choose the appropriate coordinate system in which to calculate this or that
quantity to be compared with observation. Coordinate systems are usually
chosen for mathematical convenience, not for physical appropriateness. \ This
would not matter if calculations were always carried out in a manner
independent of the coordinate system, but this is not the case. \ The result
is fruitless controversy, like that over the harmonic coordinate condition.
\end{quotation}

As he had done in Berne, but now in much more detail, Pirani continued by
advocating the use of local frames, rather than local coordinates, in
observations. \ He employed an orthonormal tetrad system along an observer's
world line, constructed using the observer's velocity vector and, to represent
most closely the Newtonian concept of non-rotation, three Fermi propagated
space-like vectors. \ Pirani then investigated the relative motion of free
particles by using the orthonormal frame and the equation of geodesic
deviation. \ He explained how a freely falling observer, by measuring the
relative accelerations of a number of neighbouring free particles, could
determine the full Riemann tensor in its neighbourhood. \ He also explained
the connection with Newtonian mechanics and the Newtonian version of the
geodesic deviation equation.

This approach was quickly taken up and became textbook material (Weber 1961;
d'Inverno 1992) and it was the part of his paper he discussed in his Chapel
Hill talk. \ In his talk he reiterated the point that the physically
meaningful way to detect gravitational effects was to measure the relative
acceleration of neighbouring free particles. He commented that one could
easily imagine an experiment for measuring the physical components of the
Riemann tensor. \ In response to a question from Bondi at the end of the talk:

\begin{quotation}
Can one construct in this way an absorber for gravitational energy by
inserting a $\frac{d\eta}{dt}$ term, to learn what part of the Riemann tensor
would be the energy-producing one, because it is that part that we want to
isolate to study gravitational waves?
\end{quotation}

Pirani replied

\begin{quotation}
I have not put in an absorption term, but I have put in a "spring". \ You can
invent a system with such a term quite easily.
\end{quotation}

In fact Pirani had inserted such a term in the equation of geodesic deviation
in his Berne talk. \ \ Peter Szekeres, one of Pirani's research students at
King's, would later expand on this idea by constructing a "gravitational
compass". \ In his work Szekeres replaced Pirani's dust cloud by a tetrahedral
arrangement of springs and gave a more detailed analysis of its response to
waves (Szekeres 1965). \ Pirani's talk and paper would, subsequently, have a
significant influence on the development of ideas for gravitational wave
detectors (Saulson 2011; Blum 2018).

In his Chapel Hill conference summary Peter Bergmann said

\begin{quotation}
The work of Pirani, which gives a simple classical observation of the
components of the Riemann-Christoffel tensor should be accepted and made part
of our equipment. \ It enables us to set up a conceptual experiment to measure
a specified component of this tensor. \ This result should be a basic
component in the design of new experiments.
\end{quotation}

By the end of the conference the doubts entertained about the physical reality
of gravitational waves may have vanished, for most at least, but a completely
satisfactory understanding of them remained to be achieved. \ No doubt Pirani
was pleased about the reception given to his work. However when asked by
Kilmister what had impressed him most about the conference and gravity
research in the United States he replied, "they have a wonderful new device,
something called a Xerox machine. We should get one as quickly as possible".

While at the conference Bondi and Pirani had discussions with Joshua Goldberg
about financial support for the King's College group. \ Goldberg had recently
joined the General Physics Laboratory of the Aeronautical Research
Laboratories (ARL), at Wright-Patterson Air Force Base in Ohio, to begin a
research and support programme for gravity. \ ARL financial and travel support
had already been provided for the Chapel Hill conference. \ Their discussions
led to the King's group receiving substantial support from the US Air Force
from about 1958 until Bondi's final report in 1966 (Bondi 1966). \ Air Force
money underpinned a period of great activity at King's and elsewhere (Goldberg
1992). \ It enabled King's to become a leading centre of gravitational
research, attracting many post-docs and visitors for both long and short term
visits. \ As Bondi later observed about that period - the place hummed.

\subsection{Plane waves}

Soon after the Chapel Hill conference, in May 1957, Bondi published a letter
in Nature (Bondi 1957b) returning to the vexed question of an exact
gravitational wave solution of Einstein's equations. \ In the paper in Nature
Bondi wrote

\begin{quotation}
...Scheidegger$^{4}$ and I$^{5}$ have both expressed the opinion that there
might be no energy carrying gravitational waves at all in the theory. \ It is
therefore of interest to point out, as was shown by Robinson$^{6}$ and has now
been independently proved by me, that Rosen's argument is invalid and that
true gravitational waves do in fact exist. \ Moreover, it is shown here that
these waves carry energy,....
\end{quotation}

As mentioned previously Scheidegger's opinion was based on his and Infeld's
research on equations of motion. In fact it had already been shown that it was
erroneous to draw that conclusion from that work (Goldberg 1955).
\ Ivor\ Robinson was a talented investigator of exact solutions and tensor
calculus who, at that time at least, was extremely reluctant to publish his
research.\textit{ }\ Based at the University of Aberwystwyth, but a frequent
visitor to King's, he had given seminars about plane waves at King's and
Cambridge in 1956. \ He had rediscovered Brinkmann's solutions, giving them
physical significance for the first time, and pointed out the position of the
Rosen solution amongst them (Pirani 1962b; Rindler and Trautman 1987).

After his recantation Bondi outlined how the flaw in Rosen's 1937 argument,
that there were no exact plane-wave metrics filling all of space-time, could
be remedied and announced that further work would soon be published. \ He also
modelled the reality of gravitational wave energy transfer by considering a
system of test particles set in motion by such a wave and noting that the
energy gained by that system could then be used by, say, letting them rub
against a rigid disc. \ Bondi expanded on this line of thought at the
Royaumont GR2 conference in 1959 (Bondi 1962)\footnote{Bondi's Royaumont talk
also included a discussion of work he had done with William McCrea on energy
transfer in Newtonian theory (Bondi, McCrea 1960). \ That line of work was
continued by Bondi's student Henry Levi who investigated transfer of energy by
gravitational induction in general relativity. \ He studied near-field
transfer of gravitational energy for quasi-static axisymmetric systems using a
perturbation method and defined a relativistic analogue of the Newtonian
Poynting vector in Bondi's paper. \ Levi concluded that a quasi-static
axisymmetric system could lose energy only in the presence of a receiver (Levi
1968).}. \ This type of so-called "sticky bead argument" had also been
mentioned by Feynman during the Chapel Hill conference in a brief remark which
he subsequently expanded (DeWitt C., Rickles 2011).

Bondi's letter in Nature was followed, about eighteen months later, by a
detailed examination of plane wave solutions. \ This joint work of Bondi,
Pirani and Ivor Robinson demonstrated once and for all that non-singular plane
wave solutions of Einstein's empty space-field equations existed and
transferred energy (Bondi 1959). \ In its way it was the first modern, global
analysis of a physically important space-time . \ Their geometrical point of
view was that space-time was a differentiable manifold as defined in modern
differential geometry. \ That is, it was a topological space covered by sets
of coordinate charts which satisfied compatibility (differentiability)
conditions in their intersections. \ They used Lichnerowicz's compatibility
requirements for a space-time (Lichnerowicz 1955). \ These were less stringent
than the assumption, commonly made in the past, that the space-time manifold
was covered by a single coordinate system. \ That assumption had led to
confusion, by Rosen and others, between coordinate singularities (like those
at the origin of polar coordinates) and physical singularities which could, in
principle at least, have observable consequences.

The first question the authors addressed and answered was - what is the
invariant, that is, coordinate independent, definition of a plane wave?
\ Confining themselves to empty space solutions of Einstein's equations they
demanded that a plane gravitational wave space-time be one having the same
degree of symmetry as a plane electromagnetic wave in Minkowski space-time.
\ This meant that a plane wave metric should admit a 5-parameter group of
motions (isometries). \ Inspecting analyses of metrics with symmetries that
had been carried out by Petrov (Petrov 1969) they found that there was one
class of empty space solutions with such a group of isometries. \ They were
thus led to consider in detail the metrics listed in Bondi's Nature paper.
\ It was sufficient for their purposes to consider the case where the plane of
polarization was fixed and the metric reduced to Rosen's metric%
\[
ds^{2}=(\exp2\varphi)(d\tau^{2}-d\xi^{2})-u^{2}\{(\exp2\beta)d\eta^{2}%
+(\exp-2\beta)d\zeta^{2}\},
\]
where $u=\tau-\xi$, $\beta=\beta(u)$, $\varphi=\varphi(u)$ and $\frac{d}%
{du}\varphi=u(\frac{d}{du}\beta)^{2}$. \ They then considered sandwich waves,
that is waves with non-zero amplitudes of finite duration, with the space-time
elsewhere being flat. \ This space-time was globally non-singular. \ They
then\ evaluated the effect of the wave, on a family of observers at relative
rest in a Minkowskian inertial frame before its arrival, and showed that such
observers were relatively in motion after the wave had passed. They showed
that the effect of the wave was to develop a relative acceleration between
freely-moving observers with their relative velocity increasing with
separation. \ This enabled them to draw the conclusion that, as Bondi and
Feynman had previously noted, gravitational waves transport energy.

They also followed up Pirani's attempt at an invariant formulation of
gravitational radiation (Pirani 1957b). \ They noted that although the plane
wave metrics considered satisfied the definition given in Pirani's paper - the
Riemann tensors of their metrics were Petrov type \textrm{II} - they
considered that definition of gravitational radiation too severe. \ They
commented that\ discussions with other workers had led them to conclude that
Pirani's definition applied only to pure radiation. \ They observed that in a
more general radiating situation, while the dominant Riemann tensor terms
might be type \textrm{II, }other terms would be present and the Riemann tensor
might actually be of type \textrm{I}$.$ \ This paper settled the long standing
debate about plane waves, for most people at least.

A further insight into plane gravitational waves was obtained by Pirani.
\ While on leave in the United States he recalled that the electromagnetic
field of a fast moving charge resembled a plane electromagnetic wave. \ Pirani
realized that he could exhibit a similar phenomenon for gravitational waves by
employing results on radiation and the algebraic structure of the Riemann
tensor from his 1957 paper. \ He showed, in an invariant manner, that the
gravitational field of a fast moving mass increasing resembled a gravitational
plane wave field the greater its speed (Pirani 1959).

\subsection{The Polish connection}

In 1957 Pirani visited Infeld who had been based at the Institute of Physics
of the Polish Academy of Sciences in Warsaw since 1950 and had formed a
relativity group there. \ Infeld was still mainly involved in improving the
approximation method of dealing with equations of motion first formulated
while he was Einstein's assistant at Princeton (Einstein 1938). \ \ The
Einstein--Infeld--Hoffmann method describes the approximate general
relativistic dynamics of a system of point-like masses due to their mutual
gravitational interactions. \ In order to avoid using energy momentum tensors,
regarded by Einstein as the not totally satisfactory part of his field
equations, and work only geometrically, particles were treated as
singularities in empty space-time. \ This in itself was not a problem, for the
method employs surface integrals surrounding the singularities.
\ Approximations and series expansions were made which are valid when speeds
are small compared to the speed of light and gravitational fields are weak.
\ The primary purpose of their work was to show that the motion of the
singularities was determined by the empty space field equations, in contrast
to electromagnetism, where the Lorentz force law is not a consequence of
Maxwell's equations. The EIH\ type expansions were ill-suited to dealing with
gravitational radiation. \ Their use in the investigation of radiation terms
in the expansions misled some, including Infeld and at times Einstein, into
concluding that gravitational radiation was not emitted by freely gravitating
bodies. \ In Warsaw Pirani gave a talk dealing not with approximation methods
like these but with his attempts to formulate an invariant definition of
gravitational waves.

Amongst Infeld's PhD students was Andrzej Trautman who so impressed Pirani
with his work on radiation that he invited him to London. \ Before joining
Infeld's group Trautman had obtained a Master's degree in radio engineering
and this must have influenced his attitude towards radiation problems (Penrose
1997). \ A hiccup followed. \ Pirani realized that the King's U.S. Air Force
grant could not be used to support visitors from Poland (and other countries).
\ Bondi had \ to hustle around and raise funds - something he was good at.
\ Trautman eventually arrived at King's and after a couple of weeks, during
which he improved his spoken English by chatting to Bondi's students, gave a
series of five lectures between May and June 1958\footnote{Trautman's visit
was his first to an English speaking country. \ He could read English but had
not had a school or university education in the language and had no experience
of speaking it. \ Before his visit, in preparation for his lectures, he took
10 or so private English lessons.\ }. \ These were mimeographed, widely
circulated, and constituted the first report published with ARL support
(Trautman 1958a). \ His first lecture was about boundary conditions for
gravitational wave theory and included material from two papers then in press
(Trautman 1958b; 1958c). \ The next three lectures covered equations of motion
and gravitational radiation, propagation of gravitational disturbances,
conservation laws and symmetry properties of space-time, and the fast
approximation method. \ The final lecture was on the equations of motion of
rotating bodies. \ While the material related to equations of motion was
considered to be of interest, and at one time Pirani thought it should be
followed up, it was not pursued at King's. \ General agreement on\ results
from approximation methods and equations of motion was not to be obtained for
many years (Kennefick 1997).

In his first lecture, probably the one that had most direct impact on the
King's group, Trautman outlined his approach to boundary conditions on
gravitational fields due to isolated matter systems. \ He reformulated
Arnold\ Sommerfeld's boundary conditions for radiative solutions of the scalar
wave equation in Minkowski space-time so that it was easier to see how to
generalize them, first to the electromagnetic field and then to the general
relativistic gravitational field. \ The conditions he imposed on space-time
metrics, in the wave zone far from spatially bounded matter sources, were
strong enough so that space-times satisfying them had finite total energy but
weak enough so that they were also satisfied by gravitationally radiating
space-times. \ The space-times were asymptotically flat, approaching Minkowski
space-time with increasing distance from any sources. \ Although asymptotic
flatness is an idealization it is a physically reasonable one for many
non-cosmological systems.

In a far field analysis Trautman evaluated the energy-momentum of the system
at infinity, using the boundary conditions and the von Freud superpotential
(von Freud 1939) for the Einstein pseudotensor. \ He showed that it was finite
and did not depend on any particular coordinate system adapted to his boundary
conditions. \ He noted that within his framework the total energy of the
system, evaluated at infinity, could only be radiated away and that,
asymptotically, the Riemann tensor of a radiating system was algebraically
special, more specifically it was type \textrm{N }(Petrov type\textrm{ II
}null). \ It was in the wave zone, where wavefronts become increasingly
planar, that the algebraically special condition came into play. \ Pirani's
algebraically special condition for radiation is strictly local, for the
Petrov type can change from point to point, but Trautman's work highlighted
the non-local nature \ of gravitational radiation.

To summarize, for systems with spatially bounded matter sources Trautman took
gravitationally radiating space-times to be ones with space-time metrics which
satisfied certain boundary conditions at infinity. \ These included the
condition of asymptotic flatness. \ It then followed that their total energy,
evaluated at infinity, was well defined and was radiated away. \ Although
there was some lack of clarity about where the boundary conditions should be
applied, and about incoming radiation (Walker 1979), Trautman's work was a
very important step forward (Hill and Nurowski 2017).

Kilmister later recalled that the lectures were very clear and both Bondi and
Pirani paid close attention. \ Bondi, Pirani and Robinson's 1959 paper was
then slowly moving towards completion and Trautman's influence is acknowledged
in it. \ His King's visit also enabled him to meet Ivor Robinson and begin a
collaboration on a class of algebraically special exact solutions of
Einstein's equations, the Robinson-Trautman solutions. \ These satisfied
Trautman's boundary conditions and could be interpreted as describing waves
coming from bounded sources (Robinson and Trautman 1960; 1962). \ Trautman's
wife R\'{o}\.{z}a Michalska-Trautman, also one of Infeld's students,
eventually convinced Infeld to change his mind about gravitational radiation
and wrote a number of papers with him on the topic.

\subsection{The early 1960s}

The total King's student population in 1960 was only a few thousand and King's
still functioned very much as a College of the University. \ Bondi's lectures
to undergraduates from that time are particularly remembered although many
found them demanding. \ He never had notes apart from scribbles on an envelope
which he would occasionally pull from his pocket. \ Bondi roamed
freely.\ \ Examinations were University based and appropriate preparation for
them was often left to a more junior member of staff. \ By now the relativity
group was well settled in. \ There were frequent interactions with groups in
continental Europe, such as those in Paris, Brussels, Hamburg and of course
Warsaw, where similar research was being undertaken, often facilitated by the
US Air Force grant. \ The relativity seminars were well established and drew
in many from outside London. Bondi, Pirani and Ivor Robinson, who was often
there, always sat in the front row at seminars and could make it a challenge
for a speaker to complete his talk. \ Mostly this arose from a desire to
understand precisely what was going on and the atmosphere was friendly, not
hostile, but occasionally it was just horseplay, particularly when they
themselves were the speaker\footnote{Apparently once when starting his talk
Ivor Robinson opened his mouth and said "I" at which point Bondi interrupted
with "Are you sure it was you?" \ "Of course, why do you ask?" \ "Because I
was determined to interrupt before you finished the first sentence."
\ Audiences usually found this sort of thing amusing but younger speakers and
students could find it unnerving.}.

By the early 1960s the work done at King's and elsewhere had been taken on
board by more than those actively involved in the theoretical and conceptual
work. \ For instance Joseph Weber, the pioneering constructor of gravitational
wave detectors, included expositions of the work of Bondi and Pirani in his
monograph (Weber 1961). \ His book included a chapter on the detection and
generation of gravitational waves but few at that time thought they would live
to witness earth based detection. \ Many of them were right.

In 1962 two influential books appeared which included reviews intended to make
the work on radiation more accessible to a wider audience. The first,
\textit{Recent Developments in General Relativity} was a festchrift volume in
honour of Infeld's 60th birthday. \ The second covered a broad range of topics
in relativity and includes a number of landmark overviews (Witten 1962).
\ Both books included two widely read articles by Pirani. These recounted not
only aspects of his own research but also the work of numerous others on what
he termed "the covariant part of gravitational wave theory" as opposed to
approaches using approximation methods (Pirani 1962b; 1962c). \ In the first
of these reviews he noted that subsequent research had changed his attitude to
his conclusion, in his 1957 paper, that gravitational radiation had to be
algebraically special. \ He explained quite forcefully that he now viewed that
conclusion as a misleading oversimplification.

Pirani's reviews were written too early to do more than mention in passing
what was one of the most compelling papers on gravitational radiation to
emerge from the King's group. \ This was number \textrm{VII} in the
"Gravitational waves in general relativity" series, the paper by Bondi, Van
der Burg and Metzner.

\subsection{Radiation from bounded sources - a new approach}

A new approach to gravitational radiation was formulated by Bondi with his
collaborators, Julian van der Burg and Kenneth Metzner. \ It resulted in the
first systematic treatment of quite general metrics describing radiation from
bounded sources (Trautman 1966) and it gave the first clear understanding of
mass loss due to gravitational radiation (M\"{a}dler and Winicour 2016).
\ Their work was highly influential and initiated a highly productive period
of research at King's. \ Both the content of their paper and the way it was
written are particularly interesting and will be discussed in some detail.

Julian van der Burg had come to King's as an 18 year old undergraduate in
1953\footnote{Material in this section makes use of \ Julian van der Burg's
reminiscences during discussions with me in 2009.}. \ With fewer than 20 new
mathematics students a year and a small staff he found the mathematics
department an intimate and friendly place. \ After he had finished his first
degree he became Bondi's second King's PhD student. \ Bondi's students, Marder
and van der Burg, would sometimes take it in turns to camp outside their
occasionally elusive advisor's office in the hope of catching him. \ Research
students always relied heavily on Pirani for back up support. \ After
submitting his PhD thesis in June (van der Burg 1959) van der Burg was on the
point of leaving to spend the summer at home when Bondi told him that he had
an idea about gravitational waves. Bondi asked if he would stay on for three
months and work on it over the summer. \ Bondi felt that there was now a good
understanding of plane and cylindrical waves and he wanted to look carefully
at the emission of waves by an isolated body. \ He was still suspicious about
the value of results from the linearized form of Einstein's equations as the
full equations are highly non-linear. \ For Bondi the essential question about
"gravitational waves" was - did they transport energy? \ In his view that was
a fundamentally non-linear phenomenon (Bondi 1990b).

Initially Bondi's idea consisted of little more than using coordinates based
on wave fronts, that is on null hypersurfaces. \ Bondi and van der Burg spent
the summer trying to find the field variables which would make Einstein's
equations tractable when such coordinates were employed. \ Bondi would come in
every so often with something scribbled on the back of an envelope and say,
"try this". \ In September van der Burg went off to the University of
Liverpool facing what was then standard for junior people, a heavy lecturing
load. \ They swapped letters fairly regularly. \ Suddenly in January 1960
there seemed to be some progress and van der Burg had a set of field equations
which looked hopeful. \ A week after sending them to Bondi he received a
letter back saying that the equations could be used to prove that
gravitational waves did not carry energy and a letter had been written to
\textit{Nature}. \ Trying to see what Bondi had done van der Burg located a
quadratic term in one of the equations implying the opposite. \ He mailed
Bondi about the equation and wrote "Have you lost the quadratic term?"; by
return \ came a postcard, "I had lost it, have cancelled the letter to
Nature". \ After that van der Burg did nothing further on the problem being
fully occupied with his teaching.

In May Bondi published a brief letter in \textit{Nature} about the results
that he and van der Burg had obtained (Bondi 1960). \ This included the
coordinates and metric form they had used and a discussion of mass loss in
various situations. \ He related one of these situations to "Infeld's result
that a set of freely moving particles does not radiate". \ Infeld's paper was
the only one cited (Infeld 1959)\footnote{Bondi was notoriously poor at
reading other people's work. \ He would often ask other people to read a paper
for him. \ Kilmister once said he didn't mind this too much because when he
was reporting on a paper Bondi would get the point so quickly and clearly that
the paper was clarified for him too.}. \ Bondi had gone to Cornell University
in March for a sabbatical term. By that time he felt that the problem was
essentially solved. \ However the transformation properties of the metric
remained unclear so at Cornell he invited a student A.W.K.(Kenneth) Metzner,
whose PhD was supervised by Philip Morrison, to investigate these.

Although he gave talks about his results at King's and other places, and had
written the brief letter to Nature, Bondi was slow to write up all the work.
\ He may have eventually been prompted to get on with it by knowing that Ted
Newman and Roger Penrose were covering similar ground but using a null tetrad
approach (Bondi 1987). \ Sometime in 1961 Bondi sent van der Burg a draft of a
lengthy paper. \ Bondi had made a substantial number of changes to their joint
work \ and the section signed by Metzner was a surprise. "To this day I have
no idea who Metzner is" van der Burg said in 2009. \ Finally, in 1962, the
research of Bondi, van der Berg and Metzner was published (Bondi 1962).

The paper is unusual in that different sections are signed by different
authors although it is clear that Bondi wrote the whole paper. \ Part A was
signed by Bondi and included discussions of causality, mass loss and Huygen's
principle. \ The method of treatment of radiation used in the paper was
illustrated by considering the much simpler case of the scalar wave equation
in Minkowski space-time. \ The paper aimed to investigate retarded solutions
from spatially bounded sources. \ The boundary conditions were therefore
proposed as ones for empty space times, outside isolated material systems,
which tended to flatness at infinity and in which only outgoing waves were present.

Bondi rehearsed some of his past concerns about previous radiation
calculations. \ He doubted that the results of the linearized theory could
always be fully trusted and observed that the non-linearity of the full
equations might well affect crucial properties of solutions. \ Furthermore it
was not clear to him that approximate solutions always corresponded to exact
solutions. \ He conceded that by then a lot was known about exact
gravitational wave solutions with planar or cylindrical symmetry. \ Whether or
not they displayed the important characteristics of waves from bounded sources
-the physically significant case - was, in his view, open to question. \ He
dwelt on the importance of mass loss and noted that a real physical wave must
convey energy. Hence outgoing waves must diminish the energy, and therefore
the mass of the source. \ Whether or not after the end of an excitation a wave
rings on, that is has tails, or wave motion ends was also a question to be investigated.

Part B of the paper was signed by both Bondi and van der Berg and contains
details of their joint work. \ First coordinates and field variable choices
which would make analyzing the field equations a tractable proposition were
detailed. \ Careful attention was paid to the interpretation of the
coordinates which were based on light-like hypersurfaces, with null normals
tangent to null geodesics ruling the hypersurfaces. \ The coordinates chosen
were a retarded time parameter, labelled $u$, where surfaces of constant $u$
were outgoing null hypersurfaces, a corresponding luminosity or areal
coordinate $r$, ranging from some finite value to future null infinity, as it
became known, along the ruling outgoing light rays and two angular coordinates
$\theta$ and $\phi$ which ranged over a two sphere and distinguished the
ruling light rays one from another. \ The underlying assumption was that a
suitable patch of the space-time manifold for far field analyses was being
considered so that the space-time topology was Euclidean there and the
topology of the null hypersurfaces was $\mathbb{R}\times S^{2}$. \ The use of
these coordinates and the focus on null hypersurfaces, as opposed to
space-like hypersurfaces, in the analysis of Einstein's equations was trail
blazing. \ One of their motivations was to avoid the appearance of terms
involving log$r$. \ Such terms had hindered other investigations.

In order to see their way through the calculations they made a couple of
simplifying assumptions which they thought would not affect their central
results. \ They assumed that the system they were considering was axially
symmetric, so metric components were independent of $\phi$, and they also
assumed that the metric was reflection symmetric, that is invariant under the
discrete transformation $\phi\rightarrow-\phi$, so reducing the number of
metric components they had to consider. \ Next they explained their choice of
metric variables and exhibited the class of metrics they were to consider.
\ These took the form%
\[
ds^{2}=(Vr^{-1}e^{2\beta}-U^{2}r^{2}e^{2\gamma})du^{2}+2e^{2\beta}%
dudr+2Ur^{2}e^{2\gamma}dud\theta-r^{2}(e^{2\gamma}d\theta^{2}+e^{-2\gamma}%
\sin^{2}\theta d\phi^{2}),
\]
where the four functions $U,V,\beta$ and $\gamma$ are functions of $u,r$ and
$\theta$.

They then presented a systematic procedure for integrating Einstein's vacuum
field equations. \ They divided the equations into groups. \ First were the
main equations, themselves divided into three hypersurface equations which had
no derivatives with respect to the retarded time coordinate $u$, and a
"standard" equation which involved such a derivative. \ Second was a trivial
equation which was a consequence of the main equations. \ Third were two
equations called supplementary conditions. \ The latter held everywhere if
they held on a hypersurface of constant $r$ and the main equations held everywhere.

They next set the boundary conditions. \ Space-time was required to have
Euclidean topology at large distances from the source and to permit
gravitational radiation, with the metric satisfying an outgoing radiation
condition similar to that of Sommerfeld. \ They assumed that space-time was
asymptotically flat, so that in their coordinates as $r\rightarrow\infty$,
\[
ds^{2}\rightarrow du^{2}+2dudr-r^{2}(d\theta^{2}+\sin^{2}\theta d\phi^{2})
\]
at future null infinity. \ It was also assumed that all metric components, and
other quantities of interest, could then be expanded in powers of $r^{-1}$as
$r\rightarrow\infty$ along each null geodesic ray in each hypersurface of
constant $u$. \ With these assumptions they deduced from the field equations
that the leading terms in the expansion of the metric\ quantities $\gamma,U$
and $V$ were $\gamma=c(u,\theta)r^{-1}$ , $\beta=-\frac{1}{4}c^{2}r^{-2}$ ,
$U=-(c_{,\theta}+2c\cot\theta)r^{-2}$, $V=r-2M(u,\theta)$, where a comma
denotes partial differentiation. \ In fact they computed more terms, from
which they were able to reach further conclusions, but those will suffice
here. \ They were then able to see that the future development was determined
by the two supplementary conditions. \ One gave the time derivative of $M$ in
terms of derivatives of the function $c$ so if $M$ was given for one value of
$u,$ and $c$ was given as a function of $\ u$ and $\theta$, its entire time
development was determined. \ Similar results followed from the second
supplementary condition. \ In the well understood empty space, asymptotically
flat, static Weyl metrics - transformed to their coordinate system - $c$ was
just an arbitrary function of $\theta$ and the function $M$ was essentially
the constant mass of the Weyl system. \ Hence they called $M$ the mass aspect
of the system. \ In a similar way they related other metric functions to the
dipole and quadrupole moments of the system.

Part C of the paper was signed by Kenneth Metzner alone. \ It contains a
computation of coordinate transformations, evaluated using an expansion in
$r$, that preserved the form of their metric and the asymptotic flatness
conditions that $U,\beta$ and $\gamma$ should tend to zero at infinity. \ The
transformations were found to be determined by a single constant and an
arbitrary function $\alpha(\theta)$. \ The constant corresponded to a Lorentz
transformation along the axis of symmetry. \ This was to be expected. \ The
unexpected result was the presence of the arbitrary function.

Part D, entitled "The Nature of the solutions", was signed by Bondi alone and
was a lengthy discussion of the meaning of the results. \ He observed, from
the results in Part B, that if changes in the source led to changes in the
field they could only do so by affecting the time derivative of the function
$c,$ and vice versa, so all the "news" was contained in that quantity, hence
he termed $c_{,u}$ "the news function". \ Furthermore, guided by a comparison
with static systems, where the definition of mass was unambiguous,\ he defined
the mass of the system at future null infinity to be the mean value over the
unit sphere of the mass aspect, that is $m(u)=\langle M\rangle$ . \ This
agreed with the usual mass for static systems and is now commonly referred to
as the Bondi mass (sometimes the Trautman-Bondi mass). \ It then followed from
one of the supplementary conditions that $m,_{u}=-\langle(c,_{u})^{2}\rangle$.
\ Bondi described this as "the central result of this paper". \ He continued

\begin{quotation}
The mass of a system is constant if and only if there is no news. \ If there
is news, the mass decreases monotonically as long as the news continues.
\end{quotation}

Unlike Trautman Bondi had avoided using pseudotensors and superpotentials to
calculate\ the energy loss presumably, in part at least, because of the
continuing uncertainty about their status and reliability. \ However,
subsequently various people used the framework of this paper to reobtain these
results employing pseudotensors and pseudopotentials. \ His approach, using
the field equations directly, gave him greater control when interpreting the
results. \ He noted that, while the mass loss result might appear to depend on
the way he had defined it, the physical significance of $m$ as mass was clear
when systems initially and finally static were considered

\begin{quotation}
a dynamic period interposed between two static periods is bound to imply a
loss of mass. \ We can ascribe this in the only physically reasonable way to
the emission of waves by the system.
\end{quotation}

Whether or not\ such transitions were possible was a question that would come
to the fore in a few years' time.

Bondi then explored and explained this new framework by discussing the
linearized equations, ways to construct solutions, non-radiative motions and
the reception of gravitational waves. \ Within a few years analyses using
coordinates based on null hypersurfaces, incorporating the general philosophy
of this paper, became commonplace.

\subsection{Radiation and geometry}

In 1960-1961 Rainer (Ray) Sachs and Josh Goldberg each spent a year visiting
Bondi's group and for two years, from 1961, Roger Penrose held a post-doctoral
position with the group. \ Goldberg and Sachs had been students of Peter
Bergmann at Syracuse University\footnote{Sachs came with partial support from
an A.E. Norman Foundation grant. \ Financial support for Pirani's activities
and interests was supplemented by this foundation until the 1990's.}.
\ \ Sachs came to King's as a post-doc after having held a similar position in
Hamburg. \ There he had been a member of Pascual Jordan's group which at
various times included people, like Engelbert Sch\"{u}cking, Wolfgang Kundt
and J\"{u}rgen Ehlers, whose work on exact solutions and other topics
influenced the King's group. \ Goldberg came on leave from the Aerospace
Research Laboratory on a National Science Foundation Senior Post-doctoral
Fellowship. \ Both Sachs and Goldberg had worked on problems related to
gravitational radiation and equations of motion. \ Goldberg and Sachs had
backgrounds in physics but Penrose's background was rather different. \ He had
been a student of the pure mathematicians William Hodge and then John Todd at
Cambridge. \ His 1956 PhD thesis was entitled "Tensor methods in algebraic
geometry".\ \ Penrose was one of the many people in Britain who had been
influenced by Bondi's BBC radio talks. \ While at Cambridge he had been
encouraged by Dennis Sciama to further his interest in physics and he attended
lectures by Bondi and Dirac. \ From the latter he had learned about
two-component spinors and he had significantly extended their previous
application, by Louis Witten and others, to general relativity (Penrose 1960).
\ He had demonstrated that the use of two-component spinors, instead of
tensors, not only simplified many calculations but also shed new light on
them, as it did on Petrov's classification of the Weyl tensor. \ Complicated
expressions, previously obtained only by very skilful use of tensors, became
almost transparently obvious. \ Sachs, Goldberg and Penrose all made
significant contributions during their time at King's.

Sachs wrote a number of papers directly related to the work of the group. \ In
his first he proposed a covariant outgoing radiation condition for gravitation
(Sachs 1961). \ His investigation included a study of the geometry of
congruences of null geodesics and in this paper he employed what he termed a
quasi-orthonormal tetrad field. \ This was a basis of three null vectors, two
real and one complex. \ He noted that bases like this were closely related to
the spinor formalism used by Penrose;\ later research, particularly by Newman
and Penrose, made considerable use of such null tetrads. \ This part of the
paper was based on his recent work with Ehlers in Hamburg (Jordan1961) and it
also contained a discussion of the optical scalars of null geodesic
congruences. \ Sachs explored the geometrical properties of light rays (null
geodesics) in terms of their expansion, twist and shear. \ These were concepts
analogous to ones in classical hydrodynamics and fluid flow. \ The latter had
previously been generalized to the relativistic domain and applied to
congruences of time-like curves by Otto Heckmann and Engelbert Sch\"{u}cking.
\ In the case of null geodesic congruences Sachs explained how to interpret
these quantities in terms of the first order change in the properties of
shadows produced by light rays falling perpendicularly on a small plane
circular disc. The expansion (or divergence) gives a measure of the expansion
of the shadow and the rotation (or twist) gives a measure of its rotation.
\ The rotation vanishes if and only if the congruence is hypersurface
orthogonal. The shear (or distortion) gives a measure of the distortion from a
circular shape to an elliptical one. \ These quantities, and the equations for
their rate of change along rays in congruences of null geodesics (along with
the analogous equations for time-like geodesics), were to play an important
role in future investigations of many aspects of general relativity. \ The
latter included exact solutions, geometrical optics, gravitational lensing,
black holes, singularity and global theorems (Stephani 2003; Ashtekar 2015).

Sachs introduced the concept of "geodesic rays" and pointed out that metrics
with geodesic rays formed a class which naturally generalized the class of
algebraically special metrics. \ He suggested that a gravitational field with
bounded sources was free of mixed (incoming and outgoing) radiation at large
distances if and only if its fall off was such as to admit, to appropriate
order, "asymptotically geodesic rays". \ His exploration of covariant
radiation conditions also led him to discuss the so-called asymptotic
"peeling" behaviour of the Riemann tensor,\ both in the linearized theory and
in certain cases of the full nonlinear theory. \ Later it was understood that
peeling behaviour, which Sachs related to geodesic and asymptotically geodesic
rays, did not in fact exclude all possible incoming radiation.

In a second paper, Sachs generalized the results of Bondi, van der Burg and
Metzner by dropping their conditions of axial and reflection symmetry (Sachs
1962a). \ He again considered the empty space equations and followed the still
novel idea of using a retarded time parameter $u$ as a coordinate, where the
level sets of $u$ were outgoing null hypersurfaces. \ In what are now often
termed Bondi, or Bondi-Sachs, coordinates the metrics he considered took the
form%
\[
ds^{2}=(Vr^{-1}e^{2\beta})du^{2}-2e^{2\beta}dudr+r^{2}h_{AB}(dx^{A}%
-U^{A}du)(dx^{B}-U^{AB}du),
\]
where
\[
2h_{AB}dx^{A}dx^{B}=(e^{2\gamma}+e^{2\delta})d\theta^{2}+4\sin\theta
\sinh(\gamma-\delta)d\theta d\phi+(\sin\theta)^{2}(e^{-2\gamma}+e^{-2\delta
})d\phi^{2}.
\]
There were now six functions of the coordinates, $V,\beta$,$U^{A}%
,\gamma,\delta$ rather than four and correspondingly more equations to
solve\footnote{\ In the Bondi et al metric $U^{2}$ and $\delta$ were zero.}.
\ Nevertheless the parametrization meant that the equations formed similar
groups to those considered by Bondi et al. \ They could be analyzed in the
same way and similar conclusions could be drawn from them. \ The
parametrization of the conformal two-metric $h_{AB}dx^{A}dx^{B}$ explicitly
exhibited the two functions $\gamma$ and $\delta$ corresponding to the two
independent modes of polarization of gravitational waves.

Sachs observed that a major conclusion of his paper was that in most arguments
there would be no essential loss of generality if attention was confined to
the axially symmetric case. \ While this was true his paper was interesting
and important in its own right. \ The metric applied to general radiating
systems and so could also apply to systems with rotating stellar bodies. \ It
had a more geometrical focus than the Bondi et al paper and it included a more
detailed analysis of various topics such as the behaviour of the Riemann
tensor. \ Sachs showed that the five leading terms in the asymptotic expansion
of the Riemann tensor satisfied the "peeling" property which he had discussed
in his previous paper (Sachs 1961). \ At any space-time event there is a
special set of four "principal null directions" defined by a non-zero Weyl
tensor and these can be used as a way of defining Petrov types (Debever 1959;
Penrose 1960). \ They are all distinct in the algebraically general case and
some or all (as in the Petrov type null case) coincide in the algebraically
special cases. \ The asymptotic expansion, as $r\rightarrow$ $\infty$, of the
Riemann tensor followed from the metric boundary conditions and took the form%
\[
R=\frac{_{0}N}{r}+\frac{_{0}III}{r^{2}}+\frac{_{0}II}{r^{3}}+\frac{_{0}%
I}{r^{4}}+\frac{_{0}I^{^{\prime}}}{r^{5}}+...
\]
This expansion, along outgoing null geodesics, exhibited the "peeling"
behaviour, that is the way in which the principal null directions "peeled off"
as$\ r$ decreased, with the asymptotically dominant $r^{-1}$ term taking the
plane wave Petrov type $N$ form\footnote{The tensors with subscripts $0$ are
parallely displaced along each ray. \ $I$ and $I^{^{\prime}}$ may be different
but both are algebraically general.}.

In this, and a subsequent paper completed while he was fulfilling his draft
requirements in the United States Army Signal Corps (Sachs 1962b), Sachs
analyzed the asymptotic symmetry group in much greater detail than had been
done before. \ His work, on what he termed the generalized Bondi Metzner (GBM)
group, led to the conclusion that the group was isomorphic to the semidirect
product of the homogeneous orthochronous Lorentz group and the
supertranslations. \ The latter were generated by an arbitrary function
$\alpha$, as in the Bondi et al paper, but now of two variables, rather than
one, $\alpha(\theta,\phi)$. The supertranslations formed an infinite
dimensional abelian normal subgroup with corresponding factor group the
homogeneous orthochronous Lorentz group. \ When $\alpha=0$ the asymptotic
symmetry transformations correspond to Lorentz transformations. Under
supertranslations, $\theta\rightarrow\theta,$ $\phi\rightarrow\phi$,
$u\rightarrow u+\alpha(\theta,\phi)$.

The asymptotic symmetry group is a generalization of the Poincar\'{e}
(inhomogeneous Lorentz group) with the four parameter subgroup of translations
being enlarged to the infinite dimensional supertranslation subgroup. \ The
supertranslations consisting of the $l=0$ and $l=1$ spherical harmonics
constitute the only invariant four dimensional sub-group and correspond to
rigid time and space translations. \ Hence there is an unambiguous definition,
at null infinity, of the total energy-momentum. \ Because the Lorentz
group\ is not a normal subgroup of the GBM group there is an ambiguity in the
definition of total angular momentum. \ Sachs noted that he was unable to
satisfactorily eliminate or restrict the function $\alpha$ by imposing
additional conditions on the metric at some retarded time. \ It was
subsequently understood that only in special cases, such as stationary
space-times, can the Poincar\'{e} group be singled out from the GBM group
(Newman and Tod 1980).

Like many of the younger relativists Sachs hoped that an understanding of
gravitational radiation would assist in the quantization of general
relativity. \ He had this in mind, and the possibility of an S-matrix theory
of gravity, when writing about the GBM group. \ In the light of his work the
group is generally known today as the BMS (Bondi, Metzner, Sachs) group. \ In
subsequent years representations of the BMS group were explored by some of
Pirani's students (Alessio and Esposito 2018); in particular by Patrick
McCarthy (1945-2005). \ McCarthy's PhD thesis contained the first of many
investigations by him (McCarthy 1971).

During their work on vacuum solutions of Einstein's equations (Robinson and
Trautman 1960; 1962) Ivor Robinson and Trautman had found that if twist-free
(and therefore hypersurface orthogonal) null geodesic congruences were shear
free then the Weyl tensor was algebraically special. \ This result helped
simplify their calculations. \ Goldberg and Sachs generalized this result.
\ They were able to prove that an empty-space solution of Einstein's equations
admits a null geodesic shear free congruence if and only if the Weyl tensor is
algebraically special (Goldberg and Sachs 1962). \ They also extended this
Goldberg-Sachs theorem, as it became known, to include certain electromagnetic
fields. \ Generalizations of their results were soon established by various
researchers and these led to significant advances in the study of exact
solutions. Before then known exact solutions were limited in number. \ Almost
all had been found by considering metrics invariant under some straightforward
symmetry group, such as plane, cylindrical and spherical symmetric metrics.
\ The Goldberg Sachs theorem, and the null tetrad formalism of Newman and
Penrose (Newman and Penrose 1962; 2009), led to the computation of many more
solutions of Einstein's equations. \ Only a very small subset of the large
number of exact solutions now known relates directly to observations. \ A
somewhat larger set has been of pedagogical interest (MacCallum 2013). \ This
may seem to be a small return on a considerable amount of research and there
is some truth in that. However some of the most significant theoretical
developments of the subject, both local and global, have been motivated by the
properties of certain exact solutions.

A radically new approach to asymptotically flat space-times and radiating
fields was introduced by Roger Penrose while he was at King's. \ At the Warsaw
conference, GR3, he observed that treating space-time from the point of view
of its conformal structure would provide a deeper understanding of infinity
(Penrose 1964a). \ He noted that from that point of view space-time points at
infinity and finite points could be treated on the same basis. \ Recalling the
well-known constructions for completing the Euclidean plane Penrose outlined
an analogous construction for Minkowski space-time. By scaling the metric he
represented Minkowski space-time as the interior of a compact subset, with
boundary, of an "unphysical" conformally related space-time. \ For Minkowski
space-time the unphysical space was a subset of Einstein's static
universe\footnote{As is not infrequently the way a similar approach had
recently been investigated by Hans Rudberg (Rudberg 1957)}. \ Penrose also
pointed out that spin $s$ zero rest-mass fields could be investigated on the
whole of the unphysical space-time because of their conformal invariance. \ He
indicated that the approach could be generalized and that general relativity
could be treated in a similar way. \ Because conformal transformations of
space-time metrics preserve null cones and null geodesics they preserve the
causal structure of space-time, a crucial point in the usefulness of the approach.

While at King's his research resulted in papers and lectures which developed
these ideas (Penrose 1963; 1964a; 1964b; 1965a). \ His results included a
geometrical definition of asymptotic flatness which avoided expansions in
powers of $r^{-1}$ and the taking of limits. \ He was able to clearly
distinguish the different types of infinities that could arise:\ future and
past time-like infinities, $\mathcal{I}^{+}$ and $\mathcal{I}^{-}$, space-like
infinity, $\mathcal{I}^{0}$, and future and past null infinity, $\mathfrak{I}%
^{+}$ and $\mathfrak{I}^{-}$ - the future and past of light-like geodesics in
the physical manifold. \ 

In his approach to asymptotic flatness a new "unphysical" metric $g$
conformally related to the physical space-time metric\ $\widetilde{g}$, where
$g=\Omega^{2}\widetilde{g}$, was introduced. \ The conformal factor $\Omega$
was positive on the physical manifold and was zero (with non-zero gradient) at
infinity. \ The conformal factor $\Omega$ could be chosen so that $\Omega
\sim\frac{1}{r}$, and consequently as $r\rightarrow\infty$ $\Omega
\rightarrow0$. \ Asymptotic behaviour could now be treated as behaviour in the
neighbourhood of the boundary hypersurface $\Omega=0$, a regular hypersurface,
denoted $\mathfrak{I}$, in a (unphysical) space-time equipped with metric $g$.
\ By requiring that the conformal structure of space-time and fields admit
extensions of appropriate smoothness across $\mathfrak{I}$ he was able to
deduce that it was a null hypersurface. \ He was also able to deduce that
$\mathfrak{I}$ was given by the disjoint union of future and past null
infinity, that is $\mathfrak{I=I}^{+}\mathfrak{\cup I}^{-}$. \ Each component
was shown to have the topology of a Minkowski space-time null cone,
$S^{2}\times E^{1}$ in a topological argument which was a forerunner to some
of his subsequent work. \ The asymptotic behaviour of fields and the peeling
property of the Riemann tensor were also deduced and he was able to construct
new geometrical formulations of energy-momentum and energy loss. \ In addition
he was able to covariantly define incoming and outgoing fields, notions with
which Trautman, Bondi et al and Sachs had wrestled. \ In short Penrose's work
encompassed and extended, in a new way, results previously obtained by Bondi
et al, Sachs, Newman, himself and others (Frauendiener 2004).

Penrose's work dealt with more than asymptotically flat space-times. \ His
lengthy 1965 paper was written in two parts. The first part dealt with
classical zero rest-mass fields, of arbitrary integer or half integer spin,
and their radiation properties. \ He demonstrated how the peeling property for
spin $s$ fields, involving $2s$ principal null directions, arose for zero
rest-mass fields in Minkowski space-time. \ The second part was devoted to his
new conformal technique, not only for asymptotically flat space-times but also
for space-times where Einstein's field equations had a non-zero cosmological
constant. \ Penrose showed how it could be concluded that when the
cosmological constant was positive (respectively negative) the hypersurface
boundary $\mathfrak{I}$ was space-like (respectively time-like) and that
$\mathfrak{I}$ was null when the cosmological constant was zero.

Wolfgang Rindler was a visiting scholar with the relativity group in 1961-62.
\ He had recently written a notable paper arising from his PhD research
conducted under the supervision of Gerald Whitrow at Imperial College (Rindler
1956). \ This paper clarified the notions of cosmological horizons. \ As
Penrose was at King's at the same time as Rindler it is not surprising that he
considered such horizons. \ He showed that once having identified the
space-like, time-like or null nature of $\mathfrak{I}$ there was a simple
diagrammatic way of representing horizons, or their absence. \ Discussions of
visual horizons, \ advanced and retarded fields in cosmological backgrounds,
and conformal representations of cosmological models, were included in his Les
Houches lectures\footnote{Penrose also discussed a representation of the
intial singularity in the Einstein-de Sitter model. \ In his Les Houches
lectures he noted that while an infinite compression $(\Omega=0)$ made
infinity finite the use there of an infinite expansion $(\Omega=\infty)$
turned a singular point into a non-singular initial space-like hypersurface.}
(Penrose 1964b) and in his talk at a conference, organized by Bondi and Gold,
on\ the problem of time (Penrose 1967)\footnote{\ At that conference an
anonymous participant was opposed to the publication of the proceedings as
Gold explained in his introduction (Gold 1967). \ He was always referred to as
Mr X. \ This attempt at anonymity was of course doomed to failure with
comments by Mr X in the proceedings (aka Richard Feynman) being particularly
sort after by readers.}.

Penrose brought to the study of relativity the creative use of topological
ideas. \ He demonstrated the importance of the conformal and causal structure
of space-time and he introduced a simple, but extremely useful, qualitative
way of describing asymptotically flat and cosmological space-time systems by
using what are now called Penrose diagrams (Wright 2014). \ His work would
significantly influence subsequent research on asymptotically flat systems
and, more generally, the understanding of the global structure of space-times.

Ted Newman was a frequent visitor to King's and while he was visiting in
1965-66, on a year's sabbatical leave from Pittsburgh University, he resumed
his collaboration with Penrose. \ With scattering problems in mind they looked
at ways in which the asymptotic symmetry group of an asymptotically flat
space-time, the infinite parameter BMS group, might be restricted to the
Poincar\'{e} group. \ Generalizing situations considered by Sachs (Sachs
1962b) they showed that if certain conditions were satisfied by the
gravitational field when the retarded time $u\rightarrow-\infty$ (or $+\infty
$) additional coordinate conditions could be imposed which would so restrict
the BMS group (Newman and Penrose 1966).

Their work at that time also included the discovery of a quite unexpected
result (Newman and Penrose1965; 1968). \ Contrary to all expectations they
found that in asymptotically flat space-times, even in the presence of
radiation, there existed exactly conserved quantities defined by surface
integrals at future null infinity. \ In the empty space case there were 10
quantities and when other fields were present additional quantities were also
conserved (Exton1969) and (van der Burg 1966; 1969). \ They found that the
analogous quantities for the linearized gravitational field vanished for
retarded fields and non-zero constants arose only when incoming radiation was
present. \ They concluded that in that case, and in the case of similar
quantities for the Maxwell and other zero rest-mass fields in Minkowski
space-time, the constants had a rather trivial interpretation; they merely
restricted the time profile of incoming fields at future null infinity.
\ However in the full non-linear theory the Newman-Penrose
(NP)\ constants\ did not necessarily vanish in the absence of incoming
radiation. \ For stationary vacuum systems for instance they could be
expressed in terms of combinations of the system's gravitational mass, dipole
and quadrupole moment.

At the time at least one person was disconcerted by this discovery. \ In his
1966 Tarner lectures, Bondi wrote

\begin{quotation}
This result I regard as horribly uncomfortable (Bondi 1967).
\end{quotation}

Elaborating a point made by Newman and Penrose Bondi noted that if a
non-radiating spheroid (with non-zero NP constant) changed its shape to a
sphere, sending out gravitational waves in the transition from old to new
equilibrium shape, its field could never be that of the sphere since that
would have a vanishing NP constant. \ Bondi concluded his discussion with a
flourish, commenting on the NP quantities

\begin{quotation}
But with the quantities never changing, I am baulked, and have to ascribe some
of the properties of the gravitational field to the dim and distant past
(Bondi 1967).
\end{quotation}

These constants, the conditions under which they exist and their physical
significance, remain subjects of continued investigation and have recently
been of renewed interest in connection with black holes (Cvetic and Satz 2018).

By the mid 1960s the foundations of gravitational radiation theory were well
established although theoretical investigations of many topics, including the
different assumptions about the asymptotic conditions and the definitions of
energy-momentum and angular momentum, would continue to be explored. \ By and
large the old controversies about gravitational waves had been resolved
although disputes about equations of motion would rumble on into the 1970's
(Kennefick 1999; 2007; 2017).\ The results of the King's research were
disseminated to wider audiences by conference talks and summer school
lectures. \ At the 1964 Brandeis summer school, one of a series organized over
a number of years by members of the physics department of Brandeis University,
Bondi, Pirani and Trautman delivered lectures which covered many of the topics
they had been studying over the previous decade (Deser and Ford 1965a).
\ Trautman's lectures included a timely introduction to modern differential
geometry in the spirit of the recently published text by the Japanese
mathematicians Kobyashi and Nomizu (Kobyashi and Nomizu 1963). \ Bondi
surveyed the work he had been involved in over the last decade and Pirani
devoted his lectures to gravitational radiation. \ For many years Pirani's
lectures were the most complete introduction to certain aspects of the subject
and to the related literature. \ They included a detailed discussion of the
application of two component spinors. \ Until the publication of the two
volume work by Penrose and Rindler (Penrose and Rindler 1984; 1986) Pirani's
lectures remained the most accessible introduction to this formalism. \ When
Pirani first heard Penrose talk about spinors his reaction had been "why use
two indices when one will do?". \ However the calculational simplifications
that resulted when spinors were used had quickly won him over. \ The other
half of the 1964 Brandeis lectures was devoted to particles and quantum field
theory. \ These included lectures by Steven Weinberg on the quantum theory of
massless particles (Deser and Ford 1965b). \ At that time the two halves
seemed to belong to different worlds but recently soft gravitons, discussed by
Weinberg in his lectures, have been related to the BMS group (Strominger 2018).

At the end of his lectures Pirani briefly mentioned gravitational radiation
from quasars and Weber's conclusion that meaningful laboratory gravitational
wave experiments were, at that time, impossible by several orders of
magnitude. \ Pirani then expressed his opinion that

\begin{quotation}
a direct observation of classical gravitational radiation is not necessary or
sufficient as a justification for the gravitational radiation theory. \ It is
my view that the primary motivation for the study of this theory is to prepare
for quantization of the gravitational field. \ The classical theory has to be
untangled first, but unless it is eventually brought into the contemporary
domain by quantization, the theory of gravitational radiation cannot have much
to do with physics.
\end{quotation}

Things did not quite turn out that way. \ In 1969 Joseph Weber claimed that he
had detected\ gravitational waves (Weber 1969). At the first public meeting,
in Cincinnati, where he spoke about this the audience recognized\ the
pioneering nature of his work. \ They greeted his announcement with enthusiasm
but also with caution. \ The caution was well founded as attempts by others to
replicate Weber's results failed (Collins 2004)\footnote{.It is salutary to
recall the conditions under which Weber worked. \ At the end of the written
account of his Cincinnati talk he noted that it had been suggested that major
progress with the detection of gravitational waves would cost a sum comparable
to the gross U.S. national product and contrasted that sum with his own
austere level of effort (Weber 1970).
\par
{}
\par
{}}. \ Nevertheless Weber's work opened up a new field of experimental and
observational work which has led, at not quite the cost of the gross national
product, to the detection of gravitational waves. \ In 1974 Russell Hulse and
Joseph Taylor discovered the first binary pulsar PSR 1913+16 and, as they
announced in 1978, this led to the first direct observational evidence of
gravitational radiation damping in binary systems as predicted by general
relativity (Kennefick 2014; Damour 2015). \ Finally, in 2016, after many years
of constructing, testing and modelling, the Laser Interferometer
Gravitational-Wave Observatory (LIGO)/Virgo collaboration announced the
GW150914 observation, the first direct land based detection of gravitational
waves (Abbott 2016; Blum 2018).

On the other hand progress on quantization of the gravitational field has been
made but has not been as conclusive and has not depended on gravitational wave
theory to the extent that perhaps Pirani and others expected.

\section{Years of change}

\subsection{Background}

Throughout the 1960s general relativity was slowly but steadily brought back
into the mainstream of physics. \ The naissance of experimental general
relativity was underway (Peebles 2017) and new astronomical observations were
changing the attitude of astronomers and physicists to general relativity.
\ Observations of quasi-stellar radio sources (to be quickly become known as
quasars) by radio telescopes were complemented by observations in the optical
domain. \ In 1963 Maarten Schmidt found a visible counterpart to the radio
source 3C 273 and obtained an optical spectrum demonstrating a redshift more
extreme than any hitherto seen in astronomy.

The advances in radio astronomy, and discoveries such as those of quasars and,
in 1964, the cosmic microwave background radiation by Arno Penzias and Robert
Wilson (Penzias and Wilson 1965) led to the curtailment of interest in the
steady state model of the universe and the establishment of the big bang
theory of the formation of the universe. \ Although they had stoutly defended
the steady state model when the observational evidence had not been
incontrovertible Bondi, Pirani (and Sciama) were now either totally converted
or ceased to voice their support for the model (Kragh 1996; Longair 2006).

The discovery of quasars accelerated investigations by relativists of general
relativistic black hole models and gravitational collapse. \ In the late 1950s
the global structure of the vacuum solution, found by Karl Schwarzschild in
1916 (Schwarzschild 1916a)\footnote{Schwarzschild also found the first
spherically symmetric perfect fluid solution which could be matched, as a
compact source, to his empty space solution. \ His fluid had constant density.
(Schwarzschild 1916b).}, and its interpretation as a black hole solution, had
become clear, particularly through the work of David Finkelstein, Martin
Kruskal and George Szekeres (Finkelstein 1959; Kruskal 1960; Szekeres 1960).
\ \ In applications the spherically symmetric, asymptotically flat and static
Schwarzschild vacuum solution had previously been used to model the exterior
of a non-rotating star and to compute the geodesic trajectories of planetary
orbits and light rays. \ Schwarzschild space-time, with metric in
Schwarzschild coordinates,%
\[
ds^{2}=-(1-\frac{2m}{r})dt^{2}+(1-\frac{2m}{r})^{-1}dr^{2}+r^{2}(d\theta
^{2}+\sin^{2}\theta d\phi^{2}),
\]
was understood to have a genuine curvature singularity at $r=0$, but there was
long thought to be a real space-time singularity at the Schwarzschild radius
$r=r_{s}$ $=2m$. \ This was despite a number of disregarded demonstrations, by
Le Ma\^{\i}tre, Synge and others, that this was not so (Israel 1987,
Eisenstaedt 1987). \ This confusion was ended when Finkelstein, Kruskal and
Szekeres showed that Schwarzschild coordinates covered only a patch of larger
space-time manifolds in which the hypersurface $r=r_{s}$ was a regular null
hypersurface. \ It could also act as an absolute event horizon constituting
the boundary of the events which could causally communicate with an
asymptotically flat region. \ The Schwarzschild radius defines the boundary of
the Schwarzschild black hole.

There was new interest in the 1930s' work of Robert Oppenheimer, and his
students George Volkoff and Hartland Snyder, on neutron stars and the
spherically symmetric collapse of a body, such as a star, to form a black hole
(Oppenheimer and Volkoff 1939; Oppenheimer and Snyder 1939; Bonolis 2017).
\ They had shown that for a sufficiently large mass there is no final stable
equilibrium state as a white dwarf or as a neutron star. \ The body collapses
through the Schwarzschild radius and continues to contract to a singular state
at $r=0$. \ The question remained: what would happen in more physically
realistic cases where spherical symmetry was not assumed and rotation and
gravitational radiation were permitted?

While there were exact general relativistic models for non-rotating
equilibrium stellar systems no asymptotically flat rotating ones had been
found, despite many attempts. \ However in 1963 Roy Kerr published a paper
entitled "Gravitational field of a spinning mass as an example of
algebraically special metrics" (Kerr 1963; 2009). \ Using the new approach to
finding solutions of Einstein's equations by studying algebraically special
metrics Kerr had found the first family of vacuum solutions which could be
identified as fields outside a rotating source. \ They were Petrov type D,
stationary, that is time independent but not invariant under time reversal,
axi-symmetric and asymptotically flat. \ The solutions formed a two-parameter
family; the parameters $m$ and $a$ determining the system's total mass $m$ and
total angular momentum $ma$ about the axis of axi-symmetry. \ When $a=0$ the
solutions reduced to the Schwarzschild solution with mass $m$.

Not much interest was shown by astrophysicists at the first Texas symposium on
relativistic astrophysics in 1963 when Kerr gave a talk about his recently
discovered metric (Kerr 1965). \ Today, as will be discussed later, Kerr's
metric is generally believed to be the unique solution of Einstein's vacuum
field equations exterior to an equilibrium rotating black hole. \ Edwin
Salpeter and Yakov Zel'dovich did suggest, in 1964, that the the enormous
amount of energy quasars would have to be radiating was due to matter in an
accretion disc falling into a supermassive black hole (Collin 2006). \ Their
suggestion was not readily accepted then but it is now the common view that
quasars are powered by the accretion of material into rotating supermassive
black holes located in the nuclei of distant galaxies.

It was in the context of these developments that January of 1965 saw the
publication of a groundbreaking paper by Roger Penrose demonstrating that
singularities existed generically when a star collapsed to form a black hole
(Penrose 1965b). \ The new approach to the analysis of the structure of
space-time being developed by Penrose was well suited to the study of
important aspects of collapse when symmetries were not assumed. From the point
of view of Penrose's formulation of asymptotic flatness, and his analysis of
space-time causal structure, the (future absolute) event horizon, or boundary
of the black hole, was the boundary of the causal past of future null
infinity. \ After a star had collapsed beyond the event horizon it would no
longer be visible to observers outside the horizon. \ For Penrose, now at
Birkbeck, another College in the University, the existence of a space-time
singularity was indicated by the occurrence in an inextendible space-time of
an inextendible incomplete timelike or null geodesic. \ Penrose introduced and
used the concept of a trapped surface - a closed space-like two-surface where
both the ingoing and outgoing orthogonal future directed null geodesics
converge. \ Such surfaces existed, for example, beyond the Schwarzschild
horizon. \ By making a number of assumptions, like the non-negativeness of the
local energy of matter, he was able to show that the formation of a trapped
surface in gravitational collapse led, in the future, to incomplete null
geodesics and hence a space-time singularity. \ The result depended on a
geometrical inequality, rather than equality, being satisfied. \ For general
relativity and other metric theories of gravity the inequality could be
interpreted as a condition on the matter energy-momentum tensor. \ That meant
both that the result was stable under perturbations and that they were not
specific to general relativity.

Space-time singularities mark the breakdown of classical theories of gravity
such as general relativity . \ While the singularity of the big bang is in the
past singularities arising as a result of gravitational collapse pose serious
problems for the predictive power of a theory. \ It is usually assumed
that\ they signal a breach of a theory's domain of validity and a quantum
theory of gravity has become applicable.

Pure mathematicians had previously used the notion of geodesic incompleteness
in Riemannian spaces (Myers 1941) but its effective use in physical situations
was new and required an understanding of the global causal and topological
properties of space-times. \ Apart from its significance for gravitational
collapse Penrose's work marked the beginning of a new era for general
relativity. \ Investigation of the global structure of space-times required a
knowledge of modern differential geometry and topology which was beyond the
standard tensor calculus familiar to relativists.

\subsection{The King's relativity group}

The middle of the 1960s at King's was marked by the holding of the fourth
international conference on general relativity and gravitation, GR4, in London
in July 1965. \ It was organized by a committee chaired by Bondi, with a lot
of the leg work being done by Kilmister. \ The latter recalled that there were
visa problems with some of the people invited from Eastern Europe which even
Bondi had difficulty resolving. \ Bondi became Chair of the International
Committee on General Relativity and Gravitation which Kilmister also joined.
\ This was an un-elected committee which coordinated the GR conferences and
associated collaborative work in general relativity. As Chair Bondi became
unavoidably involved in the, sometimes disputatious, international scientific
politics associated with this work (Lalli 2017).

GR4 attracted about 250 people. \ There were 12 invited talks and a number of
other contributions. \ These were subsequently printed in two volumes (Bondi
1965). \ The talks were at Imperial College which had an auditorium large
enough to hold the audiences. \ Invited lectures included a number on, by
then, traditional topics such as a reviews of gravitational waves and
radiation by Trautman and of exact solutions by Ehlers. Other talks dealt with
topics which had come to the fore more recently and were of increasing
interest. \ These included lectures on the gravitational lens effect by Sjur
Refsdal, on singularities of cosmological solutions by Isaak Khalatnikov, on
the physics of relativistic collapse by Igor Novikov (joint work with Yakov
Zeldovich) and a short talk by Jesse Greenstein on the new astronomical
observations of distant objects.

According to a report in the September 1965 Bulletin of the International GRG
committee, Khalatnikov's talk, based on joint work with Evgeny Lifshitz and
influenced by early work by Lev Landau (Khalatnikov and Kamenshchik 2008),
included the claim that the presence of a singularity with respect to time was
not a necessary feature of cosmological models. \ This conclusion focused
attention on cosmological singularities.

During the 1960s many visitors, like Newman, Penrose, Sachs and Schild,
contributed to the teaching at King's. \ Dennis Sciama too had been an
honorary lecturer during his post-doc. \ Now, with his own group at Cambridge
growing, he sent some of his students, including Brandon Carter, George Ellis
and Stephen Hawking, to attend relativity lectures and seminars at King's.
\ Joint seminars were also held. \ Both the Cambridge and King's relativity
groups hastily attempted to master the new global techniques that Penrose was
introducing. \ Hawking in particular, but also Ellis and Carter, went on to
make major contributions in this area. \ Hawking and Ellis quickly applied
Penrose's approach to cosmology and drew conclusions contrary to those
suggested by Khalatnikov. \ A flurry of activity in the second half of the
1960s by members of the Cambridge group, Penrose, Robert Geroch at Princeton
and others produced important results about singularities and the global
structure of space-time. \ The 1966 Adams Prize essays (named after John Couch
Adams) by Penrose and Hawking included expositions of their work (Ellis 2014,
Hawking 2014). \ Penrose's essay and Geroch's 1967 PhD thesis were both widely
distributed and in the early 1970s two publications made these new results
generally accessible (Penrose 1972; Hawking and Ellis 1973).

The reaction at King's to these new developments was mixed, with the younger
people being more enthusiastic about them than certain of the more senior
figures. \ Pirani felt that, while the ideas were mathematically interesting,
from the physical point of view the singularity theorems and studies of black
holes were pushing the theory too far. \ He remained unconvinced about black
holes until the 1990s when he rather reluctantly changed his mind. \ On the
other hand he was interested in applying modern differential geometry to
physics, in particular to classical systems and mechanics. \ However he
remained mainly interested in local rather than global analyses. \ Pirani
coauthored, with J\"{u}rgen Ehlers and Alfred Schild, a geometrically inclined
paper investigating the derivation of the Lorentzian geometry of a space-time
from the compatibility of its conformal and projective structures (Ehlers
1972). \ In physical terms it showed how measurements with light rays and
freely falling particles could be used to construct the Lorentzian geometry of
space-time. \ This paper, influenced by work by Hermann Weyl in the
1920s,\ developed a set of axioms and proofs with an emphasis on physical
plausibility rather than complete mathematical rigour. \ It was of particular
interest to those concerned with foundational questions. \ \ Pirani's interest
in differential geometry eventually led to a text book written with Michael
Crampin (Crampin and Pirani 1986).

Bondi retained his interests in stellar astronomy and astrophysics. \ He had
been instrumental in the appointment \ of Ian Roxburgh, whose interests were
in those areas, to a lectureship in the mathematics department in
1963\footnote{When, in 1959, Roxburgh had asked Bondi to take him on as a PhD
student Bondi had replied that, although he would be happy to do so, if
Roxburgh wanted to study cosmology he should first go to Cambridge and work on
stellar physics (private communication from I. Roxburgh).}. \ \ At King's,
before he moved in 1966 to another College in the University, Queen Mary
College, Roxburgh worked mainly on stellar physics although he did collaborate
on a paper related to the steady state model of the universe (Roxburgh and
Saffman 1965). \ Bondi was well aware of the new results on singularities and
global structure and their implications for gravitational collapse and
cosmology. Indeed he was one of the adjudicators of the 1966 Adams Prize,
awarded to Penrose with an auxiliary prize for Hawking. \ However as far as
his own research was concerned he remained devoted, as he once put it, to
"classical approaches".

Influenced by "the discovery of star-like radio sources", that is quasars,
Bondi returned to the study of stellar sources and the contraction of massive
objects. \ Taking the view that, as far as the processes involved were
concerned, deviations from spherical symmetry were likely to be incidental
rather than basic features, he investigated the general relativistic
spherically symmetric contraction of isotropic fluid and radiation sources
(Bondi 1964b). \ He also studied the equilibrium situation. \ The equilibrium
states of isolated self-gravitating fluid matter serve as the basic models of
stars and traditionally it has been assumed that rotating equilibrium bodies
are axially symmetric and non-rotating bodies are spherically symmetric
(Lindblom 1992). \ Bondi explored generalizations of an interesting result
which had been recently obtained by Hans Buchdahl (Buchdahl 1959).
\ \ Buchdahl had found that a general relativistic spherically symmetric
static perfect fluid sphere satisfies a bound, \ $\frac{2m}{R}\leqq n$, where
$m$ is the mass of the sphere $R$ is its boundary areal radius and $n=\frac
{8}{9.}$. \ Beyond that bound such a star would not be able to support itself
gravitationally and would collapse to form a black hole\footnote{In fact Karl
Schwarzschild had, in 1916, considered the case where a compact spherically
symmetric perfect fluid had constant energy density (Schwarzschild 1916b).
\ He had obtained the bound $\frac{2m}{R}=\frac{8}{9.}$.}. \ Buchdahl's result
was obtained by analyzing Einstein's field equations for a spherically
symmetric perfect fluid body matched to an asymptotically flat vacuum
exterior. \ In his paper Buchdahl had assumed that the energy density of the
body $\rho$ was positive and non-increasing outwards and within the body the
pressure $p$ was also positive. \ Bondi considered other combinations of the
source's pressure and density profiles with the aim of seeing how closely the
expected equilibrium limit, $n=1$, could be approached. Bondi dropped the
assumption of non-increasing energy density but retained the isotropic
pressure condition. \ With the assumptions $\rho\geqq0$, $\rho\geqq p,$
$\rho\geqq3p$ he found that $n=0.97$, $0.86$, $0.70$ respectively (Bondi 1964c).

The Buchdahl (sometimes the Buchdahl-Bondi) limit is now discussed in many of
the standard textbooks. \ Such bounds are of astrophysical importance in
determining the gravitational redshift factor of a star, they limit its
observable redshift, and consequently different configurations have continued
to be investigated. \ In the 1960s they suggested that it was problematic that
quasars were ultracompact stars.

Raymond McLenaghan, who in 1968 completed a PhD at Cambridge where his
supervisors were Hoyle and Sciama, joined the group as a research assistant
for the academic year 1966-67 after which he moved to a post at the
Universit\'{e} Libre de Bruxelles. \ Both McLenaghan and Bondi's student
K\"{u}nzle investigated fields in four dimensional space-times which satisfied
Huygen's principle (McLenaghan 1969; K\"{u}nzle 1968). \ They studied waves
and whether or not, in various space-times, they had tails; in other words
whether or not retarded wave solutions depended on their source only on the
past null cone\footnote{Such investigations had a bearing on the
Newman-Penrose constants and Bondi's comments.}. \ Equations like the
homogeneous wave equation satisfy Huygen's principle if and only if the
retarded wave solutions have no tails. \ McLenaghan, extending a result by
Paul G\"{u}nther, was able to conclude that the homogeneous wave equation on
an empty space-time satisfied Huygens' principle if and only if the space-time
was flat or a plane-wave space-time. \ \ K\"{u}nzle looked at p-form fields in
gravitational plane wave spaces and showed that Huygen's principle held for
solutions of Maxwell type equations and the scalar wave equation; while
solutions of wave equations for 1,2 and 3-forms did not satisfy Huygen's
principle, in general, and could have tails.

The U.S. Air Force contract came to an end, with the final report, which
included a collection of papers, being written in 1966 (Bondi 1966). \ Bondi
was increasingly involved in administrative and other non-research work both
inside and outside King's. \ He had discovered that he was good at meetings
and administrative tasks and others had noticed this too. \ He was invited to
join various government committees and he was asked to produce, by himself, a
report on a proposed Thames barrier to protect London in times of floods. The
report he produced in 1967 recommended the construction of a barrier and this
was subsequently built. \ Bondi rightly regarded this as one of his major
achievements. \ In 1967 he took leave from King's for a three year period to
become Director General of the European Space Research Organization - ESRO
(Bondi 1990a; Roxburgh 2007).

\ Bondi's multifaceted activities and then departure meant that Pirani carried
an increasingly heavy burden as a research supervisor. \ He kept abreast of a
wide range of research and was happy to suggest a topic and then allow a
student to pursue it, whether or not he himself retained an interest in it.
\ Some students continued to investigate topics related to radiation while
others, like Michael Crampin, Hans-Peter K\"{u}nzle and subsequently Nicholas
Woodhouse, undertook research in which modern differential geometry played a
more prominent role. \ Yet again others, like Ray d'Inverno, developed areas,
in his case algebraic computing, which had only been of passing interest to
Pirani (d'Inverno 1970). \ The seminars and joint meetings with Cambridge and
Brussels continued and attracted good audiences, including people from other
Colleges in the University like Gerald Whitrow from Imperial College, William
Bonnor from Queen Elizabeth College and Roger Penrose now at Birkbeck College.
\ Visitors like Ted Newman continued to assist with the relativity teaching
and Newman wrote a number of papers with King's students. \ He also helped by
suggesting problems to students, one example being an investigation of BMS
supertranslations in Minkowski space-time (Crampin and Foster 1966). \ Others,
like Robert Geroch who held a post-doctoral position with Penrose at Birkbeck
College from 1967 to 1969, gave series of research lectures - in Geroch's case
on the global structure of space-time and singularities. \ As was the case
with Andrzej Trautman's 1957 lectures these were reproduced and widely distributed.

The second half of the 1960s saw the staff in the relativity group augmented
by the appointment of three younger people to lectureships. \ All three either
had been, or were, Pirani's research students. \ Peter Szekeres came back from
a post-doctoral post at Cornell University in 1966 and Michael Crampin
returned from a similar position at Harvard in 1968. \ In addition, with Bondi
\ away, in 1968 Ray d'Inverno was appointed to a temporary assistant
lectureship while he was completing his PhD. \ While Crampin was becoming more
interested in applying modern differential geometry to areas like mechanics
and differential equations Szekeres and d'Inverno continued to work in general relativity.

Szekeres' PhD thesis had included a study of the propagation of gravitational
fields in matter (Szekeres 1966). \ In a \ continuation of his study of
gravitational waves he investigated colliding gravitational waves. \ Unlike
electromagnetic waves gravitational waves travelling in non-parallel
directions do not satisfy a principle of superposition and Szekeres considered
the way in which colliding plane sandwich waves diffused through each other.
\ He found that after the collision the waves ceased to be planar and a
physical space-time singularity resulted (Szekeres 1970; 1972). \ About the
same time Penrose and his student at Birkbeck College, K.A.Khan, published a
study of the scattering of two gravitational waves. They considered impulsive
gravitational plane waves and reached similar conclusions (Khan and Penrose
1971). \ Weber's recent claim to have detected gravitational radiation and its
possible future observational, as opposed to merely theoretical, importance,
gave impetus to their work. \ It influenced many subsequent investigations.

By the time d'Inverno started his PhD it had become clear that routine and
tedious calculations involving increasingly complicated space-time metrics
were taking too much time. \ Pirani was aware that computer programmes were
starting to handle such calculations. He prompted d'Inverno to construct, as
quickly as possible, an algebraic computing system which could be used to
calculate tensorial quantities such as curvature tensors. \ d'Inverno decided
to base a system, tailored to the special needs of general relativity, on a
subset of the high level programming language LISP. \ He constructed first LAM
(Lisp algebraic manipulator) and then ALAM (Atlas LAM). \ LAM was designed to
work on the Atlas 1, at that time the fastest computer available in the UK.
\ Because the Atlas version of LISP did not possess a COMPILE facility LAM was
slow and so it was replaced by ALAM written in the assembly language of Atlas.
\ D'Inverno applied ALAM to numerous problems in general relativity, including
the computation of the curvature tensors for various metrics, investigations
of energy-momentum tensors and pseudotensors and the study of variational and
characteristic initial value problems (d'Inverno 1970). \ His pioneering work
on algebraic computing in general relativity was continued at King's by Tony
Russell-Clark, another of Pirani's students who was mentored by d'Inverno.
\ When Atlas 1 was replaced by a CDC 6600 Russell-Clark wrote the successor to
ALAM - CLAM. \ CLAM, that is CDC LAM, was written in a simple command
language, eliminating the need for the user to learn LISP.

D'Inverno, in a review article (d'Inverno 1980), later recalled some
observations he and Russell-Clark had made about the results and aims of their
work. \ These included the comment that ALAM could compute, correctly, the
Einstein tensor of Bondi, van der Burg and Metzner's metric in about four
minutes (d'Inverno 1967) and their belief that, after about half an hour's
study of part 1 of the CLAM manual, a computer novice should be able to
process a large class of metrics (d'Inverno and Russell-Clark 1973). \ Their
overall aim was not only to develop programmes that could deal with a wide
variety of problems but\ also to make their use in calculations easy.
\ Algebraic computer packages of various types are now routinely used in
general relativity (MacCallum 1994, MacCallum and Skea 1994).

\section{The1970s and early 1980s: classical and quantum gravity}

\subsection{The post-Bondi era}

Further changes took place at King's in 1970. \ Peter Szekeres decided to
return to Australia at the end of the year. He took up a position in the
physics department of the University of Adelaide where he continued his work
on colliding waves. \ \ Subsequently he played an important role in the
promotion of general relativity in Australia. \ Ray d'Inverno completed his
PhD and left for a position in the mathematics department of Southampton
University. \ There he established a flourishing relativity research group.
\ He later wrote a well regarded introductory textbook on general relativity.
\ It incorporated the approach taken to the subject at King's (d'Inverno
1992). I was appointed to a lectureship in applied mathematics, to replace
Szekeres, and joined the relativity group in September 1970 after spending the
previous six years in the United States. \ The job market in physics there had
dried up as a consequence of the Mansfield amendment\footnote{The number of
physics jobs at the April meeting of the American Physical Society dropped by
a factor of four from 1968 to 1971 (Mody 2016). \ A detailed analysis has been
made by D. Kaiser (Kaiser 2012).} but the job drought was still a few years
off in the U.K. \ I had been a PhD student at Syracuse University, where Josh
Goldberg had been my research supervisor. \ My thesis was about invariant
transformations and the Newman-Penrose constants. \ After that I had spent two
years as a research fellow at the ARL in Dayton, Ohio.

By the beginning of the 1971-72 session the relativity group consisted of the
absent Bondi, Pirani, Kilmister, Crampin, myself, and a long term visiting
scholar from the Hamburg group, Henning M\"{u}ller Zum Hagen. \ In addition
there were numerous research students and short term visitors. \ Pirani was
still supervising a large number of students, however they were a fairly
self-reliant group, of necessity perhaps. \ The students continued to work on
aspects of classical general relativity and geometry, including gravitational
radiation theory. \ Pirani continued to run the weekly relativity seminar but
his own interests were moving away from relativity.\ \ He was becoming more
interested in other physical applications of differential geometry and was
distracted by various events outside academic life. \ Kilmister was busy with
other scholarly and research activities and was also carrying a heavy
administrative load.

The King's calendars for 1969-70 and 1970-71 list Bondi as head of department,
Semple having retired in 1969, but Bondi was of course absent. \ By the
beginning of 1971 the academic staff of the mathematics department, excluding
Bondi, numbered sixteen. \ The department had grown a little and the total
student body of the College in 1970-71 numbered nearly three thousand. \ There
were 52 post-graduate students in the mathematics department. \ The College
had recently started setting its own undergraduate examinations but
postgraduate courses, examinations and degrees were still University based and
inter-Collegiate activity remained important.

During the autumn of 1970 it became increasingly clear that Bondi would not
return to King's. \ When he finished at ESRO he took up a new position, in
March 1971, as Chief Scientific Advisor to the Ministry of Defence, leaving as
he put it "the austere circumstances of academic life" (Bondi 1990a). \ He
resigned his position at King's although he did retain a position as titular
professor for many years and was occasionally in the College. \ No doubt the
title was helpful in dealing with the senior ranks of the armed forces.
\ Bondi's replacement was John G. Taylor (1931-2012) whose primary interests
were elementary particle physics and quantum field theory. \ In addition he
was actively interested in neural networks and occasionally explored unusual
topics. \ Although Taylor was not a group builder his appointment meant that,
in time, the relativity group would be replaced by a group which had
supersymmetry and string theory as its main interests. \ In 1972 Michael
Crampin decided to move to the Open University and he was replaced by Paul
Davies. \ Davies had obtained a PhD in physics in 1970 from University College
London where his supervisors had been Michael Seaton and Sigurd Zienau. \ His
first research was in the field of atomic astrophysics and he had worked on
the problem of di-electronic recombination in the solar corona. \ Davies came
to King's from a post-doctoral position with Fred Hoyle at the Institute of
Astronomy at Cambridge and his main research interests by then were topics
related to cosmology. \ A number of these feature in his book on time
asymmetry which he completed after coming to King's (Davies 1974). \ This book
aimed to clarify this subject by examining it in a wide number of areas of
physics. \ Christopher Isham joined the mathematics department in 1973 and
brought with him from Imperial College a number of students, including Bernard
Kay and Jeanette Nelson. \ Isham was a quantum field theorist who had obtained
his PhD in 1969 under the supervision of Paul Matthews at Imperial College.
\ His thesis had been on twisted fields which encode topological aspects of
space-time into the quantum theory. \ He had a significant and increasing
interest in quantum gravity.

The 1960s expansion of the university system in the UK was past by the mid
1970s and post-graduate students could no longer expect to be able to obtain a
permanent lecturing position. \ The small number who were eventually able to
stay in academic life often did so by holding a series of temporary
postdoctoral positions, often for a lengthy period, before eventually
obtaining a permanent post. \ One consequence of this was King's and other
institutions were able to make some particularly outstanding post-doctoral
appointments. \ During the 1970s, as the job and financial squeeze developed,
a number of people also held one year teaching appointments in the mathematics
department, temporarily filling vacancies. \ They included Jamal Islam
(1939-2013) who was then working mainly on cosmology and relativity. Islam,
who taught in the department during the 1973-74 session, continued to pursued
this research, making notable contributions, and, in time, returned to
Bangladesh. \ Kellogg Stelle, who had just completed his PhD at Brandeis
University under the supervision of Stanley Deser, taught during the 1977-78
session before moving to Imperial College. Stephen Huggett also held a one
year mathematics lectureship in the 1979-80 session before moving to the
University of Plymouth. \ Huggett came from Roger Penrose's group at Oxford
where he had worked on twistor theory for his 1980 doctorate. \ In addition to
these temporary appointments one new permanent appointment was able to be
made. \ In 1978 Peter West, whose 1976 PhD had been supervised by Abdus Salam
at Imperial College, took up a mathematics lectureship. \ West came from a
post-doctoral position at Imperial and at that time his main interests were in
the newly developing areas of supersymmetry and supergravity.

\ During most of the 1970s the main lines of research related to gravitation
at King's were the study of classical black holes, quantum theory in curved
space-time and quantum gravity. \ Chronologically the first was classical
relativity and black holes in which M\"{u}ller Zum Hagen and I were both
involved. \ Then the research emphasis shifted to quantum fields in curved
space-times undertaken primarily by Davies, students, postdocs Stephen
Fulling, Steven Christensen and Lawrence Ford. \ Isham and his students
carried out further research on quantum gravity including the canonical
approach to quantizing general relativity and the modelling of quantum
cosmologies (Isham 1976; Isham and Nelson 1974; Blyth and Isham 1975). \ They,
together with another post-doc, Michael Duff, were the most active in quantum
gravity and related aspects of quantum field theory. \ Visitors, in particular
Stanley Deser, also participated notably in this work. \ Towards the latter
part of the 1970s activity at King's in supersymmetry and supergravity
increased significantly with Taylor, West and Stelle being active in this
area. \ The study of aspects of twistor theory was also undertaken by Huggett
and a post-doc Andrew Hodges.

\subsection{Classical gravity and classical black holes}

Since I was involved in research in these areas this section is somewhat
different from others. \ It is rather more discursive and includes some reminiscences.

\subsubsection{Early investigations of the positivity of mass}

\ During my post-doc at the ARL I had worked mostly with Jeffrey Winicour.
\ Our research had included attempts to prove that the total mass-energy of an
asymptotically flat system, as defined by Bondi et al and shown by them to be
non-increasing in time, could never become negative. \ This was important
because if it could become negative it opened up the drastic possibility of
systems with energy unbounded below and no stable ground state. It was a
difficult global problem which had previously been investigated in the case of
the total energy defined at space-like infinity, the Arnowitt, Deser, Misner
(ADM) energy (Arnowitt 1962), but apparently not in the case of the Bondi
energy. \ In both cases the mass-energy corresponded to the total (active)
gravitational mass of the system but the ADM mass was constant unlike the
Bondi mass. \ However in both cases it was expected to be non-negative when
any source was physically regular and to be zero only when space-time was
empty and flat. \ In 1968 Dieter Brill, Stanley Deser and Ludvig Faddeev had
used a variational approach to study the positivity of the ADM energy. \ They
treated the mass as a functional of asymptotically flat Cauchy data for
solutions of Einstein's equations and showed that the mass functional had only
one critical point, at flat space, and there the second variation was strictly
positive. Their work prompted us to investigate the positivity question at
future null infinity. \ We were able to show that in the weak field limit, for
space-times satisfying appropriate global conditions and positivity conditions
on the energy-momentum tensors of sources, the Bondi mass also had to be
non-negative and was zero only for flat space-time. \ For the full non-linear
theory\ we obtained further results analogous to those of Brill, Deser and
Faddeev. \ (Robinson and Winicour 1971; Brill and Jang 1980). \ These were
suggestive, however Robert Geroch explicitly demonstrated that it could not be
inferred from any of these function space results that the mass of a non-flat
space-time, either at space-like or null infinity, was
necessarily\ positive\footnote{Geroch subsequently developed a novel approach
to the positivity problem at space-like infinity where the problem could be
viewed as one in the global differential geometry of a three dimensional
Riemannian manifold\ (Geroch 1973).}

In further work, concluded after I had moved to King's, Winicour and I
constructed model vacuum space-times for which the Bondi mass could be
expressed in terms of the intrinsic and extrinsic geometry of compact
two-surfaces embedded in three dimensional Euclidean space. \ When the surface
was a two-sphere, which corresponded to flat space-time, the mass was zero.
\ By changing the two-surfaces and the parameters defining them we could track
the corresponding changes in the energy. \ These appear to have been the first
explicit calculations of the Bondi mass when the space-times were essentially
non-singular. \ Despite the form of the mass integrand suggesting that there
might be cases where the energy could become negative the energy integral
always remained non-negative and was non-zero when the families of
two-surfaces we investigated were not two-spheres (Robinson and Winicour
1972). \ It was apparent that more sophisticated attacks on these problems
were required.

In a major breakthrough two pure mathematicians, Richard Schoen and Shing-Tung
Yau, published a proof of the positivity of the ADM mass in 1979. \ An
alternative proof by Edward Witten in 1981 was more immediately accessible to
physicists. \ These results were quickly adapted to obtain proofs of the
positivity of the Bondi mass (Penrose and Rindler 1986). \ The positive mass
theorem as it became known turned out to be important not only in physics but
also in differential geometry.

\subsubsection{Classical black holes}

Gravitational waves and cosmology had been studied from the earliest days of
general relativity, but black holes were not considered widely and seriously
until the 1960s. \ This is not to say that investigations of them started from
a blank sheet then but neither observationally nor theoretically had they
previously been objects of widespread interest or study (Bonolis 2017). \ \ As
Werner Israel noted, as far as the end points of stellar evolution were
concerned, until the end of the 1950's astronomers saw no need for anything
more exotic than white dwarfs (Israel 1996). \ Not many years later things
were very different. \ The discovery of quasars was followed by the
observation of pulsars in late 1967 by Jocelyn Bell-Burnell, then a
post-graduate student at Cambridge working for a PhD with Antony Hewish
(Bell-Burnell 1977). \ This discovery and the speedy suggestion by Tommy Gold
that they were neutron stars helped push the study of black holes further into
the mainstream of physics (Gold 1968).

In 1967 Werner Israel, then on leave in Dublin from the University of Alberta,
gave one of the two talks at a half-day meeting at King's. \ Israel later
recalled that there was a large audience. \ Hermann Bondi, Felix Pirani,
Brandon Carter and Charles Misner (on sabbatical leave in Cambridge) were
among\ the people in the front row.

Motivated by a number of investigations of the effect of pertubations on event
horizons Israel had looked at static asymptotically flat solutions of
Einstein's vacuum field equations subject to conditions that a broad class of
non-rotating equilibrium black hole metrics would plausibly satisfy. \ His
striking conclusion was that the only solutions satisfying his conditions were
the positive mass Schwarzschild metrics. \ One member of the audience remarked
that the result was important - if it was correct. \ It was correct and it was
seen to be very important although initially there was some uncertainty about
how it should be interpreted\ (Israel 1987).

By the time I was actively working on the problem, influential arguments had
been advanced that the end state of stellar evolution of bodies retaining
sufficiently large mass would indeed be a black hole, rather than a naked
singularity not hidden behind a regular event horizon (Penrose 1969). \ In the
non-rotating case Israel's result suggested that, when Einstein's vacuum field
equations held, this would have to be a Schwarzschild black hole; all
multipole moments higher than the monopole being radiated away in the collapse
to the equilibrium end state. \ Within a few years perturbation calculations
were providing supporting evidence for this point of view (Price 1972a,b).

While Israel's result was clearly important it was not conclusive. \ He had
investigated static vacuum black hole solutions, which have metrics of the
form%
\[
ds^{2}=-V^{2}dt^{2}+g_{\alpha\beta}dx^{\alpha}dx^{\beta},
\]
where $V$ and the Riemannian three-metric $g_{ab}$ are regular and independent
of time $t$. \ Outside the event horizon $0<V<1$, with $V=0$ at the regular
event horizon. \ Asymptotic flatness was ensured by demanding that
$V\rightarrow1$, and $g_{\alpha\beta}$ tends to the Euclidean three-metric at
infinity in a standard way. The problem formulated by Israel was therefore a
boundary value problem for $V$ and $g_{\alpha\beta}$, determined by Einstein's
equations, on a three dimensional Riemannian manifold where $t$ was constant
(Israel 1967)\footnote{The mathematics of black holes is discussed by
Subrahmanyan Chandrasekhar (Chandrasekhar 1983a). \ Rigorous definitions and
proofs of black hole uniqueness theorems can be found in the review by Markus
Heusler (Heusler 1996). \ }.

However the metrics Israel investigated were a sub-class of the static metrics
which could satisfy the above conditions. \ They were metrics which also
admitted coordinates on the $t=$ constant surface such that the three-metric
could be written in the form
\[
g_{\alpha\beta}dx^{\alpha}dx^{\beta}=\rho^{2}dV^{2}+g_{ab}dx^{a}dx^{b},
\]
where $g_{ab}$ is a two-metric. \ In other words the function $V$ was assumed
to have no critical points and the topology of the level surfaces of $V$,
including the horizon at $V=0$, was assumed to be spherical. \ The extent to
which these additional assumptions might drive the conclusion was unclear, to
me at least. \ When I started to work on this problem, in 1970, I did so under
the assumption that Israel's conclusion would still hold when these
restrictions were not imposed. \ Although this seemed very plausible
comparatively little was known about static vacuum solutions beyond the well
studied axially symmetric Weyl metrics. \ It was always possible that some
quite unexpected black hole solution existed and \ in that case I would be
wasting my time.

Israel's overall approach was the one used in standard uniqueness proofs
involving differential equations such as Laplace's equation. \ This was first
to construct, using the field equations, appropriate identities relating
divergences and non-negative quantities whose vanishing would imply the
uniqueness result. \ Then, by integrating the identities, applying Stoke's
theorem and using the boundary conditions, to deduce that the non-negative
quantities must vanish. \ However the details of Israel's proof depended very
much on his choice of coordinates so if these could not be assumed a different
approach was needed. \ My initial aim therefore, was to find appropriate
covariant, that is coordinate independent, identities. \ But first, and quite
quickly, I was able to show that the topology of the horizon was spherical.
\ I did this by constructing an identity of the form divergence equals a
non-negative quantity where the latter was constructed by using the square of
the three dimensional Ricci tensor. \ Then, by integrating over the
three-manifold exterior to the horizon, evaluating the divergence on the
horizon and at spatial infinity and then applying the Gauss-Bonnet theorem for
two dimensional surfaces on the horizon I was able to conclude that the
topology of the $V=0$ event horizon was spherical as Israel had assumed.

It was well known that the Schwarzschild solution could be written in
isotropic coordinates where the three-metric was explicitly conformally flat.
\ It was also well known that there was a three-index tensor, sometimes called
the Cotton tensor, constructed out of the covariant derivatives of the Ricci
tensor and Ricci scalar, which was zero if and only if the three metric was
conformally flat. \ By employing the square of the Cotton tensor multiplied by
non-negative functions I was able to construct, using Einstein's equations,
covariant identities of the required form. \ However as these involved inverse
powers of terms which vanished at critical points of $V$ the critical point
problem remained. \ At this juncture I discussed the problem with my colleague
Henning M\"{u}ller Zum Hagen. \ He had been working on related topics. \ Of
particular relevance were his proofs that, put simply, static and stationary
metrics were real analytic (M\"{u}ller Zum Hagen 1970a; 1970b). \ He was very
familiar with the equations and the relevant potential theory and felt that
the critical point problem could be handled. \ He was going to Hamburg for the
summer and would work on this with his colleague there, Hans J\"{u}rgen Seifert.

During the summer of 1971 we attended the general relativity conference GR6 in
Copenhagen. \ There Stephen Hawking announced major results on classical black
holes. \ These included a proof that the topology of the surface of an
equilibrium black hole was spherical, not only in the static but also in the
stationary case applicable when black holes were rotating. \ His paper
containing the details of this work included a description of the appropriate
four dimensional framework within which to consider black holes. \ Amongst
other important results he presented a calculation leading him to claim that
an equilibrium black hole must be axially symmetric if it was rotating
(Hawking 1972). \ The proof of this last result had flaws, particularly in its
use of analyticity, rectifications of which are still being pursued.
\ Nevertheless this paper was a landmark in classical black hole theory
(Robinson 2009).

In the autumn term of 1971 Seifert visited King's and we finally completed a
rather complicated paper\ generalizing Israel's result (M\"{u}ller Zum Hagen
1973). \ Having dealt with the critical points problem in a somewhat involved
way we were able to use the equalities I had constructed to show that the
three geometry was necessarily conformally flat. \ Using the boundary
conditions and field equations it then followed very quickly that the metric
had to be Schwarzschild.

Following his paper on vacuum solutions Israel had, in short order, produced a
second paper generalizing his result to static Einstein-Maxwell black holes
(Israel 1968). \ Black holes of astrophysical interest were generally thought
to be uncharged, electrically neutral because of the presence of plasma, but
in certain theoretical contexts, particularly quantum mechanical ones, charged
black holes play a role. \ However at this time we were concerned with
macroscopic astrophysical black holes and the quantum considerations were for
the future. \ Nevertheless Israel had shown, using the same coordinate system
and methods similar to those he had used in the vacuum case, that the static
black hole solutions had to be those contained in the metrics found by Hans
Reissner and Gunnar Nordstr\"{o}m (Reissner 1916; Nordstr\"{o}m 1918). \ The
global structure of the Reissner-Nordstr\"{o}m black hole solutions is quite
different from the global structure of the Schwarzschild black holes. \ But as
far as both the asymptotically flat region exterior to the event horizon and
proofs of uniqueness theorems are concerned the underlying space-time
structures are similar. \ Israel's extension of his vacuum proof led to the
expected uniqueness result but his paper was a calculational tour de force.
\ Having completed our work on the vacuum case, we turned to generalizing it
to the electromagnetic case. \ M\"{u}ller Zum Hagen had returned permanently
to Hamburg in the summer of 1972 and as a result there were some delays. By
about the end of the year we had produced \ a paper outlining our
generalization of Israel's Einstein-Maxwell theorem (M\"{u}ller Zum Hagen 1974).

In 1973 I spent part of the summer with the Hamburg group, filling a slot
opened by Wolfgang Kundt's absence on leave. \ Our work on static black holes
having ended I had been looking at equilibrium symmetries of stellar models
and attempting to prove that static perfect fluid stellar systems,
asymptotically flat and with physically reasonable equations of state, had to
be spherically symmetric. \ This was an old problem which had been solved in
the context of Newtonian gravity early in the twentieth century and a number
of people had investigated aspects of it within the context of general
relativity (Lindblom 1992). \ The techniques used to solve the Newtonian
problem were not extendable to general relativity but I thought that the
methods we had used to handle black holes could be applied to this problem.
\ By the time I got to Hamburg I thought I had solved it but shortly after my
arrival there I found a flaw in my approach and consequently I was only able
to deal with some special cases. \ Numerous and increasingly successful
attacks on the problem have continued for a long time (Masood-ul-Alam 2007).
\ During my own work I read a paper on the topic by K\"{u}nzle (K\"{u}nzle
1971). \ His paper included a generalization of work by A. Avez and an
interesting technical result. \ If it was also assumed that the magnitude of
the gravitational field strength (the magnitude of the three dimensional
gradient of $V$) was a function of $V$ alone then it followed that static,
compact perfect fluid bodies, with asymptotically flat exteriors, were
spherically symmetric. \ This result suggested a way that the identities used
in our previous static black hole work might be generalized. \ I was able to
do this and saw that these new expressions could be used to greatly simplify
and improve our previous static black hole uniqueness calculations. \ This new
proof of the uniqueness of the Schwarzschild black hole was published some
years later (Robinson 1977).

My collaborators in Hamburg were busy working with Peter Yodzis on
constructing tests of Penrose's weak cosmic censorship hypothesis, the
conjecture that physical singularities were always hidden behind horizons.
This, together with the assumption that the exterior of the black hole was
causally well-behaved, was always assumed to be the case in the uniqueness
theorems. \ I expected that our static electromagnetic black hole result could
be simplified as our static empty space result had been, but I did not pursue
this. \ Instead I decided to look at, for me, a different and more interesting
problem, the stationary (time independent but not time reversal invariant)
analogue of the static uniqueness result. \ Since this would apply to rotating
equilibrium black holes it was very important. \ The natural conjecture was
that the equilibrium end state had to be a Kerr black hole. \ There were a
number of reasons to think that this conjecture - for a time called the
Israel-Carter or Carter-Israel conjecture - was not unreasonable despite the
fact that at that time the Kerr solution was the only known asymptotically
flat, stationary, vacuum solution of Einstein's equations. \ By then the
global structures of the known charged and uncharged, static and stationary
black hole solutions had been analyzed, by extending the approaches that had
been applied to the Schwarzschild solution, and their similarities and
differences had been understood. The Kerr family of black holes reduced to the
Schwarzschild family when the angular momentum vanished. \ Like the
Schwarzschild metric the Kerr metric was Petrov type D, it was axi-symmetric
and so on.

I could not see how to approach the conjecture by focusing on key geometrical
properties as in the static case. \ I therefore decided to follow the work
that had been done on the problem by Brandon Carter. \ A paper by Carter
published in 1971 showed that, broadly speaking, asymptotically flat vacuum
solutions of Einstein's equations corresponding to the exterior of rotating,
axi-symmetric, topologically spherical equilibrium black holes consisted of
discrete sets of families. \ Each family depended on at least one and at most
two parameters and the only family admitting the possibility of zero-angular
momentum was the Kerr family of black hole solutions (Carter 1971). \ This
result was a major step towards proving the uniqueness of the Kerr black
holes. \ I had with me in Hamburg a pre-print of some of the lectures Carter
had given at Les Houches in 1972 (Carter 1973) and I studied that. After
lengthy considerations Carter had been able to reduce the empty space problem
to a uniqueness problem for two coupled partial differential equations with
two independent and two dependent variables subject to boundary conditions on
the boundary of the black hole, the axis of symmetry and at infinity. \ This
formulation left the problem as one in partial differential equations, not
geometry, and this was quite different from the static case. \ The equations
were non-linear and Carter had obtained his result\ by employing the
linearized equations in a clever way. \ I decided to try to extend Carter's
result to the case where electromagnetic fields were present. \ I thought that
this would be a good way to acquaint myself with the formalism. \ Furthermore
it was commonly assumed, as I did, that when electromagnetic fields were
included the general nature of results, like those Carter had obtained in the
vacuum case, would continue to hold. \ Soon after Kerr's discovery of his
family of metrics the charged version of his vacuum solution was published by
Ted Newman and students taking his relativity course at the University of
Pittsburgh (Newman 1965; Newman and Adamo 2014). \ This Kerr-Newman solution
is a three parameter, asymptotically flat, axi-symmetric, stationary solution
of the Einstein-Maxwell field equations. \ In addition to the Kerr solution
parameters $m$ and $a$ there is a third parameter $q$ corresponding to the
total electric charge. \ When $q=0$ the solution reduced to the Kerr solution
and when $a=0$ the solution reduced to the static Reissner-Nordstr\"{o}m
solution. \ The black hole solutions corresponded to the cases where
$a^{2}+q^{2}\leqq m^{2}$. \ A fourth parameter $p$, the magnetic monopole
charge, was added by Carter for completeness and then the black hole solutions
correspond to the cases where $a^{2}+p^{2}+q^{2}\leqq m^{2}$\ There is no
evidence for the existence of magnetic monopoles in nature so attention is
usually restricted to the three parameter sub-family where $p$ is zero;
however in my work I followed Carter.

Including the electric and magnetic field meant that, as Carter had shown in
his lecture notes, there were now four coupled, non-linear equations to be
solved for four dependent variables which were functions of two independent
variables. \ The class of metrics that had to be considered were of the form%
\[
ds^{2}=-Vdt^{2}+2Wd\phi dt+Xd\phi^{2}+U(\frac{d\lambda^{2}}{\lambda^{2}-c^{2}%
}+\frac{d\mu^{2}}{1-\mu^{2}}).
\]
where the metric components are independent of $\phi$ and $t$ and $c$ is a
positive constant. \ The four dependent variables determining the
Einstein-Maxwell solutions were $X$, $Y$ - a potential for $W$, and two
additional functions $E$ and $B$, potentials for the electromagnetic field,
all subject to the appropriate boundary conditions. \ The non-linear equations
were complicated.

By experimenting with the linearized, but still complicated, field
equations\ I was able, somewhat to my surprise, to construct a rather fearsome
looking identity of the required form - multiples of the field equations plus
linearized field equations equal to a divergence plus non-negative terms.
\ When the field equations and boundary conditions were satisfied each of the
non-negative terms had to vanish. \ It was then a straight forward matter to
deduce the conclusion: \ pseudo-stationary, asymptotically flat,
axi-symmetric, black hole solutions of the source-free Einstein-Maxwell
equations form discrete, continuous families, each depending on at most four
parameters. \ Of such families only the Kerr-Newman family contains members
with zero angular momentum (Robinson 1974).

Buoyed by my success at being able to see my way through the complicated
Einstein-Maxwell equations, even though the final result\ was the expected
one, I decided to try to prove the uniqueness of the Kerr black hole family. I
assumed, in the light of Hawking's result, that it was reasonable to consider
only axi-symmetric systems. \ Once again this involved trying to prove a
result which might or might not be true. \ Using Carter's framework I hoped at
first that the fact that the field equations could be derived from a sigma
model type Lagrangian might be useful. \ However I only managed to recover
Carter's 1971 result by using a Noether identity and a Lagrangian based
approach (Robinson 1975a). \ Busy with teaching and so on I only occasionally
returned to think about the problem until early in 1975 a pattern suddenly
emerged which enabled me to construct an identity, analogous to the ones
Carter and I had constructed using the linearized equations but now applying
to the full non-linear equations. \ Simple arguments then led to a uniqueness
result: the family of Kerr metrics, with $\left\vert a\right\vert <m$, is the
unique axi-symmetric, pseudo-stationary family of black hole solutions of the
Einstein vacuum field equations when the event horizon is assumed to be
non-degenerate (Robinson 1975b).

It was clear to me that this result could be extended to the Einstein-Maxwell
case and the analogous uniqueness of the Kerr-Newman family of black holes
could be proven and this was done by others, systematically, in the 1980's
when\ both the static and stationary black hole proofs were improved and
extended using newly obtained results such as the positivity of the ADM mass
(Heusler 1996). \ Investigations into other asymptotically flat black holes
systems, where similar uniqueness results do not necessarily apply, and
non-asymptotically flat systems, such as cosmological ones, together with
research on increasingly rigorous and complete proofs of uniqueness results,
continue (Robinson 2009; Chru\'{s}ciel 2012). \ Today studies of ways of
testing the Kerr black hole nature of astrophysical black holes are being
carried out in preparation for the time when quantitative constraints on any
deviations from the Kerr geometry will be able to be better determined
(Krawczynski 2018, Bambi 2019). \ Be all that as it may Subrahmanyan
Chandrasekhar felt able to write in 1983

\begin{quotation}
Black holes are macroscopic objects with masses varying from a few solar
masses to millions of solar masses. \ To the extent they may be considered as
stationary and isolated, to that extent, they are all, every single one of
them, described \textit{exactly }by the Kerr solution. \ This is the only
instance we have of an exact description of a macroscopic object.
\ Macroscopic objects, as we see them all around us, are governed by a variety
of forces, derived from a variety of approximations to a variety of physical
theories. \ In contrast, the only elements in the construction of black holes
are our basic concepts of space and time. \ They are, thus, almost by
definition, the most perfect macroscopic objects there are in the universe.
\ And since the general theory of relativity provides a single unique two
parameter family of solutions for their description, they are the simplest
objects as well. (Chandrasekhar 1983b).
\end{quotation}

After years of work by very many people it was reported that the signal in the
GW150914 observation indicated that the waves were produced during the late
quasi-circular inspiral, merger and ringdown of a binary black hole system.

Increasingly in the 1970s research in the group was devoted to aspects of
quantum theory and gravity. \ Pirani's student Nick Woodhouse finished his PhD
in 1973 and then held a post-doctoral position at King's until 1975 when he
left for another post-doctoral position with John Wheeler's group at
Princeton. \ Woodhouse's thesis extended and improved the work Pirani had done
with Ehlers and Schild by rigorously deriving the differentiable and causal
structure of space-time from a set of axioms with simple and intuitively
obvious physical interpretations (Woodhouse 1973). \ Pirani's interest in
coordinate free methods applied to classical mechanics, and consequently
symplectic geometry, together with the group's increasing interest in quantum
physics, led him to invite David Simms from Trinity College Dublin to give a
series of ten lectures on geometric quantization. \ Delivered in the autumn of
1974 they outlined the programme of Bertram Kostant and Jean-Marie Souriau.
This aimed to formulate the relationship between classical and quantum
mechanics in geometrical terms, as a relationship between symplectic
manifolds, corresponding to classical phase spaces, and Hilbert spaces,
corresponding to quantum phase spaces. \ These lectures were about
quantization of classical theories as opposed to quantum theory itself and
aimed to clarify ambiguities and the role of symmetries in known approaches.
\ Woodhouse took notes and added material and the lectures were published in a
book (Simms and Woodhouse 1976). \ Pirani's interest in this area was short
lived but subsequently Woodhouse wrote one of the standard texts on geometric
quantization (Woodhouse 1979).

\subsection{Quantum gravity, black holes and quantum gravity in curved
space-times}

In February 1974 a two day symposium on quantum gravity was held at the
Rutherford Appleton Laboratory in Oxfordshire (Isham 1975). \ Organized by
Isham, Penrose and Sciama it included talks by Isham, Michael Duff, Abdus
Salam and Stephen Hawking amongst others. \ The idea that the gravitational
field should be quantized arose soon after the development of quantum
mechanics in the 1920s and the first technical papers on quantizing the
gravitational field were written in the 1930s (Blum and Rickles 2018). \ Since
then the quest\ to construct a completely satisfactory reconciliation of
quantum theory and general relativity has remained unsuccessful. Different
approaches have convinced adherents. Currently the dominant one is associated
with string theory and its developments. \ The quest has had its fair share of
hopes raised, dashed, then raised again\footnote{One time line of research
into quantum gravity from the 1930s onward can be found in an article by Carlo
Rovelli (Rovelli 2001).}.

Isham's talk, "An introduction to quantum gravity", was an overview of
approaches to quantum gravity that were then current (Isham 1975a). \ This was
one of many review talks he was invited to give over the next couple of
decades. Isham discussed, among others, the two main approaches, canonical and
covariant quantization. \ The former encompassed developments of the
Hamiltonian formulation of general relativity. \ This was the line of research
Pirani had pursued, in its early days, for his first doctorate. \ In it
gravity and gauge fields are regarded as infinite dimensional analogues of
constrained mechanical systems. \ In the covariant approach gravity is treated
analogously to other Lagrangian field theories, in particular
electromagnetism. \ The metric tensor is separated into a classical background
metric plus a quantum correction term. \ Quantization then proceeds via the
methods that had proven successful in the quantization of the electromagnetic
field, that is, as a perturbation calculation making use of Feynman diagrams.
\ Compared with canonical quantization this was the natural route for people
with a particle physics background to take. \ Unfortunately neither the
covariant or canonical approach was proving successful and, unlike the
situation with quantum electrodynamics, there were no guiding experimental results.

Isham also included an introduction to a topic that was just coming to the
fore, quantum field theory in curved space-time. \ In this a field, such as a
scalar field, is quantized, but the space-time metric is not. \ The matter
field equations are taken from equations for fields in special relativity with
the Minkowski metric replaced by the space-time metric. \ This is a half-way
house to a full theory of quantum gravity. \ The underlying assumption is that
physically meaningful results can be deduced in the appropriate contexts, as
they are in quantum mechanics when the electromagnetic field is included but
is itself left unquantized. \ Isham first explored some of the problems that
had to be dealt with, even for linear field theories, when the usual Minkowski
space-time background, with its Poincar\'{e} symmetry group, is replaced by a
non-flat space-time. These included the non-uniqueness of the choice of
positive frequency solutions, \ the apparent observer dependence of the notion
of the no-particle state or vacuum state, wave-particle duality and the
meaning of a particle. \ He also discussed the problem of back reaction, when
the quantum field itself acted as the source of the (classical) background
metric field, and the modified Einstein's equations of semi-classical gravity.
\ The latter are the equations obtained by replacing Einstein's equations,
$G_{\mu\nu}(g)=T_{\mu\nu}(matter,g)$, for classical, that is, non-quantized
fields, by $G_{\mu\nu}(g)=\langle T_{\mu\nu}(\widehat{matter},$ $g)\rangle$.
\ Here $\langle\rangle$ denotes the expectation value of the quantized system
in some suitable state, $\widehat{matter}$ indicates that the quantized matter
fields are used, and $g$ denotes the classical metric tensor. \ Although not
without its own problems this half way house approach seemed to be offering
the possibility of progress lacking in the attempts to develop a full quantum
theory of gravity.

Duff's talk at the conference was devoted to covariant quantization (Duff
1975). \ It included discussions of the comparatively new method of
dimensional regularization. \ Regularization involves separating the divergent
part of the integrals that arise in quantum field theories from the finite
parts of the integrals in a gauge invariant and covariant manner so that the
divergent part can be dealt with by renormalization\footnote{In
renormalization all the divergences are consistently removed by re-defining
physical parameters in terms of bare parameters and the regularization.}.
\ Dimensional regularization had been successfully applied to gauge theories
and it was natural to investigate its use in the case of gravity. \ Duff also
discussed conformal invariance and anomalies, including work he had recently
done with Derek Capper on discovering the gravitational conformal (sometimes
Weyl or trace) anomaly using dimensional regularization.\ This topic was to
continue \ to engage him, and others, for years as its significance for
gravity became increasingly understood. \ In classical theories of massless
fields in interaction with gravity, such as Maxwell's electrodynamics in
(only) four dimensions, invariance of the action under conformal
transformations $g_{\mu\nu}\rightarrow\Omega^{2}g_{\mu\nu}$, where $\Omega$ is
a non-zero function, is reflected in the vanishing of the trace of the
energy-momentum tensor $T_{\mu\upsilon}$, that is $g^{\mu\upsilon}%
T_{\mu\upsilon}=0$. \ Duff pointed out that dimensional regularization
respects only identities which are valid in all dimensions and do not involve
the dimension explicitly otherwise perturbation theory anomalies will occur.
\ So in the renormalized quantum theory the classical trace-free property of
$T_{\mu\upsilon}$ will be lost and there will be an anomalous trace, that is
$g^{\mu\upsilon}\langle T_{\mu\upsilon}\rangle\neq0$. \ In fact such anomalies
are not artifacts of this particular regularization scheme. \ Consistent
results are obtained when other methods of regularization are employed.

Neither the canonical nor the covariant approaches to quantum gravity in vogue
at this time have been as successful as was once hoped. \ In time it was
conclusively demonstrated that quantum gravity, dealt with via the covariant
approach, is not renormalizable. \ However methods developed in its
investigation proved useful in dealing with gauge theories.

Salam's talk was entitled "Impact of quantum gravity theory on particle
physics" in response to the organizer's request for a talk on this topic
(Salam 1975). \ He started his talk by saying\ "...there has been very little
impact...". \ He then went on to say that the particle physics community
believed - erroneously in his view - that the energies at which quantum
gravity effects would manifest themselves would be in excess of 10$^{19}$ BeV
and so need not be considered. \ This was a fair enough assessment at that
time but it would not be many years before that situation had changed. \ This
occurred for a number of reasons but a contributing factor was the result
presented in Hawking's talk (Hawking 1975a; Hawking 1975b).

Hawking's talk, on particle creation by black holes, was the first public
presentation of his famous result that a black hole could emit particles as
well as absorb them\footnote{The interesting background to Hawking's
calculation and the different strands of research which motivated him can be
found in a review by one of Hawking's former research assistants Don Page
(Page 2005).}. \ To Hawking's surprise his calculation confirmed a suggestion
by Jacob Bekenstein (Bekenstein 1973) that a black hole would have a
temperature which was a non-zero multiple of its surface gravity and a finite
entropy proportional to its area. \ Hawking had regarded Bekenstein's proposal
as just an analogy as it had been thought that black holes could only absorb
and not emit and so would have zero temperature. \ However Hawking was able to
demonstrate that thermal radiation was emitted, the black hole temperature was
given, in Planck units, by $T=\kappa/2\pi\approx10^{-6}(\frac{M_{o}}{M})$
$\mathrm{K}$, and the black hole entropy was given by $S=A/4$. \ In these
expressions $\kappa$ is the black hole's surface gravity\footnote{The surface
gravity is a measure of the acceleration needed to keep a particle on the
horizon}, $A$ is its area, $M$ its mass and $M_{o}$ is the mass of the sun.
\ For the Schwarzschild black hole $\kappa=\frac{1}{4M}$ where $M$ is the
Schwarzschild mass.

In his talk Hawking outlined his calculation of the emission of particles in
the formation of a black hole at late times when the collapse had settled down
to a stationary black hole. \ His approach was to investigate quantized matter
fields in the (classical) Schwarzschild and other classical black hole
background space-times. \ He found particle emission at a steady rate, in all
modes, coming from the black hole. \ One consequence of this was that when
quantum effects were taken into account the area of a black hole would not
always increase. \ This was in contrast to the situation for classical black
holes where Hawking's earlier demonstration that the area of a black hole
would never decrease had been one of the motivations for Bekenstein's entropy
conjecture. \ Hawking's paper connected quantum theory, thermodynamics,
geometry and gravity in an unprecedented way. \ Viewed retrospectively this
was a notable moment and the one for which the conference is probably most
remembered. \ It immediately stimulated an intense period of research into
quantum theory in curved space-times. \ Today approaches to quantum gravity
are viewed sceptically if they do not incorporate the Bekenstein-Hawking black
hole/thermodynamic connection.

For those at the symposium hearing the details for the first time Hawking's
calculation was difficult to assess but the results were clearly important -
if they were right. By that time Hawking's speech was not easy to understand
and an assistant simultaneously projected a written version of what he was
saying onto a screen. \ Moreover his use of a combination of geometrical\ and
quantum mechanical arguments was new to most of the audience \footnote{Paul
Davies drove Abdus Salam, Tetz Yoshimura and myself to and from the conference
and, as I recall, there was not much more discussion of Hawking's talk than
the other talks while we were travelling. \ Everyone needed more time to
absorb it.
\par
Rumours of Hawking's work had been floating around before the conference -
there had been talk of exploding black holes which sounded very strange - so I
had asked Roger Penrose about it after a seminar in London. \ \ He had
discussed the result with Hawking and told me he thought that Hawking was
right and that influenced my own attitude towards Hawking's result.
\par
At some point during the conference Salam did say something to the effect that
he hoped the field (general relativity) would remain a "friendly \ pursuit" as
opposed to other more competitive areas such as his own. \ However the laid
back attitude to research in gravity was on the way out.}.

Hawking's first publication about his result came out in March, a brief note
in Nature entitled "Black hole explosions?" (Hawking 1974). \ Although the
effect Hawking had found was tiny for solar mass and larger black holes, like
those at the centre of galaxies, for very small black holes it was not
insignificant. \ Hawking wrote

\begin{quotation}
As a black hole emits this thermal radiation one would expect it to lose mass.
\ This in turn would increase the surface gravity and so increase the rate of
emission. \ The black hole would therefore have a finite life of the order of
$10^{71}(M_{0}/M)^{-3}s$. \ For a black hole of solar mass this is much longer
than the age of the Universe$^{2}$. \ There might, however, be much smaller
black holes which were formed by fluctuations in the early Universe. \ Any
such black hole of mass less than 10$^{15}g$ would have evaporated by now.
\end{quotation}

So ending its life in an explosion.

As the question mark in the paper's title suggests its conclusions were
cautiously presented. \ Regarding Bekenstein's suggestion Hawking noted that

\begin{quotation}
...Bardeen, Carter and I considered that the thermodynamical similarity
between $\kappa$ and temperature was only an analogy. \ The present result
seems to indicate, however, that there may be more to it than this. \ Of
course this calculation ignores the back reaction of the particles on the
metric, and quantum fluctuations of the metric. \ These might alter the picture.
\end{quotation}

Hawking's result was not immediately universally understood or even thought to
be correct and at that point Hawking himself was still not completely certain
about it. \ Upon hearing Hawking's talk John Taylor apparently thought that
its conclusions were incorrect. \ In July, together with Davies, he published
a rebuttal to Hawking's paper in Nature,\ also with a question mark in its
title, "Do black holes really explode?" (Davies and Taylor 1974). \ The
publication of the details by Hawking in 1975 together with various different
confirmations of his conclusions eventually led to his result being generally
understood and established although its quantum mechanical implications are
still being debated.

Davies' own assessment changed quite quickly and a highly productive period of
work on quantum field theory in curved space-times began at
King's\footnote{The term quantum theory in curved space-times is used to
include quantum theory in (subsets of) Minkowski space-time and in space-times
of different dimensions.}. \ In August 1974 he submitted to the Journal of
Physics A a paper entitled "Scalar particle production in Schwarzschild and
Rindler metrics" (Davies 1975). \ With the aim of understanding Hawking's
result Davies considered the quantum field theory of a massless scalar field
in a subset of two dimensional Minkowski space-time which he termed the
Rindler wedge. \ \ In two dimensions and in Minkowski coordinates the
Minkowski metric is given by%
\[
ds^{2}=dt^{2}-dx^{2}.
\]
\ Changing coordinates to $X=\sqrt[2]{x^{2}-t^{2}}$, $T=\tanh^{-1}(t/x)$
gives
\[
ds^{2}=X^{2}dT^{2}-dX^{2}.
\]
\ The lines of constant $X$ correspond to the world lines of an observer
undergoing a uniform acceleration of magnitude $\frac{1}{X}$. \ The Rindler
wedge is then defined by $0<X<\infty$, $-\infty<T<\infty$, $x^{2}\geqq t^{2}$.
\ For uniformly accelerated observers the two asymptotes $X=0$, $T=-\infty$
and $X=0$, $T=\infty$ behave as past and future event horizons
respectively\footnote{Such coordinates and space-times are now termed Rindler
coordinates and Rindler space-times although they were introduced in the first
decade of the twentieth century and subsequently used by numerous people
including Bondi in his discussion of uniform acceleration in 1957. \ Rindler,
amongst others, had noted the close similarity of the wedge to the static
exterior region of the Schwarzschild black hole. \ His discussion of them in
the context of the Kruskal extension brought them to contemporary attention
(Rindler 1966).}.

Davies noted that an analysis of flat space-time quantum field theory, in two
dimensions and\ in Rindler coordinates, might provide a conceptually and
calculationally simple test for the black hole case with the advantage that
the standard quantization scheme using Minkowski coordinates was available for
comparison. \ Furthermore, by equipping the space-time with a perfectly
reflecting mirror placed at a fixed distance to the right of the origin the
properties of the Schwarzschild black hole static exterior could be well
replicated. \ The role of the mirror was to turn incoming (left moving) waves
into outgoing (right moving) waves just as incoming waves are changed into
outgoing waves on passing through a body collapsing to form a black hole.
\ Aspects of this general framework: two dimensions, simple linear field
equation, Rindler space-time and coordinates and reflecting mirrors were soon
to be employed in a number of calculations by members of the King's group in
their attempts to understand the Hawking effect and to further develop quantum
theory in curved space-times.

Davies applied Hawking's black hole argument using this simple model.
\ Previously Davies had found Hawking's result surprising as earlier work had
indicated that particle conservation would normally be expected in the static
region of Schwarzschild space-time. \ He now realized that Hawking's result
hinged on the event horizon. \ Davies found the hitherto little suspected
result that the fixed reflecting mirror appeared, to a uniformly accelerating
observer, with acceleration $\alpha$, to radiate at a constant temperature of
$\frac{\alpha}{2\pi}$ in geometrical or Planck units. \ Comparing this with
Hawking's black hole temperature of $\frac{\kappa}{2\pi}$ the acceleration
equates to the surface gravity.

In his interpretation of his results, and in comparing them with Hawking's,
Davies wrote,

\begin{quotation}
the apparent production of particles in this case is somewhat paradoxical
because there is no obvious source of energy for the production. \ Such
emission of radiation is, of course, absent when the system is quantized in
conventional Minkowski coordinates, so the result demonstrates how the concept
of a particle is ill-defined and observer dependent (Davies 1975).
\end{quotation}

Subsequently much work, in the quantum mechanical context, would be done by
the King's group on energy-momentum computations and the observer dependence
of events. \ Some years later Davies would feel able to write a paper entitled
"Particles do not exist" (Davies 1984).

Over the next few years four post-docs who were to play important roles in the
work on quantum field theory and gravity came to the mathematics department.
Stephen Fulling and Mike Duff took up two year positions in 1974. \ Fulling
came from a post-doctoral position at the University of Wisconsin Milwaukee.
There he had worked with Leonard Parker, one of the founding fathers of
quantum field theory calculations in cosmological backgrounds (Parker 2017).
\ Their work had included the development and use of the concept of a
Bogoliubov transformation - a linear transformation of creation and
annihilation operators (Bogoliubov 1959) which was central to Hawking's
calculations and other work on quantum theory in curved space-times.
\ Fulling's 1972 PhD was from Princeton University where his supervisor had
been Arthur Wightman. \ He brought with him an expertise on quantum field
theory in curved space-times which was unusual at that time.\ \ He had already
written an important paper on the non-uniqueness of canonical field
quantization in curved space-times (Fulling 1973). \ Davies' result about
radiation detected by an accelerating observer, together with that earlier
work by Fulling, can be regarded as precursors of what is now known as the
"Unruh effect" (sometimes the "Fulling-Davies-Unruh effect"). \ In William
Unruh's seminal work model detectors with acceleration $\alpha$ were
constructed. \ Unruh showed that the detectors "clicked" at a rate consistent
with their observation of a gas of particles with temperature $\frac{\alpha
}{2\pi}$ (Unruh 1976). \ The effect is now recognized as required for the
consistency of flat space-time quantum field theory in inertial and
accelerated frames and its descriptions of observed phenomena such as particle
decay (Fulling and Matsas 2014).

Mike Duff came after having held post-doctoral positions at Trieste and
Oxford. \ He had been one of Abdus Salam's and Chris Isham's PhD students at
Imperial College\footnote{\ Years later he recalled that the topic of his PhD
research "was greeted with hoots of derision when I announced it at the
Cargese Summer School en route to my first post-doc in Trieste. \ The work
originated with a bet between Abdus Salam and Hermann Bondi about whether you
could generate the Schwarzschild solution using Feynman diagrams. \ You can
(and I did, but I never found out if Bondi ever paid up)." \ (After dinner
talk at the Workshop on Frontiers in Field Theory, Quantum Gravity and String
Theory, Puri, India 1996.)}. \ Duff's primary interest was in quantum field
theory and he pursued that while at King's.

The third post-doc, Steven Christensen, came to King's in 1975 for one year
after completing his PhD at the University of Texas. \ His thesis supervisor
there had been Bryce DeWitt who was also in England during the academic year
1975-76. \ De Witt, one of the leading figures in quantum gravity, was based
in Oxford for the year and during that time there was a lot of interaction
between the groups at King's and Oxford as well as the group at Cambridge.
\ In addition another leading figure, Stanley Deser, visited King's from
Brandeis University for an intense period during 1976.

Larry Ford's arrival in 1977 to take up a two-year postdoctoral position gave
a new impetus to the research. \ Like Fulling and Christensen his expertise
was also in quantum theory in curved space-times. \ He too had been a student
at Princeton but the supervisor of his 1974 PhD thesis had been John Wheeler.
\ Again like Fulling he came from a post-doc with Leonard Parker. \ Throughout
this period post-graduate students like Davies' students, Tim Bunch and
Nicholas Birrell, and Isham's students, Bernard Kay and Jeanette Nelson, did a
lot of the calculating and made notable contributions themselves.

The lines of research pursued at King's, and elsewhere of course, included
both technical ones aimed at enabling calculations to be carried out on a firm
footing as well the investigation of models - in particular models in two
dimensions where unilluminating calculations could be stripped away. \ The
research led to insights into Hawking's result and to a much better
understanding of the general theory of quantum field theory in curved
space-time. \ Much time was devoted to the development of a particular method
of dealing with the ultraviolet divergences that arise in quantum field
theory, that is, divergences related to the short distance behaviour of the
vacuum expectation values of products of field operators. \ This method of
regularization, termed "point splitting", was a technical matter of some
importance. \ In certain applications it was found to be better than other
methods. \ It was used in the computation of the vacuum expectation value of
the energy momentum tensor entering the semi-classical Einstein equation\ as
well as in the computation of other physically significant quantities. \ Built
on previous work by Julian Schwinger and Bryce DeWitt this approach to
regularization had been used by Christensen in his PhD thesis. \ In the paper
based on his thesis work Christensen thanks DeWitt for the encouragement he
received in getting through the tedious calculations the work involved
(Christensen 1976). \ Christensen arrived at King's to find that Davies,
Fulling and the students were already sharing in this necessary tedium. \ The
method can be illustrated by outlining some of the steps in the computation of
the vacuum expectation value of the energy momentum tensor of a scalar field
in a background gravitational field. \ Consider, in that example, the product
of two field operators appearing in each term of the stress tensor.
\ Evaluation at the same space-time point gives a result which is divergent.
\ This problem is then remedied by evaluating each field operator at a
different space-time point, the points being infinitesimally separated by a
proper distance $\varepsilon$ along the unique geodesic connecting them, so
obtaining a finite bi-tensor object. \ This procedure leads to the regularized
form of $\langle T_{\mu\upsilon}\rangle$. \ After performing an expansion in
$\varepsilon$ the terms which diverge as $\varepsilon$ tends to zero are
displayed in the form $\varepsilon^{-4}$, $\varepsilon^{-2}$ and $\ln$
$\varepsilon$ and can be isolated. \ Covariance is maintained by introducing
additional parameters - the components of the vector tangent to the geodesic
at the point given by $\varepsilon=0$. \ Subtraction of the divergent terms as
well as the subtraction of certain finite terms in the vacuum expectation
value of the stress tensor, so that conservation laws are satisfied, leaves a
renormalized finite vacuum expectation value. \ Point splitting regularization
was used in many applications and a number of general formulae were found
using it.\ 

Christensen and Fulling shared an office with Duff. \ Although Isham and Duff
were largely working on other topics in quantum field theory their insights
were always influential. \ There was a certain amount of good natured joshing
from the quantum field theorists about the virtues of dimensional
regularization as opposed to the work of the "point splitters". \ However when
the results came through they changed their minds. \ Snapshots of a few of the
large number of papers produced at King's in the second half of the 1970s give
an idea of the work that was done.

Fulling and Davies used point splitting regularization to compute the
energy-momentum tensor, $\langle T_{\mu\upsilon}\rangle$, of a massless scalar
field in two dimensions influenced by the motion of a perfectly reflecting
mirror.\ They showed that there was a flux of energy radiated which could be
either positive or negative depending on the instantaneous mirror velocity and
its changes. \ When the acceleration is increasing the flux is
negative.(Fulling and Davies 1976).

Duff and Isham, together with Deser during his 1976 visit, undertook further
work on the conformal anomaly discussed by Duff in his Rutherford Laboratory
talk (Deser 1976a). \ They calculated the most general form of the trace of
the energy momentum tensor in various dimensions. \ In two dimensions they
showed that $g^{\mu\upsilon}\langle T_{\mu\upsilon}\rangle=aR$;\ where $a$ is
a constant and $R$ is the Ricci scalar, the only non-zero component of the two
dimensional curvature of the metric. \ An expression, in terms of geometrical
quantities, for $g^{\mu\upsilon}\langle T_{\mu\upsilon}\rangle$ was also
obtained in four dimensions. \ In that case, as they pointed out, only one of
the geometrical terms could be removed by finite local counter terms .
\ Conformal anomalies turned out to be important in a number of different
contexts, not least in the work at King's (Duff 1994).

Davies, Fulling and Bill Unruh, then at McMaster University but a frequent
visitor to London, investigated Hawking's work in further detail by
considering a general two dimensional space-time and a two dimensional model
of gravitational collapse (Davies 1976). They calculated $<T_{\mu\nu}>$, the
vacuum expectation value of the energy-momentum tensor of a massless scalar
field, regularizing the energy-momentum tensor by point splitting, and found
that quantum radiation production was incompatible with a conserved and
traceless $<T_{\mu\nu}>$. \ In an Eureka moment in the bathtub Fulling
realized that the conservation law could be rescued by adding a $Rg_{\mu\nu}$
term, and that resembled what Deser, Duff and Isham were doing. Consequently
they required conservation but allowed a trace;\ the trace term acted as a
source in the conservation law so that radiation could be created. \ In their
paper they concluded that, in their collapse model, black-hole evaporation
occurred with pairs of particles being created outside the horizon and not
entirely in the collapsing matter, negative energy being carried into the
future horizon of the black hole by one particle of such a pair while the
other particle of the pair contributed to the thermal flux at infinity. \ This
was in contrast to the flat- space mirror systems where all radiation
originated at the mirrors.

Fulling and Christensen made a direct link between the conformal anomaly and
Hawking radiation. \ Using point splitting regularization they showed that the
analogue of the Hawking effect in two space-time dimensions is entirely due to
the existence of the trace anomaly. \ They noted that the magnitude of the
Hawking black body effect at infinity was directly proportional to the
magnitude of the anomalous trace (in two dimensions a multiple of the
curvature scalar) and they observed that, in the final state of collapse, a
knowledge of either number completely determined the stress tensor outside a
body. \ In other words no conformal anomaly no Hawking radiation. \ They also
found that in the four dimensional case the trace anomaly determined the
energy momentum-tensor, $\langle T_{\mu\upsilon}\rangle$, up to one function
of position (Christensen and Fulling 1977; Christensen 1984b).

The thermodynamics of black holes revealed in Bekenstein and Hawking's work
continued to be studied at King's. \ For example\ Davies found that if a black
hole spins faster than a certain rate it undergoes a phase transition beyond
which, instead of radiating and getting hotter by the Hawking effect, it cools
as it radiates like a normal body. \ He found similar results for black holes
carrying a sufficiently large charge (Davies 1977).

Quantization in cosmological backgrounds was also actively investigated in a
number of papers by Davies, Ford, Birrell and Bunch. \ This research included
studies by Davies and his student Tim Bunch of quantum field theory in a de
Sitter space-time background. \ Their work became of more immediate relevance
than can have been expected with the emergence, towards the end of the King's
programme, of the ideas about inflation and the inflationary universe.
\ During the inflationary period the universe resembles de Sitter space and
the tiny variations in temperature superimposed on the uniform cosmic
microwave background could be quantum fluctuations generated in that period.
\ The vacuum state now considered to be appropriate in the discussion of these
turned out to be one investigated by Bunch and Davies (Bunch and Davies 1978).
\ It is the zero-particle state seen by a observer in free fall in the
expanding universe and possesses no quanta at asymptotically past infinity.
\ Because of their work it is now known as the Bunch-Davies vacuum although
others had investigated it previously.

Ford, in a discussion of important points of principle, investigated negative
energy densities and fluxes due to quantum coherence effects (Ford 1978;
1997). \ \ Negative energy fluxes arose in the moving mirror model discussed
by Fulling and Davies. \ Negative energy plays an important role in the
Hawking effect. \ Recall that Davies, Fulling and Unruh concluded that pairs
of particles were created outside the event horizon with one of the pair
escaping to infinity and the other falling into the horizon, the latter
particle carrying negative energy as measured at infinity. \ Negative energy
densities and fluxes also arise in flat space-time as, for example, in the
Casimir effect and can also arise, as Ford discussed, as a result of quantum
coherence effects. \ Ford observed that if arbitrary fluxes of negative energy
were available and if negative energy was shone onto a hot object, resulting
in a net decrease in entropy, the second law of thermodynamics could be
violated. \ He considered various examples which led him to suggest that if a
negative energy flux $F$ was constrained by an inequality of the form
$\left\vert F\right\vert \lesssim\tau^{-2}$, where $\tau$ is a characteristic
time over which the negative energy flux occurs, such a violation will not
occur. \ This work initiated the difficult study of quantum energy
inequalities. \ Classical energy inequalities are used in proofs of
singularity and global theorems but, as Ford observed, may break down in the
quantum regime.

The plethora of work on quantum theory in curved space-times carried out at
King's and other institutions, including work on spaces with non-trivial
topologies, is more fully discussed in the book Birrell and Davies wrote
towards the end of the decade (Birrell and Davies 1982). \ At that time at
least this was the "go to" book for anyone wanting to learn about the new
developments in quantum theory in curved space-time just as Hawking and
Ellis's book had been for those wanting to learn about space-time global structure.

\ By 1980 work on this area at King's was starting to wind down. \ It had been
an exciting period for all involved. \ A few years later Christensen, for
example, was to comment that it had been a very inspiring time and he had not
seen it repeated since (Christensen 1984b).

Isham returned to Imperial College in 1976. \ The post-docs all moved on, Duff
to post-doctoral positions first at Queen Mary College and then at Brandeis
University, Fulling to a faculty position at Texas A\&M, Christensen to post
doctoral positions in Utah and then at Harvard and Ford to the University of
North Carolina and then Tufts. \ They all continued to make major
contributions to research in this area. Some years later Fulling wrote a
textbook on the subject (Fulling 1989). \ It included an appendix containing
his well-cited but previously unpublished pre-print, "Varieties of Instability
of a Boson Field in an External Potential and Black Hole Klein Paradoxes"
written while he was at King's. \ While Duff was at Brandeis and Christensen
was at Harvard they collaborated on further work on the conformal anomaly.

Some years later, after the dust had settled, Christensen edited a festchrift
volume in honour of Bryce DeWitt (Christensen 1984a). \ A number of the
articles included reflections on the work that had been done at King's. \ In
his article, entitled "What have we learned from quantum field theory in
curved space-time?" (Fulling 1984),\ Fulling commented that quantum field
theory in curved space-time was still a scene of confusion and controversy in
early 1977 while later in the article he noted that 1978 \ saw "the days of
glory" end. \ He listed substantial issues whose resolution he considered to
have been agreed. \ They included the following: an accelerating observer in
empty space will detect particles in the sense that its detector will click or
its thermometer will get hot; the expectation value of the stress tensor in a
particular quantum state is well defined independently of the motion of the
observer; the renormalized stress tensor of a conformally invariant quantum
field has a non-vanishing trace; this trace anomaly is well defined except for
the coefficient of any term equal to the trace of a covariant, local,
polynomial, conserved functional of the metric tensor involving derivatives of
order less than or equal to four - which is an arbitrary renormalization
constant; \ point splitting is better than dimensional regularization for
calculating the expectation value of the stress tensor but not for calculating
an effective Lagrangian, and which is better to calculate depends on its
intended use.

Fulling also observed that the semi-classical coupling of gravity and the
stress tensor in the semi-classical Einstein field equations as above or as in
their modification by the addition of geometrical terms to take into account
gravitational back reaction, were generally regarded as just a stop-gap,
hopefully physically reasonably accurate in at least some situations. \ In a
comment on the unclear relationship between quantum gravity proper and quantum
field theory in a classical background Fulling mentioned the critique of Duff
(Duff 1981). \ Amongst other things Duff had pointed out that any
gravitationally induced quantum process that produces particles would also
produce gravitons and the quantization of the matter fields would either be
trivial or physically incomplete unless the gravitational field was also quantized.

On a lighter note, Christensen, in his article in the DeWitt festchrift
volume, wrote that when he submitted his first paper to the Physical Review an
editor objected to the term "point splitting" on the grounds that points can't
be split. \ Hence the term in the published paper was changed to "covariant
point-separation method". A similar change was made by a Physical Review
editor to a paper that Stanley Deser wrote with Pirani and myself during his
1976 visit. \ We had investigated an interesting approach to gravity that had
been developed, in analogy with an early string model of elementary particles,
by Tullio Regge and Claudio Teitelboim (now Claudio Bunster) (Regge and
Teitelboim 1977). \ In their work the basic fields were taken to be not the
components of a metric but functions describing the embedding of four
dimensional space-time in a ten or possibly higher dimensional manifold. \ Our
paper, a critique of their work, showed that their theory suffered from a
gauge-dependence which appeared to be physically unacceptable. \ We also
demonstrated that their field equations were inequivalent to Einstein's by
showing that they admitted solutions which were not solutions of Einstein
equations. \ We termed their theory "G-string theory" for short and entitled
our preprint "Imbedding the G-string". \ A Physical Review editor objected to
our terminology and so in the published paper the new theory was termed \ the
new embedding model of general relativity and the title of the paper was
changed to "New embedding model of general relativity" (Deser 1976b).

Paul Davies left for the physics department of Newcastle University in 1980
where he continued his work on quantum field theory in curved space-times.
\ Early in the 1970s Davies had begun writing reviews for Nature and he
steadily extended this activity to the writing of books about gravity,
cosmology and other subjects. \ He became well-known beyond academic circles
for both these and his journalistic and broadcasting activities.

\subsection{Supergravity and twistor theory}

In the late 1970s and early 1980s work on two other significant lines of
research were also undertaken at King's. \ Their aims included, in different
ways, the formulation of satisfactory quantum theories of gravity. \ This
research was rather different from the work to which most of this essay is
devoted, and its full consideration is outside its scope. However aspects of
it warrent mentioning not only for their own importance but also because\ they
illustrate the changing nature of gravitational research at King's in the
period immediately before the College itself underwent major changes.
\ Contributions to the study of on gravity and supergravity, by Kelly Stelle
and Peter West, and to twistor theory, by Stephen Huggett and Andrew Hodges,
are sketched.

\subsubsection{Supergravity and gravity}

The discovery of supersymmetry, an invariance of a theory under the
interchange of fermions and bosons, is usually regarded as taking place in the
early 1970s (Kane and Shifman 2000). \ The application of these ideas to
gravity, where Einstein's general relativity is extended by accompanying
general coordinate transformations with supersymmetry as a local symmetry,
took place in 1976 (Freedman 1976; Deser and Zumino 1976)\footnote{A personal
account of the early days of supergravity has been given by S. Deser.
(Deser2018).}. \ In supergravity, in addition to the usual spin 2 graviton of
quantum gravity, there is a new type of particle, of gravitational origin, a
spin 3/2 particle called the gravitino\footnote{This is the contemporary use
of the word "gravitino" coined by Pirani (Pirani 1955a).
\par
An early contributor to super-mathematics and physics, where anti-commuting
variables are used, was John Martin who, from 1966, was a member of the King's
physics department (Martin.1959 a; 1959b).
\par
{}}.

Interest in these developments amongst the quantum field theorists at King's
such as John Taylor, his post-docs and students, was immediate and research on
supersymmetry began. \ In 1977 Kelly Stelle and Peter West began working on
developing supergravity. \ Their research included the exploration of the
first supergravity theory where the supersymmetry algebra closed only when the
equations of motion of the theory were satisfied. \ A consequence of this was
that it was very difficult to couple supergravity to supersymmetric matter and
to quantize. \ Stelle and West were able to formulate a supergravity theory
possessing a symmetry algebra that closed in the usual way without the use of
the equations of motion (Stelle and West 1978a). \ This allowed them to
construct the analogue of the tensor calculus of general relativity for
supergravity. \ It led to the construction of the most general supersymmetric
theory (Stelle and West 1978b; 1978c). \ The latter theory provides the
framework for all discussions trying to realize a supersymmetric model of
nature\footnote{Related work was also undertaken by Sergio Ferrara and Peter
van Nieuwenhuizen.}.

Stelle and West also worked on the formulation of gravity as a gauge theory.
\ The general idea of this approach was to bring gravity into line with the
theories of the other fundamental forces of nature which are formulated in
terms of gauge theories. \ \ As previously mentioned Dennis Sciama had been an
early contributor to aspects of this line of research in developing the
Einstein-Cartan or ECSK theory. \ Stelle and West's aim was to completely
formulate gravity as a Yang-Mills theory rather than to just gauge translation
symmetries as other work had done. \ They built on previous work in which, by
adopting one constraint, gravity and supergravity had been formulated as gauge
theories of, respectively, the Poincar\'{e} and super-Poincar\'{e} groups
(Chamsedine and West 1977)\footnote{Although that constraint broke the
Yang-Mills symmetry the derivation and simplicity of the construction led to
its use in the construction of conformal supergravity theories. \ It underpins
the more recent construction of higher spin theories.}. \ Stelle and West
showed that the construction for gravity could be made completely invariant
under the Poincare gauge symmetries without adopting the constraint. \ Much
more satisfactorily they introduced scalar fields and spontaneous symmetry
breaking (Stelle and West 1979; 1980).

Supporting experimental evidence for supersymmetry and supergravity has yet to
be found but the ideas underpin contemporary string theory and its developments.

\subsubsection{Twistor theory}

The years 1979 to 1983 saw a small but significant amount of research done at
King's on twistor theory - a theory which Roger Penrose had developed from his
work on two-component spinors, the behaviour of zero rest-mass fields under
conformal transformations and the geometry of null geodesics. \ One of the
aims of the twistor programme was to deal with gravity and its quantization.
\ During his year at King's Stephen Huggett gave a post-graduate course on the
theory. \ Together with Paul Tod, who had given a similar course at Oxford,
Huggett turned the lectures into a standard introduction to the subject
(Huggett and Tod 1985). \ This included some of the then most recent
developments such as the construction of anti-self dual solutions of the
classical Yang-Mills equations and the non-linear graviton. \ After Huggett
left for the University of Plymouth work in the mathematics department on
twistor theory continued. \ Andrew Hodges, whose PhD had been supervised by
Penrose at Birkbeck College, held a postdoctoral position between 1981 and
1983. \ His and Huggett's work at King's refined and kept alive what was for a
long time the totally marginal and unfashionable study of twistor diagrams
(Hodges and Huggett 1980). \ One of the main aims of this aspect of twistor
theory was the production of a manifestly finite theory of scattering in
quantum field theory. \ At King's Hodges investigated massless M\o ller and
Compton scattering (Hodges 1983a; 1983b). \ \ He also studied the
regularization of divergences and began an attack on a long standing problem
in twistor diagram theory by introducing an idea for dealing with massive
states (Hodges 1985). \ When his post-doc ended so too did this line of work
at King's. \ Subsequently Hodges joined Penrose's group at Oxford where he
continued to study twistor diagrams. \ Twenty first century investigations of
gauge theories, led by string theorists, have revived interest in them and his
work has contributed significantly to this activity (Atiyah 2017). \ While at
King's Hodges also completed years of work on his acclaimed biography of Alan
Turing (Hodges 1983c).

\section{Conclusion}

In 1981 Kilmister had taken over from the pure mathematician Albrecht
Fr\"{o}hlich as head of the mathematics department and became centrally
involved in administration. \ By then Pirani had all but ceased to be
significantly engaged in research. \ Quantum field theory and supersymmetry
had become the main areas of activity as far as theoretical physics in the
mathematics department was concerned. \ The relativity seminar had become
first a general theoretical physics seminar and then a quantum field theory
seminar. \ Alice Rogers, whose research area was supersymmetry, joined the
department in 1983, initially as a research associate, and after holding long
term fellowships became a permanent member of staff in 1994. \ Davies' post
was not filled until 1984 when Paul Howe was appointed under a special
nationwide "new blood scheme". \ This aimed to ameliorate the job drought by
providing support for a small number of new permanent university positions.
\ Howe's research interests were quantum field theory, supersymmetry and
string theory. \ In time, a large string theory group, under the leadership of
Peter West, became a leading centre of research.

In the early 1980s a re-organization of the University of London was proposed
with a number of Colleges merging or closing down. \ The proposal included
changes which would affect King's. \ Two of the smaller Colleges of the
University, Queen Elizabeth College and Chelsea College, were to be
incorporated into Kings to form a College with over 5,000 students and about
500 full-time\ academic staff. \ The merger took place in 1985 and the King's
mathematics department, which was down to 14 full-time academic staff in
1983-84, re-formed with 41 full-time members of \ staff from the merging
colleges. \ The quantum field theorist Raymond Streater and three other
mathematical physicists from the closed down Bedford College also joined the
department\footnote{The new head of department, Peter Saunders, came from
Queen Elizabeth College. \ He had been Pirani's student in the 1960s, one of
the very few to produce a thesis in cosmology, in his case on non-isotropic
universe. \ }. \ It took a long time for the department to rebalance and new
appointments could not be made for many years. \ A similar situation held
across the physical sciences and engineering departments and recovery took a
long time. \ Today the College is much larger, it has over thirty thousand
students and there are well over 50 full time academic members of staff in the
mathematics department. \ After the time of the merger the structure of the
University began to change with its Colleges increasingly functioning as
separate universities, both financially and academically.

\section{Postscript}

The three original members of the King's relativity group had all retired by
the mid 1980s and become emeritus professors. \ After their retirements they
continued to be active. \ Bondi contined to publish papers on topics in
general relativity and to mull over the aspects of the subject which had most
engaged him. \ He also continued to be actively interested in, amongst other
things, education. \ Bondi's inaugural lecture had been entitled "Science as
an education". \ He was interested in ways of teaching relativity (Bondi 1959)
but he was also seriously concerned about all levels of education throughout
his life\footnote{Bondi would regularly give talks in schools. \ He would
provide two titles and let a school choose between them. \ One day a school
teacher called up to ask what would be the difference between the two talks.
\ He was promptly informed by Bondi's secretary that there were two titles but
just one talk.}. \ He encouraged Pirani and Kilmister to be involved in
various educational matters and, to different degrees, they
were\footnote{Kilmister recalled that once upon a time they had been concerned
about the shortage of mathematics teachers in England. \ They had trooped
along to lobby the civil servant with responsibilities in that area. \ When
they arrived they were told that the problem had been solved. \ The civil
sevice had re-defined "mathematics teacher" and there was no longer a
shortage.}. \ Pirani's various retirement activities included the writing and
reviewing of books for children and young people and the coauthoring of an
illustrated book "The Universe For Beginners" (Pirani and Roche 1993; Pirani
2011). \ He ceased to do any academic work after completing his differential
geometry book with Crampin, with one exception. \ Bondi persuaded him to
investigate plane waves again and they published two papers on this topic
(Bondi and Pirani 1988; 1989)\footnote{In the old days, at least, Bondi and
Pirani used to converse on the telephone about calculations involving tensor
calculus by indicating superscripts with high treble and subscripts with deep
bass.}. \ Kilmister continued to pursue his interest in Eddington's work and
wrote a number of books related to it (Bondi 1995).\newpage

Conversations with Felix Pirani and Clive Kilmister were recorded in 2005 and
2006 and are warmly remembered.

I benefited greatly from extended conversations with Michael Crampin, Ray
d'Inverno, Josh Goldberg, Ted Newman, Jim Ritter, Peter Saunders and Julian
van der Burg. \ June Barrow-Green carefully read and commented on a draft. \ I
thank them all.

I am grateful for all the help and information I received, particularly from
Steve Christensen, Paul Davies, Stanley Deser, Mike Duff, Larry Ford, Stephen
Fulling, Robert Geroch, Andrew Hodges, Stephen Huggett, Bernard Kay, Pawel
Nurowski, Dean Rickles, Andrew Robinson, Barbara Robinson, Ian Roxburgh, Roger
Schafir, John Silvester, John Stachel, Nick Woodhouse, and Andrzej Trautman.

I profited from numerous suggestions by the helpful referees and journal
editor Wolf Beiglb\H{o}ck.

Last, but not least, I must acknowledge my debt to George Papadopoulos and
Peter West whose encouragement and assistance resulted in this essay being
completed.\newpage{\LARGE Gravitational waves in general relativity I-XVI:
published by the Royal Society of London\bigskip}

The first thirteen and the last two of these papers were published \ in the
Proceedings of the Royal Society of London A. \ All the authors were, at one
time or another, based at King's College London except for Ivor Robinson who
was a regular visitor. \ The fourteenth paper is by authors who were not at
King's and was added to the series with Hermann Bondi's permission. \ Unlike
the others, it was published in the Philosophical Transactions of the Royal
Society of London A.

\textrm{I}. \ Marder L. 1958. \ Cylindrical waves \textbf{244}: 524-537.

\textrm{II}. Marder L. 1958. \ The reflexion of cylindrical waves
\textbf{246}: 133-143.

\textrm{III}. Bondi H., Pirani F.A.E., Robinson I. 1959. \ Exact plane waves
\textbf{251}: 519-533.

\textrm{IV}. Pirani F.A.E. 1959. \ The gravitational field of a fast moving
particle \textbf{252}: 96-101.

\textrm{V}. Marder L. 1961. An exact spherical wave \textbf{261}: 91-96.

\textrm{VI}. Sachs R.K. 1961. \ The outgoing radiation condition \textbf{264}: 309-338.

\textrm{VII}. Bondi H., Van der Burg M.G.J., Metzner A.W.K. 1962. Waves from
axisymmetric isolated systems \textbf{269}: 21-52.

\textrm{VIII}. Sachs R.K. 1962. \ Waves in asymptotically flat space-time
\textbf{270}: 103-126.

\textrm{IX}. van der Burg M.G.J. 1966. \ Conserved quantities \textbf{294}: 112-122.

\textrm{X}. van der Burg M.G.J. 1969. \ Asymptotic expansions for the
Einstein-Maxwell field \textbf{310}: 221-230.

\textrm{XI}. Marder L. 1969. \ Cylindrical spherical waves \textbf{313}: 83-96.

\textrm{XII}. Marder L. 1969. \ Correspondence between toroidal and
cylindrical waves \textbf{313}: 123-130.

\textrm{XIII}. Bondi H., Pirani F.A.E. 1989. \ Caustic properties of plane
waves. \textbf{421}, 395-410.

\textrm{XIV}. Chru\'{s}ciel P.T., MacCallum M.A.H., Singleton D.B. 1995.
\ Bondi expansions and the 'polyhomogeneity' of $\mathfrak{I}.$
\textit{\ Philosophical Transactions of the Royal Society of London A
}\textbf{350}: 113-141.

\textrm{XV. }Bondi H. 2000. The loss-free case \textbf{456: }2645-2652.

\textrm{XVI. }Bondi H. 2004 Standing Waves \textbf{460}: 463-470.\newpage

Abbott B.P. et al. (LIGO Scientific Collaboration and Virgo Collaboration)
2016. \ Observation of Gravitational Waves from Binary Black Hole Merger.
\ \textit{Physical Review Letters }\textbf{116}: 061102.

Adams W.G. 1871. \ Report of Professor W.G. Adams on observations of the
eclipse of December 22, 1870, made at Augusta, in Sicily. \ \textit{Monthly
Notices of the Royal Astronomical Society }\textbf{31}: 155-161.

Alessio F., Esposito G. 2018. \ On the structure and applications of the
Bondi-Metzner-Sachs Group. \ \textit{International Journal of Geometric
Methods in Modern Physics }\textbf{15}: 1830002.

Anon. 1921. \ Prof. Einstein's Lectures at King's College, London and the
University of Manchester. \ \textit{Nature }\textbf{107}: 504.

Arnowitt R., Deser S., Misner C.W. 1962. \ The Dynamics of General Relativity.
\ In Witten pp227-265.

Ashtekar A., Berger B.K., Isenberg J., MacCallum M.  2015. \ \textit{General
Relativity and Gravitation: A Centennial Perspective.} Cambridge, Cambridge
University Press.

Atiyah M., Dunajski M., Mason L.J. \ 2017. \ Twistor theory at fifty: from
contour integrals to twistor strings. \ \textit{Proceedings of the Royal
Society A }\textbf{473: }2017.0530.

Baldwin O.R., Jeffery G.B. 1926. \ The Relativity Theory of Plane Waves.
\ \textit{Proceedings of the Royal Society of London A} \textbf{111}: 95-104.

Bambi C. 2019. \ Astrophysical Black Holes: A Review. \ \textit{ArXiv:
1906.03871 [astro-ph].}

Barbour J.B., Pfister H. Eds. 1995. \ \textit{Mach's Principle: From Newton's
Bucket to Quantum Gravity. \ }Boston,\textit{ }Birkh\"{a}user.

Beck G. 1925. \ Zur Theorie Bin\"{a}rer Gravitationsfelder.
\ \textit{Zeitschrift f\"{u}r Physik} \textbf{33}: 713-728.

Bell Burnell S.J. 1977. \ Petit Four. \ \textit{Annals of the New York Academy
of Sciences }\textbf{302}: 685-689.

Bekenstein J.D. 1973. \ Black Holes and Entropy. Physical Review D \textbf{7}:
2333 -- 2346.

Birrell N.D., Davies P.C.W. 1982. \ \textit{Quantum fields in curved space.
\ }Cambridge, Cambridge University Press\textit{.}

Blum A., Giulini D., Lalli R., Renn J. 2017. \ Editorial introduction to the
special issue "The Renaissance of Einstein's Theory of Gravitation".
\textit{The European Physical Journal H }\textbf{42}: 95-105.

Blum A., Lalli R., Renn J. 2018. \ Gravitational waves and the long relativity
revolution. \ \textit{Nature Astronomy} \textbf{2}: 534-543.

Blum A., Lalli R., Renn J. 2015. \ The Reinvention of General Relativity: A
Historiographical Framework for Assessing One Hundred Years of Curved
Space-Time. \ \textit{Isis} \textbf{106}: 598-620.

Blum A., Rickles D. Eds. 2018. \ \textit{Quantum Gravity in the First Half of
the Twentieth Century}. \ Berlin, Edition Open Sources.

Blyth W.F., Isham C.J. 1975. \ Quantization of a Friedman Universe filled with
a scalar field. \ \textit{Physical Review D} \textbf{11}: 768-\emph{778.}

Bogoliubov N.N., Tolmachev V.V., Shirkov D.V. 1959. \ \textit{A New Method in
the Theory of Superconductivity}. \ New York, Consultants Bureau New York.

Bondi H. 1942. \ On the generation of waves on shallow water by
wind.\ \ \textit{Proceedings of the Royal Society of London} A \textbf{181:} 67-71.

Bondi H. 1947. \ Spherically Symmetric Models in General Relativity.
\ \textit{Monthly Notices of the Royal Astronomical Society }\textbf{107}:
410-425. \ Reprinted in 1999 with an editor's note by A.Krasi\'{n}ski in
\textit{General Relativity and Gravitation} \textbf{31}: 1777-1805.

Bondi H. \ 1948. \ Review of Cosmology. \ \textit{Monthly Notices of the Royal
Astronomical Society }\textbf{108:} 104-120.

Bondi H. 1952a. \ On Spherically Symmetric Accretion. \ \textit{Monthly
Notices of the Royal Astronomical Society }\textbf{112}: 195-204.

Bondi H. 1952b. \ \textit{Cosmology. \ }Cambridge, Cambridge University Press.

Bondi H. 1956a. \ The steady-state Theory of Cosmology and Relativity. \ In
Mercier and Kervaire.  pp152-154.

Bondi H. 1956b. \ The electromagnetic field due to a uniformly accelerated
charge, with special reference to the case of gravitation. \ In Mercier and
Kervaire.  p 98.

Bondi H. 1957a. \ Negative Mass in General Relativity. \ \textit{Reviews of
Modern Physics }\textbf{29}: 423-438.

Bondi H. 1957b. \ Plane Gravitational Waves in General Relativity.
\ \textit{Nature }\textbf{179}: 1072-73.

Bondi H. 1959. \ The teaching of special relativity. \ \textit{Reports of
Progress in Physics }\textbf{22}: 97-120.

Bondi H. 1960. \ Gravitational Waves in General Relativity. \ \textit{Nature
}\textbf{186}: 535.

.Bondi H. 1962. \ On the physical characteristics of gravitational waves. \ In
Lichnerowicz \& Tonnelat (1962) pp 129-134.

Bondi H. 1964a. \ Radiation from an Isolated System. \ In Infeld pp 115-122.

Bondi H. \ 1964b. \ The contraction of gravitating spheres.
\ P\textit{roceedings of the Royal Society of London} A\textbf{ 281}: 39-48.

Bondi H. \ 1964c. \ Massive spheres in general relativity.
\ P\textit{roceedings of the Royal Society of London} A\textbf{ 282}: 303-317.

Bondi H. ed. 1965. \ \textit{Abstracts/Proceedings: International Conference
on Relativistic Theories of Gravitation}, \textit{2 volumes}. London, ARL-0032
King's College London.

Bondi H. 1966. \ ARL 66-0075 Research Program in Relativity Physics.

Bondi H. 1967. \ \textit{Assumption and Myth in Physical Theory.} Cambridge,
Cambridge University Press.

Bondi H. 1978. \ Interview of Hermann Bondi by David DeVorkin on 1978 March
20, Niels Bohr Library \& Archives, American Institute of Physics, College
Park, MD USA, www.aip.org/history-programs/niels-bohr-library/oral-histories/4519.

Bondi H. 1987. \ Gravitating toward wave theory. \ \textit{The Scientist}%
\textbf{ 1: }17.

Bondi H. 1990a. \ \textit{Science, Churchill \& Me. \ }Oxford, Pergamon
Press\textit{.}

Bondi H. 1990b. \ This week's Citation Classic. \ CC/Number 30, July 23, 1990. http://garfield.library.upenn.edu/classics1990/A1990DN22600001.pdf.

Bondi H. 1995. \ Essay review: How clever are we? \ \textit{Studies in the
History and Philosophy of Modern Physics} \textbf{26}: 333-337.

Bondi H., Gold T. 1948. \ The Steady State Theory of the Expanding Universe.
\ \textit{Monthly Notices of the Royal Astronomical Society }\textbf{108:} 252-270.

Bondi H., Gold T. 1955. \ The field of a uniformly accelerated charge, with
special reference to the problem of gravitational acceleration.
\ \textit{Proceedings of the Royal Society of London A} \textbf{229}: 416--424.

Bondi H., Hoyle F. 1944. \ On the mechanism of accretion by stars.
\ \textit{Monthly Notices of the Royal Astronomical Society }\textbf{104}: 273-282.

Bondi H., Kilmister C.W. 1959. \ Review: The impact of Logik Der Forschung.
\ \textit{The British Journal for the Philosophy of Science }\textrm{X}%
\textit{: }55--57.

Bondi H., McCrea W.H. 1960. \ Energy transfer by gravitation in Newtonian
theory. \ \textit{Proceedings of the Cambridge Philosophical Society
}\textbf{56}: 410-413.

Bondi H., Pirani F.A.E., Robinson I. 1959. \ Gravitational waves in general
relativity III. \ Exact Plane Waves. \ \textit{Proceedings of the Royal
Society of London} A\textbf{ 251}: 519-533.

Bondi H., Pirani F.A.E. 1988. \ Energy conversion by gravitational waves.
\textit{Nature }\textbf{332}: 212.

Bondi H., Pirani F.A.E. 1989. Gravitational waves in general relativity
\textrm{XIV}, P\textit{roceedings of the Royal Society of London} A\textbf{
421}: 395-410.

Bondi H., Samuel J. 1997. \ The Lense-Thirring Effect and Mach's Principle.
\ \textit{Physics Letters }A \textbf{228}: 121-126.

Bondi H., Van der Burg M.G.J., Metzner A.W.K.\ 1962. \ Gravitational waves in
general relativity VII. \ Waves from axi-symmetric isolated systems.
P\textit{roceedings of the Royal Society of London} A\textbf{ 269}: 21-52.

Bonolis L. 2017. \ Stellar structure and compact objects before 1940: Towards
relativistic astrophysics. \ \textit{European Physical Journal H }\textbf{42}: 311-393.

Brill D.R., Jang P.S. 1980. \ The Positive Mass Conjecture. \ In Held Vol.1,
pp 173-193.

Brinkmann H.W. 1923. \ On Riemann spaces conformal to Einstein space.
\ \textit{Proceedings of the National Academy of Sciences }\textbf{9: }172-174.

Brinkmann H.W. 1925. \ Einstein spaces which are mapped conformally on each
other. \ \textit{Mathematische Annalen }\textbf{94}: 119-145.

Buchdahl H. 1959. \ General Relativistic Fluid Spheres. \textit{\ Physical
Review} \textbf{116:}1027-1034.

Bunch T.S., Davies P.C.W. 1978. \ Quantum field theory In de Sitter Space:
renormalization by point splitting. \ \textit{Proceedings of the Royal Society
of London A} \textbf{360}: 117-134.

Campbell L., Garnett W. 1882.\textit{ \ The Life of James Clerk Maxwell.
\ }London, MacMillan.

Carter B. \ 1971. \ Axisymmetric black hole has only two degees of freedom.
\textit{Physical Review Letters} \textbf{26}: 331-332.

Carter B. 1973. \ Black Hole Equilibrium States Part \textrm{II} General
Theory of Stationary Black Hole States\ In DeWitt \& De Witt pp. 125-214.
\ Reprinted with an editorial note by M. Abromowicz in \textit{General
Relativity and Gravitation} \textbf{42}: 647-744.

Carmeli M., Fickler S., Witten L. 1970. \ \textit{Relativity}. \ New York,
Plenum Press.

Cervantes-Cota J.L., Galindo-Uribarri S., Smoot G.F. 2016. \ A Brief History
of Gravitational Waves. \ \textit{Universe }\textbf{2:} 22.

Chamseddine A.H., West P.C. 1977. \ Supergravity as a Gauge Theory of
Supersymmetry. \ \textit{Nuclear Physics B }\textbf{129}: 39-44.

Chandrasekhar S. 1979. \ Einstein and general relativity: Historical
perspectives. \ \textit{American Journal of Physics}\textbf{ 47}: 212-217.

Chandrasekhar S. 1983a. \ \textit{The Mathematical Theory of Black Holes.}
\ Oxford, Clarendon Press.

Chandrasekhar S. 1983b. \ On stars, their evolution and their stability.
\ \textit{Nobel lecture 8 December 1983. }Nobel Prize.org. Nobel Media AB
2019. Thu. 18 Apr 2019. https://www.nobelprize.org/prizes/physics/1983/chandrasekhar/lecture/.

Chen C-M., Liu J-L., Nester J.M. 2018. \ Gravitational energy is well defined.
\textit{\ International Journal of Modern Physics D}: \textbf{27}:1847017.

Chen C-M., Nester J.M., Ni W-T. 2017. \ A brief history of gravitational wave
research. \ \textit{Chinese Journal of Physics }\textbf{55}: 142-169.

Christensen S.M. 1976. \ Vacuum expectation value of the stress tensor in an
arbitrary curved background. \ The covariant point-separation method.
\ \textit{Physical Review D} \textbf{14}: 2490-2501.

Christensen S.M. Ed. 1984a. \ \textit{Quantum Theory of Gravity. \ }Bristol,
Adam Hilger Ltd.

Christensen S.M. 1984b. \ The world of the Schwinger-DeWitt Algorithm and the
magical a$_{2}$ coefficient. \ In Christensen 1984a.  pp53-65.

Christensen S.M., Fulling S.A. 1977. \ Trace anomalies and the Hawking effect.
\ \textit{Physical Review }D \textbf{15}: 2088-2104.

Chru\'{s}ciel P.T., Costa J.L., Heusler M. 2012. \ Stationary Black Holes:
Uniqueness and Beyond. \ \textit{Living Reviews in Relativity} \textbf{15}: 7.

Clark R.W. 1973. \ \textit{Einstein: The Life and Times}. \ London, Hodder and Stoughton.

Clifford W.K. 1863 \ The Analogues of Pascal's Theorem. \textit{Quarterly
Journal of Pure and Applied Mathematics}

Clifford W.K. 1873. \ On the Hypotheses which Lie at the Bases of Geometry.
\ \textit{Nature }\textbf{183: }14-17, 36, 37.

Clifford W.K. 1876. \ On the space-theory of matter. \ \textit{Proceedings of
the Cambridge Philosophical Society }\textbf{2:} 157-158.

Clifford W.K. 1885. \ \textit{The Common Sense of the Exact Sciences. \ }New
York,\textit{ }Appleton \& Company.

Collin S. 2006. \ Quasars and Galactic Nuclei, a Half-Century Agitated Story.
\ \textit{AIP Conference Proceedings \textbf{861}}: 587-595\textit{.}

Collins H.M. 2004. \ \textit{Gravity's Shadow. \ }Chicago, University of
Chicago Press.

Combridge J.T. 1965. \ \textit{Bibliography of Relativity and Gravitation 1921
to 1937. \ }London, King's College.

Crampin M., Foster J. 1966.\ \ Supertranslations in flat space-time.
\ \textit{Mathematical Proceedings of the Cambridge Philosophical Society
}\textbf{62: }269-276\textbf{.}

Crampin M., Pirani F.A.E. 1986. \ \textit{Applicable Differential Geometry.
\ }Cambridge, Cambridge University Press.

Cvetic M., Satz A. 2018. \ General relation between Aretakis charge and
Newman-Penrose charge. \textit{Physical Review D} \textbf{98}: 124035.

Damour T. 2015. \ 1974: The discovery of the first binary pulsar.
\ \textit{Classical and Quantum Gravity }\textbf{32}: 124009.

Davies P.C.W. 1974. \ \textit{The Physics of Time Asymmetry. \ }Leighton
Buzzard, Surrey University Press.

Davies P.C.W. 1975. \ Scalar particle production in Schwarzschild and Rindler
metrics. \ \textit{Journal of Physics A \textbf{8}: 609-616.}

Davies P.C.W. 1977. \ The thermodynamic theory of black holes.
\ \ \textit{Proceedings of the Royal Society of London A }\textbf{353:} 499-521.

Davies P.C.W. 1984. \ Particles do not exist. \ In Christensen. \ pp 66-77.

Davies P.C.W., Fulling S.A., Unruh W.G. 1976. \ Energy-momentum tensor near an
evaporating black hole. \ \textit{Physical Review D} \textbf{13}: 2720-2723.

Davies P.C.W., Taylor J.G. 1974. \ Do black holes really explode?
\textit{Nature} \textbf{250: }37-38.

Debever R. 1959 Sur la tenseur de super-energie. \ \textit{Comptes rendus
}\textbf{249:} 1324-1326.

Deser S. 2018. \ A brief history (and geography) of supergravity: the first
three weeks and after. \ \textit{The European Physical Journal H} \textbf{43}: 281-291.

Deser S., Duff M.J., Isham C.J. 1976a. \ Non-local conformal anomalies.
\ \textit{Nuclear Physics B} \textbf{111}: 45-55.

Deser S., Ford K.W. (Eds.) 1965a. \ \textit{Brandeis Summer Institute in
Theoretical Physics 1964, volume 1, lectures on general relativity.}
\ Englewood Cliffs, Prentice-Hall Inc.

Deser S., Ford K.W. (Eds.) 1965b. \ \textit{Brandeis Summer Institute in
Theoretical Physics 1964, volume 2, lectures on particles and field theory.}
\ Englewood Cliffs, Prentice-Hall Inc.

Deser S., Pirani F.A.E., Robinson D.C. \ 1976b. \ Imbedding the G-string
pre-print, published as New embedding model of general relativity.
\ \textit{Physical Review D} \textbf{14}: 3301-3303.

Deser S., Zumino B. \ 1976. \ Consistent Supergravity. \ \textit{Physics
Letters} B \textbf{62}: 335-337.

DeWitt B. 1957. \ Introductory Note. \ \textit{Reviews of Modern Physics}
\textbf{29: }351.

De Witt B. \ 2005. \ God's Rays. \ \textit{Physics Today }\textbf{58}: 32-34.

DeWitt B. 2009. \ Quantum gravity: yesterday and today. \ \textit{General
Relativity \& Gravitation }\textbf{41}: 413-419.

DeWitt C. 1957 \ \textit{Conference on the role of gravitation in physics}.
\ Ohio, Wright Air Development Center Technical Report 57-216. ASTIA Document
No. AD118180.

DeWitt C., DeWitt B. Eds. 1964. \ \textit{Relativity, groups and topology Les
Houches 1963.} \ New York, Gordon and Breach Inc.

DeWitt C., DeWitt B. Eds. 1973. \ Les Astres Occlus, \textit{Black Holes, Les
Houches 1972}. \ New York, Gordon \& Breach.

DeWitt C., Rickles D. Eds. 2011. \ \textit{The Role of Gravitation in Physics.
\ Report from the 1957 Chapel Hill Conference. \ }Berlin, Edition Open Sources..

d'Inverno R.A. 1970. \ \textit{The application of algebraic manipulation by
computer to some problems in general relativity.} \ PhD Thesis, University of
London, King's College.

d'Inverno R.A. 1980. \ A Review of Algebraic Computing in General Relativity.
\ In Held (1980). Vol.1 pp491-537.

d'Inverno R.A. 1992. \ \textit{Introducing Einstein's Relativity. \ }Oxford,
Clarendon Press.

Dirac P.A.M. 1950. \ \textit{Generalized Hamiltonian dynamics}. \ Canadian
Journal of Mathematics \textbf{2, }129-148.

Domb C. (Ed.) 1963. \ \textit{Clerk Maxwell and Modern Science. }London, The
Athlone Press.

Domb C. 1980. \ James Clerk Maxwell in London 1860-1865. \textit{Notes and
Records of the Royal Society of London }\textbf{35}: 67-103.

Duff M.J. 1975. \ Covariant quantization. \ In Isham, Penrose, Sciama. \  pp 78-135.

Duff M.J. 1981. \ Inconsistency of quantum field theory in curved space-time.
\ In Isham C.J., Penrose R., Sciama D.W. 1981. \textit{Quantum Gravity 2: A
Second Oxford Symposium. \ Oxford, }Clarendon Press. pp81-105.

Duff M.J. 1994. \ Twenty Years of the Weyl Anomaly. \ \textit{Classical and
Quantum Gravity }\textbf{11:} 1387-1404.

Dyson F.W., Eddington A.S., Davidson C. 1920. \ A Determination of the
deflection of Light by the Sun's Gravitational Field. \ \textit{Royal Society
of London, Philosophical Transactions A }\textbf{220}: 291-333.

Eddington A.S. 1922. \ The Propogation \ of Gravitational Waves.
\ \textit{Royal Society of London. \ Proceedings A }\textbf{102: }268-282.

Eddington A.S. 1923. \ \textit{The Mathematical Theory of Relativity.}
\ Cambridge, Cambridge University Press.

Eddington A.S. 1924. \ A comparison of Whitehead's and Einstein's formulae.
\ \textit{Nature }\textbf{113}: 192.

Edgar R. 2004. \ A Review of Bondi-Hoyle-Lyttleton Accretion. \ \textit{New
Astronomy Review }\textbf{48}: 843-859.

Ehlers J. ed. 1979. \ \textit{Isolated Gravitating Systems in General
Relativity. \ Amsterdam, }North-Holland Publishing Company.

Ehlers J., Pirani F.A.E., Schild A. 1972. \ The geometry of free fall and
light propogation. \ In O'Raifeartaigh. \ pp 63-84. \ Reprinted in 2012, with
an editorial note by A.Trautman in \textit{General Relativity and Gravitation}
\textbf{44}: 587--1609.

Einstein A. 1915. Die Feldgleichungen der Gravitation. \ \textit{K\~{o}niglich
Preussiche Academie der Wissenshaften Zu Berlin, Situngsberichte: 844--847.}

Einstein A. 1916a. Die Grundlagh der Allgemeinen Relativit\"{a}tstheorie.
\ \textit{Annalen der Physik}\textbf{ 49}: 769-822.

Einstein A. 1916b. \ N\"{a}herungsweise Integration der Feldgleichungen der
Gravitation \textit{K\~{o}niglich Preussiche Academie der Wissenshaften Zu
Berlin, Situngsberichte 688-696.\ }

Einstein A. 1918. \ \"{U}ber Gravitationswellen. \ \textit{K\~{o}niglich
Preussiche Academie der Wissenshaften Zu Berlin, Situngsberichte: 154-167.}

Einstein A. 1921a. \ A Brief Outline of the Development of the Theory of
Relativity. \ \textit{Nature }\textbf{Feb 17}: 782-784.

Einstein A. 1921b. \ King's College Lecture. \ \textit{The Collected Papers of
Albert Einstein, volume 7, 2002.} \ Princeton, Princeton University Press.
\ Engel A. (trans.), Schucking E. (consult.): Document 58, pp238-240.

Einstein A. \textit{1922. \ Sidelights on relativity\ }(Jeffery G.B., Perrett
W., Transl.)\textit{. \ }London, Methuen and Company Ltd.

Einstein A. 1949. \ Notes for An Autobiograpy. \ \textit{Saturday Review Nov.
26,1949, }pp9-12. \ https://archive.org/details/EinsteinAutobiography

Einstein A. 1954. \textit{\ Albert Einstein (Princeton) to Felix Pirani
(Cambridge, Engl.) 2 February 1954}. \ Albert Einstein Archives, The Hebrew
University of Jerusalem, Call no. 17-447 [3 typed sheets].

Einstein A. 1987ff. \ \textit{The Collected Papers of Albert Einstein}%
,\textbf{ }Princeton (CPAE), Princeton University Press. \ Originals and
English translation available online at http://einstein.papers.press.princeton.edu.

Einstein A., Infeld L., Hoffmann B. 1938. \ The Gravitational Equations and
the Problem of Motion. \ \textit{Annals of Mathematics }\textbf{39: }65-100.

Einstein A., Rosen N. 1937. \ On gravitational waves. \ \textit{Journal of the
Franklin Institute }\textbf{223}: 43-54.

\ Eisenstaedt J. 1986. \ La relativit\'{e} g\'{e}n\'{e}rale \`{a}
l'\'{e}tiage: 1925-1955. \ \textit{Archive for History of Exact Sciences
}\textbf{35}: 115-185.

\ Eisenstaedt J 1987. \ Trajectoires et Impasses de la Solution de
Schwarzschild. \ \textit{Archive for History of Exact Sciences }\textbf{37}: 275-357.

\ Eisenstaedt J. 1989a. \ The Low Water Mark of General Relativity, 1925-1955.
\ In Howard and Stachel.  pp 277-292.

Eisenstaedt J. 2006. \ \textit{The Curious History of Relativity: how
Einstein's theory of Gravity was Lost and Found Again. \ }Princeton, Princeton
University Press.

Eisenstaedt J., Kox A.J. Eds. 1992 \ \textit{Studies in the History of General
Relativity. \ }Boston, Birkh\"{a}user.

Ellis G.F.R. 2014. \ Stephen Hawking's 1966 Adams Prize Essay. \ \textit{The
European Physical Journal H} \textbf{39}:403-411.

Ellis G.F.R., Penrose R.\ 2010. \ \ Dennis William Sciama: 18 November 1926-19
December 1999. \ \textit{Biographical Memoirs of Fellows of the Royal Society
of London }\textbf{56}: 401-422.

Exton A.R., Newman E.T., Penrose R. 1969. \ Conserved quantities in the
Einstein-Maxwell theory. \ \textit{Journal of Mathematical Physics
}\textbf{10}, 1566-1570.

Finkelstein D. 1959. \ Past-Future Asymmetry of the Gravitational Field of a
Point-Particle. \ \textit{Physical Review} \textbf{110}: 965-967.

Ford L.H. 1978. \ Quantum Coherence Effects and the Second Law of
Thermodynamics. \ \textit{Proceedings of the Royal Society of London A
}\textbf{364}: 227-236.

Ford L.H. 1997. \ Quantum Field Theory in Curved Spacetime. \ \textit{ArXiv}:
gr-qc/9707062 v1.

Foster A.W., Pirani F.A.E. 1948. \ Use of the Hartman Formula.
\ \textit{American Journal of Physics }\textbf{16}: 56.

Frauendiener J. 2004. \ Conformal Infinity. \ \textit{Living Reviews in
Relativity} \textbf{7}: 1. https://doi.org/10.12942/lrr-2004-1

Freedman D.Z., van Nieuwenhuizen P., Ferrara S. 1976. \ Progress toward a
theory of supergravity. \ \textit{Physical Review D }\textbf{13}: 3214-3218.

Fulling S.A. 1973. \ Non-uniqueness of canonical field quantization in
Riemannian space-time. \ \textit{Physical Review }D \textbf{7}: 2850-2862.

Fulling S.A. 1984. \ What have we learned from quantum field theory in curved
space-time? \ In Christensen 1984, pp 42-51.

Fulling S.A. 1989. \ \textit{Aspects of Quantum Field Theory in Curved
Space-Time}. \ Cambridge, Cambridge University Press.

Fulling S.A., Davies P.C.W. 1976. \ Radiation from a moving mirror in two
dimensional space-time: conformal anomaly. \ \textit{Proceedings of the Royal
Society of London A }\textbf{348}: 393-414.

Fulling S.A., Matsas G.E.A. 2014. \ Unruh effect. \ \textit{Scholarpedia}
\textbf{9} (10): 31789.

Gale G. 2015. \ Cosmology: Methodological Debates in the 1930's and 1940's.
\ \textit{Stanford Encyclopedia of Philosophy.}

https://plato.stanford.edu/entries/cosmology-30s/.

Galindo S., Cervantes-Cota J.L. 2018. \ Clifford's Attempt to test his
Gravitational Hypothesis. \ \textit{Revista Mexicana de Fisica E }\textbf{64:
}162-168.

Geroch R. \ 1973. \ Energy extraction. \ \textit{Annals of the New York
Academy of Sciences }\textbf{224}: 108-117.

Gibbons G., Will C.M. 2008. On the Multiple Deaths of Whitehead's Theory of
Gravity. \ \textit{Studies in the History and Philosophy of Modern Physics}
\textbf{39}:41-61.

Goenner H., Renn J., Ritter J., Sauer T. 1999. \ \textit{The Expanding Worlds
of General Relativity}. \ Boston, Birkhauser Verlag.

Gold T. Ed. 1967. \ \textit{The Nature of Time. \ }Ithaca, Cornell University Press.

Gold T. 1968 Rotating Neutron Stars as the Origin of the Pulsating Radio
Sources. \textit{Nature}.\textbf{ 218} : 731-732.

Goldberg J.N. 1955. \ Gravitational Radiation. \ \textit{Physical Review
}\textbf{99}: 1873-1883.

Goldberg J.N. 1992. \ U.S. Air Force Support of General Relativity: 1956-1972.
\ In Eisenstaedt \& Kox. \ pp89-102.

Goldberg J.N., Sachs R.K. 1962. \ A theorem on Petrov types. \ \textit{Acta
Physica Polonica Supplement }\textbf{22: }13-23. Reprinted in 2009
\textit{General Relativity and Gravitation }\textbf{41}: 433-444 with an
editorial note by A. Krasi\'{n}ski and M. Przanowski pp421-432.

Gray J. 2006. \ Overstating their case? \ Reflections on British mathematics
in the nineteenth century. \ \textit{BSHM Bulletin }\textbf{21: }178-185.

Greaves W.M.H. 1940. Obituary. \ \textit{Monthly Notices of the Royal
Astronomical Society }\textbf{100}: 258-263.

Havas P. 1979. Equations of motion and radiation reaction in the special and
general theory of relativity. \ In Ehlers. \ pp 74-155.

Hawking S.W. 1972. \ Black holes in general relativity. \textit{Communications
in Mathematical Physics} \textbf{25}, 152-166.

Hawking S.W. 1974. \ Black hole explosions. \ \textit{Nature} \textbf{248}: 30-31.

Hawking S.W. 1975a. \ Particle creation by black holes. \ In Isham, Penrose,
Sciama.  pp 219-267.

Hawking S.W. 1975b. \ Particle Creation by Black Holes.
\ \textit{Communications in Mathematical Physics }\textbf{43:} 199-220.

Hawking S.W. 2014. \ Singularities and the geometry of space-time.
\ \textit{The European Physical Journal H} \textbf{39}:413-503.

Hawking S.W., Ellis G.F.R. 1973. \ \textit{The large scale structure of
space-time}. \ Cambridge, Cambridge University Press.

Hawking S.W., Israel W. Eds. 1987. \ \textit{300 years of gravitation.
\ }Cambridge, Cambridge University Press.

Hearnshaw F.J.C. 1929. \textit{\ The centenary history of King's College
London 1828-1928. \ }London, George G. Harrap \& Company Ltd.

Held A. (Ed.) 1980. \ \textit{General Relativity and Gravitation Vols. 1 \& 2.
\ }New York, Plenum Press.

Heusler M. 1996. \ \textit{Black Hole Uniqueness Theorems. \ }Cambridge,
Cambridge University Press.

Hill C.D., Nurowski P. 2017. \ How the Green Light Was Given for Gravitational
Wave Search. \ \textit{Notices of the American Mathematical Society
}\textbf{64}, 686-692.

Hodges A.P. 1983a. \ Twistor Diagrams And Massless M\o ller Scattering.
\ \textit{Proceedings of the Royal Society of London A} \textbf{385}: 207-228.

Hodges A.P. 1983b. \ Twistor Diagrams And Massless Compton Scattering.
\ \textit{Proceedings of the Royal Society of London A} \textbf{386}: 185-210.

Hodges A.P. 1983c. \ \ \textit{Alan Turing: the enigma}. \ London, Burnett
Books Ltd.

Hodges A.P. 1985. \ Mass Eigenstates in Twistor Theory. \ \textit{Proceedings
of the Royal Society of London A} \textbf{397}: 375-396.

Hodges A.P., Huggett S. 1980. \ Twistor Diagrams. \ \textit{Surveys of High
Energy Physics }\textbf{1}: 333-353.

Hoffmann B. Ed. 1966. \ \textit{Perspectives in Geometry and Relativity.
\ }Bloomington, Indiana University Press.

Howard D., Stachel J. Eds. 1989. \ \textit{Einstein and the History of General
Relativity. \ }Boston, Birkh\"{a}user.

Hoyle F. 1948.\ A New Model for the Expanding Universe. Monthly Notices of the
Royal Astronomical Society \textbf{108}: 372-382.

Hoyle F., Lyttleton R.A. 1939. The effect of interstellar matter on climate
variation. \ \textit{Proceedings of the Cambridge Philosophical Society
}\textbf{35}: 405-415.

Huelin G. 1978. \ \textit{King's College London 1828-1978. }London, University
of London King's College.

Huggett S.A., Tod K.P. 1985. \ \textit{An Introduction to Twistor Theory.
\ }Cambridge, Cambridge University Press.

Infeld L., Scheidegger A.E. 1951. \ Radiation and Gravitational Equations of
Motion. \ \textit{Canadian Journal of Mathematics }\textbf{3}: 195-207.

Infeld L. Ed. 1964. \ \textit{Relativistic Theories of Gravitation}. Oxford,
Pergamon Press.

Isham C.J. 1975a. \ An introduction to quantum gravity. \ In Isham, Penrose,
Sciama pp 1-77.

Isham C.J. 1976. \ Some quantum field theory aspects of the superspace
quantization of general relativity. \ P\textit{roceedings of the Royal Society
of London} A\textbf{ 351}: 209-232.

Isham C.J., Nelson J.E. 1974. \ Quantization of a Coupled Fermi Field and
Robertson-Walker metric. \ \textit{Physical Review D}\textbf{ 10}: 3226-3234.

Isham C.J., Penrose R., Sciama D.W. eds. 1975. \ \textit{Quantum Gravity.
\ }Oxford, Clarendon Press.

Israel W. \ 1967. \ Event Horizons in Static Vacuum Space-Times.
\ \textit{Physical Review }\textbf{164: }1776-1779.

Israel W. 1968. \ Event Horizons in Static Electrovac Space-Times.
\ \textit{Communications in Mathematical Physics} \textbf{8}: 245- 260.

Israel W. 1987 \ Dark stars: the evolution of an idea. \ In Hawking \&
Israel.  pp 199-277.

Israel W. 1996. \ Imploding stars, shifting continents and the inconsistency
of matter. \ \textit{Foundations of Physics }\textbf{26}: 595-616.

Jeffery G.B. 1924. \ \textit{Relativity for physics students. \ }London,
Methuen \& Co. Ltd..

Jordan P., Ehlers J., Sachs R.K. 1961. \ English translation in 2013,
\ Contribution to the theory of pure gravitational radiation. \ Exact
solutions of the field equations of the general theory of relativity
\textrm{II}, in \textit{General Relativity and Gravitation }2013, \textbf{45}: 2691-2753.

Kaiser D. 2012. \ Booms, Busts and the World of Ideas: Enrollment Pressures
and the Challenge of Specialization. \ \textit{Osiris }\textbf{27}: 276-302.

Kaloper N., Kleban M., Martin D. 2010. \ M$^{\text{c}}$Vittie's Legacy: Black
Holes in an Expanding Universe. \ \textit{arxiv 1003.4777v.3 [hep-th].}

Kane G.L., Shifman M. 2000. \ \textit{The Supersymmetric World}:\textit{ The
Beginning of The Theory. \ }Singapore, World Scientific.

Kennefick D. 1999. \ Controversies in the History of the Radiation Reaction
problem in General Relativity. In Goenner, Renn, Ritter and Sauer.  pp207-234.

Kennefick D. 2005. \ Einstein Versus the Physical Review. \ \textit{Physics
Today }\textbf{58}: 9, 43-48.

Kennefick D. 2007. \ \textit{Traveling at the Speed of Thought: Einstein and
the Quest for Gravitational Waves. \ }Princeton\textit{, }Princeton University Press.

Kennefick D. 2014. \ Relativistic Lighthouses: The Role of the Binary Pulsar
in proving the existence of Gravitational Waves. \ \textit{arXiv: 1407.2164
[physics.hist-ph].}

Kennefick D. 2017. \ The binary pulsar and the quadrupole formula controversy.
\ \textit{The European Physical Journal H }\textbf{42}: 293-310.

Kerr R.P. 1963. \ Gravitational field of a spinning mass as an example of
algebraically special metrics. \ \textit{Physical Review Letters }\textbf{11}: 522-523.

Kerr R.P. 1965. \ Gravitational Collapse and Rotation. \ In
\textit{Quasistellar Sources and Gravitational Collapse. }Eds. Robinson I.,
Schild A., Schucking E. pp 99-102\textit{. \ }Chicago, University of Chicago Press.

Kerr R.P. 2009 The Kerr and Kerr-Schild metrics. \ In Wiltshire, Visser,
Scott.  pp38-72.\textit{ }

Khalatnikov I.M., Kamenshchik A.Yu. \ 2008 Lev Landau and the problem of
singularities in cosmology. \ \textit{arXiv: 0803.2684v1 [gr-qc].}

Khan K.A., Penrose R. 1971. \ Scattering of two impulsive gravitational plane
waves. \ \textit{Nature} \textbf{229}: 185-186.

Kilmister C.W 1949. \ The use of quaternions in wave-tensor calculus.
\ \textit{\ Proceedings of the Royal Society of London} A\textbf{ 199}: 517-532.

Kilmister C.W. 1951. \ Tensor identities in Wave-Tensor Calculus.
\ P\textit{roceedings of the Royal Society of London} A\textbf{ 207}: 402-415.

Kilmister C.W. 1966. \ Alternative Field Equations in General Relativity. \ In
Hoffmann.  pp 201-216.

Kilmister C.W.K. 1973. \ \textit{General Theory of Relativity. \ }Oxford,
Pergamon Press Ltd.

Kilmister C.W. 1988. \ Obituary J.T.Combridge. \ \textit{Bulletin of the
London Mathematical Society }\textbf{20}:156-158.

Kilmister C.W. 1994. \ George Frederick James Temple 1901-1992.
\ \textit{Biographical Memoirs of Fellows of the Royal Society of London
}\textbf{40}: 385-400.

Kilmister C.W. 1995. \ Obituary George Frederick James Temple.
\ \textit{Bulletin of the London Mathematical Society }\textbf{27}: 281-287.

Kilmister C.W., Newman D.J. 1961. \ The use of algebraic structures in
physics. \ \textit{Proceedings of the Cambridge Philosophical Society}
\textbf{57}: 851-864.

Klein O. 1928. Zur f\"{u}nfdimensionalen Darstellung der
Relativit\"{a}tstheorie. \textit{Zeitschrift f\"{u}r Physik }\textbf{46}:188-208.

Kobayashi S., Nomizu K. 1963. \ \textit{Foundations of Differential Geometry
Volume 1. \ }New York, John Wiley \& Sons Inc.

Kox A.J., Eisenstaedt J. Eds. 2005 \ \textit{The Universe of General
Relativity. \ }Boston, Birkh\"{a}user.

Kragh H. 1996. \ \textit{Cosmology and Controversy.} Princeton, Princeton
University Press.

Kragh H. 2012 Geometry and Astronomy: Pre-Einstein Speculations of
Non-Euclidean Space. \ \textit{arXiv 1205.4909} \textit{[physics.hist-ph].}

Krawczynski H. 2018. \ \ Difficulties of Quantitative Tests of the
Kerr-Hypothesis with X-Ray Observations of Mass Accreting Black Holes.
\textit{\ General Relativity and Gravitation }\textbf{50}: 100.

Kruskal M.D. 1960. \ Maximal extension of Schwarzschild metric.
\ \textit{Physical Review }\textbf{119}: 1743-1745.

K\"{u}nzle H.P. 1968. \ Maxwell fields satisfying Huygen's principle.
\ \textit{Proceedings of the Cambridge Philosophical Society }\textbf{64}: 779-785.

K\"{u}nzle H.P. 1971. \ \ On the spherical symmetry of a static perfect fluid.
\ \textit{Communications in Mathematical Physics }\textbf{20}: 85-100.

Lake K., Abdelqader M. 2011. \ More on M$^{\text{c}}$Vittie's Legacy: A
Schwarzschild-de Sitter black and white hole embedded in an asymptotically
flat $\Lambda CDM$ cosmology. \ \textit{Physical Review D}: \textbf{84}:
044045\textit{.}

Lalli R. 2017. \ \textit{Building the General Relativity and Gravitation
Community During the Cold War. \ }Cham Switzerland, University of Rochester,
Springer International Publishing.

Lehner C., Renn J., Schemmel M. 2012. \ \textit{Einstein and the Changing
Worldviews of Physics.} \ Boston, Birkh\"{a}user.

Levi H. 1968. \ Gravitational Induction. \ \textit{Proceedings of the
Cambridge Philosophical Society }\textbf{64}: 1081-1087.

Lichnerowicz A. 1955. \ \textit{Th\'{e}ories R\'{e}lativistes de la
Gravitation et de l'Electromagn\'{e}tisme. }\ Paris, Masson et Cie.

Lichnerowicz M.A., Tonnelat M.A. Eds. 1962. \textit{Les Th\'{e}ories
Relativistes de la Gravitation Royaumont (21-27 Juin 1959}). \ Paris, Centre
National de la Recherche Scientifique.

Lichnerowicz A. 1992. \ Mathematics and General Relativity: A Recollection.
\ In Eisenstaedt, Kox (1992) pp103-108. \ 

Lindblom L. 1992. \ On the symmetries of equilibrium stellar models.\textit{
\ Philosophical Transactions of the Royal Society of London A }\textbf{340: }353-364.

Longair M. 2006. \ \textit{The Cosmic Century}. \ Cambridge, Cambridge
University Press.

\textit{Measuring and Modeling the Universe}, ed. W. L. Freedman, pp 1-18.
\ Cambridge, Cambridge University Press.

MacCallum M.A.H. 1989. \ George Cunliffe M$^{\text{c}}$Vittie (1904-1988)
obituary. \ \textit{Quarterly Journal of the Royal Astronomical Societ}y
\textbf{30: }119-122.

MacCallum M.A.H. 2013. \ Exact solutions of Einstein's equations.
\ \textit{Scholarpedia}, 8(12):8584.

MacCallum M.A.H., Skea J.E.F., M$^{\text{c}}$Crea J.D., M$^{\text{C}}$Lenaghan
R.G. 1994. \ \textit{Algebraic Computing in General Relativity}. \ Oxford,
Clarendon Press.

MacCallum M.A.H., Skea J.E.F. 1994. \ SHEEP:\ A computer algebra system for
general relativity. \ In MacCallum M.A.H., Skea J.E.F., M$^{\text{c}}$Crea
J.D., M$^{\text{C}}$Lenaghan R.G. (1994) pp 1-172.

McCarthy P.J. 1971. \ \textit{Properties and Representations of the
Bondi-Metzner-Sachs group. }\ PhD thesis, University of London, King's College London.

McCrea W. 1955. \ Jubilee of relativity theory, conference at Berne.
\ \textit{Nature }\textbf{176}: 330-331.

McLenaghan R.G. 1969. \ An explicit determination of the empty space-times on
which the wave equation satisfies Huygens' principle. Proceedings of the
Cambridge Philosophical Society \textbf{65: }139-155.

McVittie G.C. 1929a. \ On Einstein's unified field theory.
\ \textit{\ Proceedings of the Royal Society of London} A\textbf{ 124}: 366-374.

McVittie G.C. 1929b. \ On Levi-Civita's Modification of Einstein's Unified
Field Theory. \ \textit{Philosophical Magazine }\textbf{8}: 1033-1044.

McVittie G.C. 1933. \ The mass particle in an expanding universe.
\ \textit{Monthly Notices of the Royal Astronomical Society }\textbf{93}: 325-339.

McVittie G.C. 1937. \ \textit{Cosmological Theory}. \ London, Methuen \& Co. Ltd.

McVittie G.C. 1939. \ Observations and Theory in Cosmology.
\ \textit{Proceedings of the Physical Society (London) }\textbf{51}: 529-537.

McVittie G.C. 1946. \ The Regraduation of Clocks in Spherically Symmetric
Space-times of General Relativity. \ \textit{\ Proceedings of the Royal
Society of Edinburgh Ser. }A\textbf{ 62}: 147-155.

McVittie G.C. 1955. \ Gravitational Waves and One-dimensional Einsteinian
Gas-Dynamics. \textit{Journal of Rational Mechanics and Analysis} \textbf{4}, 201-220.

McVittie G.C. 1956. \ \ \textit{General Relativity and Cosmology. \ }New York,
John Wiley \& Sons.

McVittie G.C. 1978. \ Interview of George McVittie by David DeVorkin on 1978
March 21, Niels Bohr Library \& Archives, American Institute of Physics,
College Park, MD, USA, www.aip.org/history-programs/niels-bohr-library/oral-histories/4774.

M\"{a}dler T., Winicour J. 2016. \ Bondi-Sachs formalism.
\ \textit{Scholarpedia \textbf{11}} (12): 33528\textit{.}

Mahon B. 2003. \ \textit{The Man who changed everything: The Life of James
Clerk Maxwell}. \ Chichester, John Wiley \& Sons Ltd.

Martin J.L. 1959a. \ Classical Dynamics, and the 'Classical Analogue' of a
Fermi Oscillator. \ \textit{Proceedings of the Royal Society of London A
}\textbf{251}: 536-542.

Martin J.L. 1959b. \ The Feynman Principle for a Fermi System.
\ \textit{Proceedings of the Royal Society of London A }\textbf{251}: 543-549.

Masood-ul-Alam A.K.M. 2007. \ Proof that static stellar models are spherical.
\ \textit{General Relativity and Gravitation }\textbf{39}: \ 55-85.

Mavrid\`{e}s S. 1973. \textit{\ L'Univers relativiste}. \ Paris, Editions
Albin Michel.

Maxwell J.C. 1856. \ On Faraday's Lines of Force. \ \textit{Cambridge
Philosophical Transactions }\textbf{10: }27-83.

Maxwell J.C. 1861. On physical lines of force. \textit{Philosophical Magazine}
\textbf{90}: In four parts, part \textrm{I:}161-223,part \textrm{II }281-291
and 338-348, part \textrm{III }12-24, part \textrm{IV }85-95.

Maxwell J.C. 1865. A dynamical theory of the electromagnetic field.
\ \textit{Philosophical Transactions of the Royal Society of London}
\textbf{155}: 459--512.

Mercier A., Kervaire M. (Eds.) 1956. \ \textit{F\"{u}nfzig Jahre
Relativit\"{a}tstheorie/Cinquantenaire de la Th\'{e}orie de la Relativit\'{e}/
Jubilee of Relativity Theory, Helvetica Physica Acta, Supplementum IV}, pp.
286, Birkhauser Verlag, Basel (1956). \ https://www.e-periodica.ch.

Mody C.C.M. 2016. \ Santa Barbara Physicists in the Vietnam Era. \ In
\textit{Groovy Sciences: Knowledge, Innovation and American Counterculture.
\ }Kaiser D., McCray W.P. Eds. \ Chicago, The University of Chicago Press. pp 70-108.

Moseley H. 1839. \ \textit{Lectures on Astronomy delivered at King's College,
London. \ }London, John W, Parker.

M\"{u}ller zum Hagen H. \ 1970a. On the analyticity of static black hole
vacuum solutions of Einstein;s equations. \ \textit{Proceedings of the
Cambridge Philosophical Society} \textbf{67}: 415-421.

M\"{u}ller zum Hagen H. \ 1970b. On the analyticity of stationary black hole
vacuum solutions of Einstein;s equations. \ \textit{Proceedings of the
Cambridge Philosophical Society} \textbf{68}: 199 -201.

M\"{u}ller zum Hagen H., Robinson D.C., Seifert H.J. \ 1973. \ Black holes in
static vacuum space-times, \textit{General Relativity and Gravitation}
\textbf{4}: 53-78.

M\"{u}ller zum Hagen H., Robinson D.C., Seifert H.J. \ 1974. \ Black holes in
static electrovac space-times, \textit{General Relativity and Gravitation}
\textbf{5}: 61-72.

Myers S.B. 1941. \ Riemannian manifolds with positive mean curvature.
\textit{Duke Mathematical Journal} \textbf{8}: 401-404.

Newman E.T. 2005. \ A Biased and Personal Description of GR at Syracuse
University 1951-1961 in Kox \& Eisenstaedt (2005) pp 373-383.

Newman E.T., Adamo T. 2014. \ Kerr-Newman metric. \ \textit{Scholarpedia}
\textbf{9}: 31791.

Newman E.T., Couch E., Chinnapared K., Exton A., Prakash A. Torrence R. 1965.
\ \textit{Journal of Mathematical Physics }\textbf{6}: 918-919.

Newman E.T., Penrose R. 1962. \ An approach to Gravitational Radiation by a
Method of Spin Coefficients. \ \textit{Journal of Mathematical Physics
}\textbf{3}: 566-578.

Newman E.T., Penrose R. 1965. \ 10 exact gravitationally-conserved quantities.
\ \textit{Physical Review Letters }\textbf{15}: 231-233.

Newman E.T., Penrose R. 1966. \ Note on the Bondi-Metzner-Sachs Group.
\ \textit{Journal of Mathematical Physics }\textbf{7}: 863-870.

Newman E.T., Penrose R. 1968. \ New conservation laws for zero rest-mass
fields in asymptotically flat space-time. \ \textit{Proceedings of the Royal
Society of London A }\textbf{305}: 175-204.

Newman E.T., Penrose R. 2009. \ Spin-coefficient formalism.
\textit{Scholarpedia}, \textbf{4}(6):744.

Newman E.T., Tod K.P. 1980. \ Asymptotically Flat Space-Times. \ In Held Vol.2
(1980) pp 1-36.

Nolan B.C. 2017. \ Local properties and global structure of M$^{\text{c}}%
$Vittie space-times with non-flat FLRW backgrounds. \ \textit{Classical and
Quantum Gravity }\textbf{34}: 225002.

Nordstr\"{o}m, G. 1918. On the Energy of the Gravitational Field in Einstein's
Theory. \textit{Verhandl. Koninkl. Ned. Akad. Wetenschap., Afdel. Natuurk.,
Amsterdam }\textbf{26}: 1201--1208.

Niven W.D. Ed. 1890. \ \textit{The Scientific Papers of James Clerk Maxwell.
\ }Cambridge, Cambridge University Press.

Oppenheimer J.R., Snyder H. 1939. \ On Continued Gravitational Contraction.
\ \textit{Physical Review }\textbf{56}: 455-459.

Oppenheimer J.R., Volkoff G.M. 1939. \ On Massive Neutron Cores.
\ \textit{Physical Review }\textbf{55}: 374-381.

O'Raifeartaigh L. Ed. 1972. \textit{General Relativity Papers in Honour of
J.L.Synge}. \ Oxford, Clarendon Press.

Page D. 2005. \ Hawking radiation and black hole thermodynamics. \ \textit{New
Journal of Physics }\textbf{7}:203-235. \ \textit{arXiv: hep-th/0409024 v3.}

Parker L., Navarro-Salas J. 2017. \ Fifty years of cosmological particle
creation. \ \textit{arXiv: 1702.07132v1 [physics}. \textit{hist-ph]}.

Pearson K. 1892. \ \textit{The Grammar of Science. \ }London, Walter Scott.

Peebles P.J.E. 2017. \ Robert Dicke and the naissance of experimental gravity
physics 1957-1967. \ \textit{The European Physical Journal H }\textbf{42}, 177-259.

Penrose R. 1960. \ A spinor approach to general relativity. \ \textit{Annals
of Physics.} \textbf{10}:171-209.

Penrose R. 1962. \ Calculating GR in spinor form. \ In Lichnerowicz \&
Tonnelat pp428-431.

Penrose R. \ 1963. \ Asymptotic properties of fields and space-times.
\ \textit{Physical Review Letters \ }\textbf{10}: 66-68.

Penrose R. 1964a. \ The light cone at infinity. \ In Infeld\ pp 369-373.

Penrose R. 1964b. \ Conformal treatment of Infinity. \ In DeWitt C. \& DeWitt
B. (1964) \ pp565-584.

Penrose R. 1965a. \ Zero rest-mass fields including gravitation: asymptotic
behaviour. \ \textit{Proceedings of the Royal Society of London A
}\textbf{284}: 159-203.

Penrose R. 1965b. \ Gravitational collapse and space-time singularities.
\ \textit{Physical Review Letters \ }\textbf{14}: 57-59.

Penrose R. \ 1967. \ Cosmological boundary conditions for zero rest-mass
fields. \ In Gold (1967) pp42-54.

Penrose R. 1969. \ Gravitational Collapse: the Role of General Relativity.
\ \textit{Revista del Nuovo Cimento }\textbf{1}: 252-276. \ Reprinted with
editor's note by A. Kr\'{o}lak in \textit{General Relativity and Gravitation}
\textbf{34}: 1135-1163 (2002).

Penrose R. 1972. \textit{\ Techniques of differential topology in relativity.
\ }Philadelphia. \ Society for Industrial and Applied Mathematics.

Penrose R., Rindler W. 1984. \ \textit{Spinors and space-time, volume 1}.
\ Cambridge, Cambridge University Press.

Penrose R., Rindler W. 1986. \ \textit{Spinors and space-time, volume 2}.
\ Cambridge, Cambridge University Press.

Penrose R., Robinson I., Tafel J. 1997. \ Andrzej Mariusz Trautman.
\ \textit{Classical and Quantum Gravity }\textbf{14: \ }A1-A8.

Perrett W., Jeffery G.B. \textit{1923. \ The Principle of Relativity.
\ }London, Methuen and Company Ltd.

Petrov A.Z. 1954, Classification of spaces defining gravitational
fields\textit{. Jubilee collection, Uchenye Zapiski Kazanskogo Universiteta,
Kazan State University, Kazan}, \textbf{114, no. 8}: \ 55--69. \ English
translation with editor's note by M.A.MacCallum and short biography by P.A.
Gusev 2000. \ \textit{General Relativity and Gravitation}\textbf{ 32}: 1661-1685.

Petrov A.Z. 1969. \ \textit{Einstein spaces. } Oxford, Pergamon Press Ltd.

Penzias A.A., Wilson R.W. 1965. \ A Measurement Of Excess Antenna Temperature
At 4080 Mc/s. \ \textit{Astrophysical Journal Letters} \textbf{142}: 419--421.

Pirani F.A.E. 1951. \ \textit{On the quantization of the gravitational field
of general relativity.} \ D.Sc. thesis, Carnegie Institute of Technology.

Pirani F.A.E. 1955a. \ On the Energy-Momentum Tensor and the Creation of
Matter in Relativistic Cosmology. \ \textit{Proceedings of the Royal Society
of London} A\textbf{ 228}: 455-462.

Pirani F.A.E. 1955b. \ Review of M$^{\text{c}}$Vittie G.C. 1955.
\ \textit{Mathematical Reviews} \textbf{16}, 1165.

Pirani F.A.E. 1956a. \ On the Definition of Inertial Systems in General
Relativity. \ In Mercier and Kervaire (1956) pp198-203.

Pirani F.A.E. 1956b. \ On the physical significance of the Riemann tensor.
\ \textit{Acta Physica Polonica }\textbf{15}, 389-405. \ Reprinted in
\textit{General Relativity and Gravitation }2009 \textbf{41}: 1215-1232 with
an accompanying paper by J.L.Synge and an editorial note by A. Trautman pp1195-1203.

Pirani F.A.E. 1957a. \ \textit{The relativistic basis of mechanics. \ }PhD
thesis, Cambridge University.

Pirani F.A.E. 1957b. \ Invariant formulation of gravitational radiation
theory. \ \textit{Physical Review }\textbf{105}: 1089-1099.

Pirani F.A.E. \ 1959. \ Gravitational waves in general relativity \textrm{IV.}
\ The gravitational field of a fast moving particle. \ \textit{Proceedings of
the Royal Society of London }A \textbf{252}: 96-101.

Pirani F.A.E. 1962a. \ Gauss's theorem and gravitational energy. \ In
Lichnerowicz \& Tonnelat pp 85-91.

Pirani F.A.E. 1962b. Survey of gravitational radiation theory. \ In
\textit{Recent developments in general relativity. } New York, Pergamon Press. \ pp89-105.

Pirani F.A.E. 1962c. \ Gravitational Radiation. \ In Witten. pp 199-226.

Pirani F.A.E. 2011. \ Interview of Felix Pirani by Dean Rickles on 2011 June
23, Niels Bohr Library \& Archives, American Institute of Physics, College
Park, MD USA, www.aip.org/history-programs/niels-bohr-library/oral-histories/34463.

Pirani F.A.E., Roche C. 1993. \ \textit{The Universe For Beginners.
\ }Cambridge, Icon Books.

Pirani F.A.E., Schild A. 1950. \ On the Quantization of Einstein's
Gravitational Field Equations. \ \textit{Physical Review }\textbf{79}: 986-991.

Pirani F.A.E., Schild A., Skinner R. 1952. \ Quantization of Einstein's
Gravitational Field Equations II. \ \textit{Physical Review }\textbf{87}: 452-454.

Price R.H. 1972a. Non-spherical pertubations of relativistic gravitational
collapse \textrm{I} scalar and gravitational pertubations. \ \textit{Physical
Review D} \textbf{5}: 2419-2438.

Price R.H. 1972b. Non-spherical pertubations of relativistic gravitational
collapse \textrm{II} Integer spin, zero rest-mass fields. \ \textit{Physical
Review D} \textbf{5}: 2439-2454.

Randall J. \ 1963. \ Aspects of the Life and Work of James Clerk Maxwell. \ In
Domb (1963) pp1-25.

Regge T., Teitelboim C. 1977. \ General Relativity \`{a} la string: a progress
report. In\ \textit{Proceedings of the First Marcel Grossmann Meeting
(Trieste, Italy 1975), }Ruffini R. ed., Amsterdam, North-Holland, pp77-88.

Reissner H. 1916. \"{U}ber die Eigengravitation des elektrischen Feldes nach
der Einsteinschen Theorie. \textit{Annalen der Physik }\textbf{50}: 106--120.

Rice A. 1996. \ Mathematics in the Metropolis: A Survey of Victorian London.
\ \textit{Historia Mathematica }\textbf{23}: 376-417.

Rice A. 2006. \ British mathematics 1837-1901. \ \textit{BSHM Bulletin
}\textbf{21: }164-177

Rindler W. 1956. \ Visual horizons in world models. \ \textit{Monthly Notices
of the Royal Astronomical Society }\textbf{116}: 662-677. \ Reprinted in 2002
in \textit{General Relativity and Gravitation} \textbf{34: }133-153 with an
editorial note by A. Krasi\'{n}ski pp131-132.

Rindler W. 1966. \ Kruskal space and the uniformly accelerated frame.
\ \textit{American Journal of Physics }\textbf{34}: 1174-1178.

Rindler W., Trautman A. 1987. \ \ Introduction. \ \textit{Gravitation and
Geometry (a volume in honour of Ivor Robinson): 9-19. \ Napoli, Bibliopolis.}

Robinson D.C. 1974. Classification of black holes with electromagnetic fields,
\textit{Physical Review D }\textbf{10}: 458-460.

Robinson D.C. 1975a. \ Applications of variational principles to classical
perturbation theory in general relativity. \ \textit{Mathematical Proceeding
Cambridge Philosophical Society} \textbf{78}: 351-356.

Robinson D.C. 1975b. \ Uniqueness of the Kerr black hole, \textit{Physical
Review Letters} \textbf{34}: 905-906.

Robinson D.C. 1977. \ A simple proof of the generalization of Israel's
theorem. \textit{General Relativity and Gravitation} \textbf{8}: 695-698.

Robinson D.C. 2009. \ Four Decades of Black Hole Uniqueness Theorems. In
Wiltshire, Visser \& Scott (2009), pp 115-143. \ With addendum: https://nms.kcl.ac.uk/david.robinson/web\_page/blackholes.pdf.

Robinson D.C. 2016. \ Felix Arnold Edward Pirani. \textit{Physics Today
}\textbf{69} 8: 66.

Robinson D.C., Winicour J. 1971. \ Scaling behaviour of gravitational energy.
\textit{Journal of Mathematical Physics} \textbf{12}: 995-999.

Robinson D.C., Winicour J. 1972. \ Energy of gravitational shock
waves.\textit{\ Journal of Mathematical Physics} \textbf{13}: 1435-1441.

Robinson I., Trautman A. 1960. \ Spherical Gravitational Waves.
\ \textit{Physical Review Letters }\textbf{4: }431-432.

Robinson I., Trautman A. \ 1962. \ Some spherical waves in general relativity.
\ \textit{Proceedings of the Royal Society of London A }\textbf{265}: 463-473.

Rosen N. 1937. \ Plane Polarized Waves in the General Theory of Relativity.
\ \textit{Physikalische Zeitschrift der Sowjetunion }\textbf{12}: 366-372.

Rosen N. 1956. \ On Cylindrical Gravitational Waves. In Mercier \& Kervaire
(1956) pp 171-175.

Rovelli C. 2001. \ Notes for a brief history of quantum gravity.
\ \textit{arXiv: gr-qc/0006061}.

Roxburgh I.W. \ 2007. \ Sir Hermann Bondi KCB. \ \ \textit{Biographical
Memoirs of Fellows of the Royal Society. \ }\textbf{53}: 45-61.

Roxburgh I.W., Saffman P.G. 1965. \ The Growth of Condensations in
a\ Newtonian Model of the Steady State Universe. \ \textit{Monthly Notices of
the Royal Astronomical Society} \textbf{129}: 181-189.

Rudberg H. 1957. \ \textit{The compactification of a Lorentz space and some
remarks on the foundation of the theory of conformal relativity.
\ }Dissertation, University of Uppsala. \ \textit{Physics Abstracts No. 30,
}\textbf{61} (1958).

Ruse H.S. 1946. A.G.D.Watson's principal directions for a Riemannian V$_{4}$.
\ \textit{Proceedings of the Edinburgh Mathematical Society }\textbf{7}: 144-152.

Russell-Clark R.A. 1973. \ \textit{The application of algebraic manipulation
by computer to some problems in gravitational radiation theory}. \ PhD Thesis,
University of London.

Sachs R.K. 1961. \ Gravitational waves in general relativity VI. \ The
outgoing radiation condition. \ \textit{Proceedings of the Royal Society of
London} A\textbf{ 264}: 309-338.

Sachs R.K. 1962a. \ Gravitational waves in general relativity VIII. \ Waves in
asymptotically flat space-time. \ \textit{Proceedings of the Royal Society of
London} A\textbf{ 270}: 103-126.

Sachs R.K. 1962b. \ Asymptotic symmetries in Gravitational Theory.
\ \textit{Physical Review }\textbf{128}: 2851-2864.

Sachs R.K. 1964a. \ The characteristic initial value problem for gravitational
theory. \ In Infeld (1964) pp 93-105.

Sachs R.K. 1964b. \ Gravitational Radiation. \ In DeWitt C. \& DeWitt B.
(1964) pp523-562.

Salam A. \ 1975. \ Impact of quantum gravity theory on particle physics. \ In
Isham, Penrose, Sciama pp 500-537.

S\'{a}nchez-Ron J.M. 1992. \ The Reception of General Relativity Among British
Physicists and Mathematicians (1915-1930). \ In Eisenstaedt \& Kox \ pp57-88.

S\'{a}nchez-Ron J.M. 2005 \ George M$^{\text{c}}$Vittie: The Uncompromising
Empiricist. \ In Kox \& Eisenstaedt (2005) pp189-221.

Sauer T. 2004. \ Albert Einstein's 1916 Review Article on General Relativity.
\ \textit{arXiv: physics/0405066v1.}

Saulson P.R. 2011. \ Josh Goldberg and the physical reality of gravitational
waves. \textit{\ General Relativity \& Gravitation} \textbf{43}: 3289-3299.

Scheidegger A.E. 1953. \ Gravitational Motion. \ \textit{Reviews of Modern
Physics }\textbf{25}: 451-468.

Schilpp P.A. Ed. \ 1970 \textit{Albert Einstein, philosopher-scientist.}
\ Illinois, Open Court-La Salle-Illinois.

Schutz B.F. 2012. \ Thoughts About a Conceptual Framework for Relativistic
Gravity. In Lehner, Renn and Schemmel (2012), pp259-272.

Schwarzschild K. 1916a. \ \"{U}ber das Gravitationsfeld eines Massenpunktes
nach der Einstein'schen Theorie. \textit{Reimer, Berlin Sitzungsberichte der
K\"{o}niglich-Preussischen Akademie der Wissenschaften.} S. 189 ff.

English translation 2008. \ On the Gravitational Field of a Point-Mass,
According to Einstein's Theory. \textit{The Abraham Zelmanov Journal}\textbf{
1}: 10-19.

Schwarzschild K. 1916b. \ \"{U}ber das Gravitationsfeld einer Kugel aus
inkompressibler Fl\"{u}ssigkeit. \textit{Reimer, Berlin Sitzungsberichte der
K\"{o}niglich-Preussischen Akademie der Wissenschaften. \ S. 424-434.}

English translation 2008. On the Gravitational Field of a Sphere of
Incompressible Liquid, According to Einstein's Theory. \ \textit{The Abraham
Zelmanov Journal}, 2008 \textbf{1}: 20-32.\textit{ }

Sciama D. 1959. \ \textit{The unity of the Universe.} London, Faber and Faber.

Sciama D. 1962. \ On the analogy between charge and spin in general
relativity. \ In \textit{Recent developments in general relativity. } New
York, Pergamon Press. \ pp415-439.

Silvester J.R. 2010. \ C.W.Kilmister 1924-2010. \ \textit{The Mathematical
Gazette }\textbf{94: }529-531.

Simms D.J., Woodhouse N.M.J. 1976. \textit{Lectures on geometric quantization.
\ }Berlin, Springer-Verlag.

Stelle K.S. \ 1977. \ Renormalization of higher derivative quantum gravity.
\ \textit{Physical Review D }\textbf{16: }953-969.

Stelle K.S. 1978. Classical Gravity with Higher Derivatives. \ \textit{General
Relativity and Gravitation} \textbf{9}: 353--371.

Stelle K.S., West P.C. 1978a. \ Minimal Auxiliary Fields for Supergravity.
\ \textit{Physics Letters B} \textbf{74}: 330-332.

Stelle K.S., West P.C. 1978b. \ Tensor Calculus for the Vector Multiplet
coupled to Supergravity. \ \textit{Physics Letters B} \textbf{77}: 376-378

Stelle K.S., West P.C. 1978c. \ Relation between vector and scalar Multiplets
and gauge invariance in supergravity. \ \textit{Nuclear Physics }B
\ \textbf{145}: 175-188.

Stelle K.S., West P.C. 1979. \ de Sitter gauge invariance and the geometry of
the Einstein-Cartan theory. \ \textit{Journal of Physics A }\textbf{12}: L205-L210.

Stelle K.S., West P.C. 1980. \ Spontaneously broken de Sitter symmetry and the
gravitational holonomy group. \ \textit{Physical Review D }\textbf{21: }1466-1488.

Stephani H., Kramer D., Herlt E., MacCallum M., Hoenselaers C. 2003.
\ \textit{Exact Solutions of Einstein's Field Equations. \ }Cambridge,
Cambridge University Press.

Strominger A. \ 2018. \ \textit{Lectures on the Infrared Structure of Gravity
and Gauge Theory. \ }Princeton, Princeton University Press.

Synge J.L. 1955. \ Relativity: \textit{The Special Theory}. \ Amsterdam,
North-Holland Publishing Company.

Synge J.L. 1960. \ Relativity: \textit{The General Theory}. \ Amsterdam,
North-Holland Publishing Company.

Synge J.L., Schild A. 1949. \ \textit{Tensor Calculus. \ }Toronto, University
of Toronto Press.

Szekeres G. 1960. \ On the singularities of a Riemannian manifold.
\ Publicationes Mathematicae Debrecen \textbf{7}: 285-301. \ Republished with
an editorial note by his son Peter Szekeres in \textit{General Relativity and
Gravitation }\textbf{34: }1995-2016 (2002).

Szekeres P. 1965. \ The Gravitational Compass. \ \textit{Journal of
Mathematical Physics }\textbf{6: }1387-1391.

Szekeres P. 1966. \ On the Propogation of Gravitational Fields in Matter.
\ \textit{Journal of Mathematical Physics }\textbf{7: }751-761.

Szekeres P. 1970. Colliding gravitational waves. \ \textit{Nature
}\textbf{228}: 1183-1184.

Szekeres P. 1972. \ Colliding gravitational waves. \ \textit{Journal of
Mathematical Physics }\textbf{13}: 286-294.

Temple G. 1923. \ A generalisation of professor Whitehead's theory of
relativity. \textit{Proceedings of the Physical Society of London}
\textbf{36}: 176-193.

Temple G. 1924. \ Central orbits in relativistic dynamics treated by the
Hamilton-Jacobi method. \ \textit{Philosophical Magazine} \textbf{48}: 277-292.

Temple G. 1938 \ New systems of normal coordinates for relativistic optics.
\textit{Proceedings of the Royal Society of London A }\textbf{168}: 122-148.

Temple G. 1939. \ Relativistic cosmology. \ \textit{Proceedings of the
Physical Society (London) }\textbf{51: }465-478.

Temple G. 1981. \ \textit{100 years of mathematics}. London, Duckworth.

Temple G., Flint H.T. 1967. \ William Wilson 1875-1965. \ \textit{Biographical
Memoirs of Fellows of the Royal Society }\textbf{13}: 386-391.

Thompson A.H. 1962. \ \textit{The investigation of a set of weakened field
equations for general relativity. \ }PhD thesis, University of London, King's College.

Thompson F.M.L. Ed. 1990. \ \textit{The University of London and the World of
Learning 1836-1986. }\ London\textit{, The Hambledon Press.}

Titchmarsh E.C. 1958. \ George Barker Jeffery 1891-1957.
\ \textit{Biographical Memoirs of Fellows of the Royal Society }\textbf{4}: 128-137.

Tolstoy I. 1981 \textit{James Clerk Maxwell \ A Biography. \ }Edinburgh, Canongate.

Trautman A. 1958a. \ \textit{Lectures on General Relativity, mimeographed
notes, King's College, London May-June 1958}. Reprinted with an editorial note
by P.Chru\'{s}ciel in \textit{General Relativity and Gravitation 2002
}\textbf{34}:\textbf{ }721-762.

Trautman A. 1958b. \ Boundary conditions at infinity for physical theories.
\ \textit{Bulletin De l'Academie Polonaise Des Sciences. Serie Des Science
}\textbf{6}: 403-406.

Trautman A. 1958c. \ Radiation and boundary conditions in the theory of
gravitation. \textit{\ Bulletin De l'Academie Polonaise Des Sciences. Serie
Des Science }\textbf{6}: 407-410.

Trautman A. 1962. Conservation laws in\textit{ }General Relativity. \ In
Witten pp 69-198.

Trautman A. 1965. \ Foundations and current problems of general relativity.
\ In Deser and Ford, Vol. 1 pp1-248.

Trautman A. 1966. \ The General Theory of Relativity. \ \textit{Soviet
Physics, a translation of Uspekhi Fizicheskikh Nauk }\textbf{89}: 319-399.

Unruh W.G. 1976. \ Note on black hole evaporation. \ \textit{Physical Review
D} \textbf{14}: 870-892.

van der Burg M.G.J. 1959. \ \textit{Axisymmetric solutions in general
relativity. \ }PhD thesis, University of London, King's College.

van der Burg M.G.J. 1966. \ Gravitational Waves in General Relativity
\textrm{IX}: Conserved Quantities. \ \textit{Proceedings of the Royal Society
of London }\textbf{294}: 112-122.

van der Burg M.G.J. 1969. \ Gravitational Waves in General Relativity
\textrm{X}: Asymptotic Expansions for the Einstein-Maxwell field.
\ \textit{Proceedings of the Royal Society of London }\textbf{310}: 221-230.

von Freud P. 1939. \ Uber die Ausdrucke der Gesamtenergie und des
Gesamtimpulses eines Materiellen Systems in der Allgemeinen
Relativitatstheorie. \ \textit{Annals of Mathematics} \textbf{40}: 417-419.

Walker M. 1979 \ Remark on Trautman's Radiation Condition. \ In Ehlers. \ pp 61-62.

Weber J. 1961. \ \textit{General Relativity and Gravitational Waves. \ }New
York, Interscience Publishers Inc.

Weber J. 1969. \ Evidence for discovery of gravitational radiation.
\ \textit{Physical Review Letters }\textbf{22}: 1320-1324.

Weber J. 1970. \ Gravitational Radiation Experiments.  In Carmeli, Fickler,
Witten.  pp133-143.

Weyl H.\ 1922. \ \textit{Space-time-matter}. \ London, Methuen and Company Ltd.

Whitehead A.N. 1922. \ \textit{The principle of relativity, with applications
to physical science.} Cambridge, Cambridge University Press.

Will C. M. 1993. \ \textit{Theory and Experiment in Gravitational
Physics}\textbf{.} Cambridge, Cambridge University Press, 2nd edition.

Wilson R.J. 2017. \ The Gresham Professors of Geometry Part \textrm{I:} the
first one hundred years. \ Part \textrm{II:} the next three hundred years.
\ \textit{Bulletin of the British Society for the History of Mathematics
}\textbf{32}: 125-148.

Wilson W. 1918 \ Relativity and gravitation. \ \textit{Proceedings of the
Physical Society of London }\textbf{31}: 69-78.

Wilson W. 1928 \ Relativity and wave mechanics. \textit{\ Proceedings of the
Royal Society of London} A\textbf{ 118}: 441-448.

Witten L. 1962. \ \textit{Gravitation: an introduction to current research}.
\ New York, John Wiley and Sons Inc.

Wiltshire D.L., Visser M., Scott S.M. Eds. 2009. \ \textit{The Kerr Spacetime:
Rotating Black Holes in General Relativity. \ }Cambridge, Cambridge University Press.

Woodhouse N.M.J. 1973. \ The differential and causal structures of space-time.
\ \textit{Journal of Mathematical Physics}\textbf{ 14: }495-501.

Woodhouse N.M.J. 1979. \ \textit{Geometric Quantization}. \ Oxford, Clarendon Press.

Wright A.S. 2014. \ The Advantages of Bringing Infinity to a Finite Place:
Penrose Diagrams as Objects of Intuition. \ \textit{Historical Studies in the
Natural Sciences} \textbf{44}: 99-139.

Yang C.W. 1974. \ Integral formalism for gauge field. \ \textit{Physical
Review Letters} \textbf{33}: 445-447.

\end{document}